\documentclass[aps,pre,twocolumn,showpacs,epsfig,epsf,amssymb,floatfix]{revtex4-1}
\usepackage{graphicx}
\usepackage{amssymb}
\usepackage{amsmath}
\usepackage{subfigure}
\usepackage{booktabs}
\usepackage{float}
\usepackage{longtable}

\begin{document}

\title{Role of special cross-links in structure formation of bacterial DNA polymer.}
\author{Tejal Agarwal$^{1}$, G.P. Manjunath$^2$, Farhat Habib$^3$, Pavna Lakshmi$^4$, Apratim Chatterji$^{1,5}$}
\email{apratim@iiserpune.ac.in}
\affiliation{
$^1$ IISER-Pune, 900 NCL Innovation Park, Dr. Homi Bhaba Road,  Pune-411008, India.\\
$^2$ IISER Mohali, Knowledge city, Sector 81, SAS Nagar, Manauli-140306, India.\\
$^3$ Inmobi, Cessna Business Park, Outer Ring Road, Bangalore-560103, India.\\
$^4$ Cologne Graduate School of Ageing Research, University Hospital-Cologne, Germany.\\
$^5$ Center for Energy Science, IISER-Pune,  Dr. Homi Bhaba Road,  Pune-411008, India.
}

\date{\today}
\begin{abstract}
Using data from contact maps of the DNA-polymer of {\em E. Coli}  (at kilobase pair resolution) as an input to our model,
we introduce cross-links between monomers in a bead-spring model of a ring polymer at very specific points along
the chain. By suitable Monte Carlo Simulations, we show that the presence of these cross-links leads to a
particular architecture and organization of the chain at large (micron) length scales of the DNA.
We also investigate the structure of a ring polymer
with an equal number of cross-links at random positions along the chain. We find that though the polymer does
get organized at the large length scales, the nature of the organization is quite different from the organization
observed with cross-links at specific biologically determined positions. We used the contact map of {\em E. Coli}  bacteria
which has around $4.6$ million base pairs in a single circular chromosome. In our coarse-grained flexible 
ring polymer model, we used $4642$ monomer beads and observed that around $80$ cross-links are enough 
to induce the large-scale organization of the molecule accounting for
statistical fluctuations caused by thermal energy.
The length of a DNA chain of an even simple bacterial cell such as {\em E. Coli} is much longer than typical proteins, hence
we avoided methods used to tackle protein folding problems.
We define new suitable quantities to identify large scale structure of a polymer chain with a few cross-links.

\end{abstract}

\maketitle
\section{Introduction}
The organization of chromatin at mesoscopic length ($>30 nm$) scales has been a topic of intense
research in this decade \cite{dekker,aiden,joyeux,job,Wendy,dixon,mirny1,debashish2,kremer,alexgros,hofmann,rama,geoffrey,rosa,nicodemi,nicodemi2,balaji}, 
specially after the work of Liebermann Aiden et. al. \cite{aiden} where the authors 
mapped out the spatial proximity maps of DNA segments of human genome (each segment of length 1 Mega Base Pairs) 
inside nucleus using a technique called Hi-C: high-throughput sequencing.
The experimental studies provide a {\em contact map} of DNA segments \cite{aiden,laub,Wendy,dixon,cagliero,jennifer}.
A contact map is a color map that shows which DNA segments, numbered $i= 1,2,3...N_D$ are spatially close to other
DNA segments j ($j=1,...,N_D$) with high/low frequency.
The question is whether with this information, can one predict the spatial organization of the DNA chain which is expected
to help in identifying the biological consequences.

The physics approach is of course to consider chromatin as a polymer chain, and the
chromatin within the nucleus as a collapsed polymer coil \cite{job,mirny1,mirny2,mirny3,alexander,nicodemi,nicodemi2,sachs,davide1,bohn,mariano,geoffrey,geoffrey1,anton,colby}. 
The resolution of Hi-C experiments has increased to 1 kilo-BP (kilobase pair) which is still above 
the persistence length of naked ds-DNA (approximately, $50$ nm with 150 BPs) as reported for bacterial cells
\cite{laub}.  However, inside cells around 150 BPs
of DNA wrap around histone-like proteins (for bacterial cells) to form higher order structure \cite{rob}, 
and the persistence length of a DNA polymer   
chain is still debated in vivo \cite{Maeshima}. 

 Anyways,  DNA organization at {\em large} length scales can be viewed as the 
organization of a flexible polymer.  The {\em large} length scales of ds-DNA in question are $100$nm-microns, 
such that a DNA-segment consisting of a kilo to mega BPs can be considered as a coarse-grained monomer in 
a bead-spring model of the polymer chain.  Our work aims to elucidate the structure of chromatin at this length scale, or 
equivalently that of a flexible polymer with added constraints.  
Generically we  show  that adding spatial constraints by cross-linking  a minimal number of
{\em specific}  monomers along the length of chain can lead to the organization of an entire long ring-polymer  
into a specific structure, but there are structural fluctuations due to $k_BT$.  

Much of the research is focussed on structure and organization of the chromatin during interphase stage 
\cite{mirny1,mirny2,mirny3,kremer,rosa,mateos}: the stage of the cell cycle when the cell does not divide 
into daughter cells.  It is also known that the individual chromosome is not arranged as a random 
walk polymer.  From the data of contact map, we observe that some DNA segments have a much
higher spatial association with other chain segments  and show up as
the presence of so-called Topologically Associated Domains (TADs) \cite{aiden,laub,dixon} in the contact map.
An actual chromatin  is not just a long polymer chain within a nucleus or within a cell (for bacteria)
but there are also various proteins and enzymes doing various functions of the cell.
For example, there are  DNA-binding proteins which attach two different and specific segments of DNA chain together
and the enzyme topoisomerase which allows chains to cross each other by suitably cutting and rejoining chains \cite{alexander}.
Polymer physics principles are suitably adapted to incorporate effects of proteins and enzymes 
related activity when investigating the origin or reasons of formation of 
TADs \cite{mirny1,mirny2,mirny3,kremer,rama,dekker}. Furthermore, when studying DNA-polymer organization in the 
interphase stage, physicists (and we) assume the system to be in a state of local equilibrium so that 
principles of statistical mechanics can be applied.

Studies have shown that the organization of chromatin is a fractal globule
rather than an equilibrium coil \cite{mirny2,mirny3}. 
The understanding is that segments of DNA get locally collapsed to form {\em unentangled}
crumpled sections of the coil, such that within a segment there are many contacts, and fewer number of contacts
between collapsed neighboring segments.  These then show up as TADs in the contact map. 

In the last  years, there  have been more detailed polymeric models which can reproduce the 
experimentally measured TADs for sections of DNA. The most successful of them are 
the SBS (Strings and Binders) model \cite{nico1,marenduzzo} and the loop extrusion model
\cite{mirny_cell_reports,alipour,naumova_13}. 
In the SBS model, monomers along a chain have the same size but
have distinct affinities of attraction for freely diffusing  binder-molecules. There are as many distinct 
kinds of binder molecules as  different kinds of monomers. Monomers of the same kind but separated along the chain contour
can get attached to the same binder molecule to result in the formation of loops. Some parameters,
such as the number of different kind of monomers or the number of monomers of each kind,  can be optimally chosen
to reproduce and fit the TADs of a particular segment of a DNA by solving a multidimensional optimization 
problem. In the loop extrusion model, there are boundary elements (BE monomers) at specific 
sections along the chains. A pair of special monomers (LE-monomers), which probabilistically bind with each other 
to extrude loops of variable lengths by diffusing/translocating along the length of the chain, 
but LE-monomers are constrained to remain bounded between the  BE-monomers. Again, a search through a large  
parameter space leads to optimal TADs with a quantitative match with  experimental data. 
Both models seem to crucially depend on the formation and contact between suitably sized loops at appropriate 
locations, which in turn results in a match with experimental contact map data.
Other researchers \cite{mathieu,davide} use optimization algorithms with weighted constraints to get an
idea of the large scale structure of bio-molecules.

Instead of investigating the origin of TADs where some headway has already been made , we ask a different question.
Given the contact map, can we predict the global spatial organization of a polymer? Is there even 
any organization of the entire macromolecule? Note that the contact map does 
not give information of {\em spatial} organization of polymer, it just gives the frequency  of finding 
different DNA-sections in proximity.  In this study, We assume that polymer-sections with the highest 
frequencies of contact to be permanently cross-linked  and investigate if cross-links (CLs) above a minimal number and at 
special biologically well-determined locations along the chain (as determined from the contact map) play a vital
role in giving shape and structure to the entire DNA-polymer. 
We compare the organization of a polymer with CLs at biologically determined positions along the 
contour (Bio-cross-links: BC) with  polymer organization with 
equal number of cross-links between monomers, but the monomers to be cross-linked are chosen at random (random cross-links
RC). The question is whether an equal number of minimal constraints at random positions give "structure" to the polymer?
We generate ten independent random configurations of cross-links (CLs) and compare "structure" of polymers
for ten independent RCs and 1 BC.

It is easier to work with simpler
systems, e.g. DNA of bacteria such as Escherichia coli ({\em E. coli}) which is a ring polymer.
Bacterial cells have no nucleus, the number of chromosomes is typically $1$ or $2$ per cell, and the DNAs are much
shorter. Bacterial DNA also shows TADs, and they have DNA-binding proteins \cite{dixon}.  
We choose a bead-spring flexible polymer model of {\em E. Coli}, a bacteria with a single chromosome for our 
studies.  The question is how do we determine if a polymer, which is expected to 
be unstructured,  is structurally organized or not.

We start our Monte Carlo (MC) simulations from $9$ independent initial configurations of ring polymers,
without taking into account that cross-linked monomers should be in proximity. However,  
cross-link potentials are applied between   monomer pairs from a particular CL-set. 
We then allow the chain to relax to equilibrium for each case 
in  $9$ independent MC runs.  If we observe that the DNA polymer relaxes to almost the
same ``structure'' within statistical fluctuations in each case, then we could claim that the 
polymer is organized. Also, we try out with fewer CLs and check the minimal number of CLs required
to achieve organization of the polymer. To check for the organization, we had to come up with different structural 
quantities which we describe in the next sections.
Our hypothesis of special cross-links in DNA-polymers can be further established by testing it on more 
bacterial chromosomes. 

We have not included confinement effects due to cell walls due to two reasons: (a) the bacterial
DNA is within the nuceloid region which occupies only $15$\% to $25$\% of the cell volume \cite{joyeux} (b) we wanted
to focus only on the effects of having CLs at specific locations unencumbered by any other competing effects.
Also, we do not put effects of supercoiling in our simple bead-spring polymer model because proximity
effects between segments due to supercoiling, if any, should show up in the contact maps.

The organization of the manuscript is as follows: the next section, viz., section II discusses the model
of DNA-polymer,  the computational method by which we generate initial conformations, methodology
 to relax polymers to local equilibrium using after which we calculate ensemble averages of statistical quantities
to identify spatial organization of the polymer. The next section, section III, discusses the statistical
quantities and our results by which we arrive at our conclusions.  We end with
summarizing our conclusions section IV.

\section{Model and simulation method.}
We use Monte Carlo simulations to explore the different microstates of the DNA chain. The DNA of bacteria 
{\em E. coli} is a ring polymer.
We model {\em E. Coli} DNA with a bead-spring model of ring polymer with  $N_{EC} = 4642$ number of monomers in the 
ring.  Thus, each coarse grained monomer bead  in our model represents  $10^3$ BPs. The  DNA model-polymer is placed 
near the center of the simulation box of size $(200 a)^3$ with periodic boundary conditions (PBC). The quantity 
$a$ is the unit of length and is the average distance between two neighboring monomer beads along the chain contour.
The box size $L=200 a$ is chosen to be much larger than the expected maximum diameter of the polymer coil.
The diameter $\sigma$ of monomer bead is chosen to be $0.2a$. The Lennard-Jones potential, suitably truncated 
at $r= 2^{1/6} \sigma$ and shifted (the Weeks Chandler Andersen-WCA potential), is used to model the excluded
volume interaction between the monomers. A harmonic spring potential connects adjacent monomers along the chain contour.
$V(r) = \kappa (| \vec{r} | - a)^2$,  where $| \vec{r} | $ is the distance between 
two monomers and $a$ is the unit of length. 

We have chosen $\kappa = 200 k_BT$, where the thermal energy $k_BT=1$ is the unit of energy in our simulations. 

Using data from \cite{cagliero} and subsequent analysis methods which are described in detail in the Appendix-1, 
we obtained the frequency of finding two segments of {\em E. Coli}-DNA spatially close to each other.
We use this data from contact maps as an input to our simulations.
The experimental resolution of the size of segments is $10^3$ base pairs. The model monomer in our simulations represents
a DNA segment exactly of the same size as the experimental resolution. 
We cross-link  monomers  whose frequency  of being in spatial proximity is greater than
threshold frequency $p_c$. 
Depending on the value of threshold frequency that we choose, we can have (a) $ N_{CL} =47$ or (b) 
$N_{CL}=159$ pairs of  monomers of the DNA-polymer that we cross-link. 
We bind these pairs of monomers together by an additional spring potential $V_c=\kappa_c(r-a)^2$ with $\kappa_c=
20k_bT$. The cross-linked monomers  are held together at a distance of $a$, but the different CLs can  move 
with respect to each other as the chain explores different conformations.

The set of $47$ CLs from biological contact-map data, which we refer to as BC-1 in the rest of the paper,
are a subset of $159$ CLs which we call BC-2.
To analyze whether the overall mesoscale organization of the chain is determined primarily by a
particular choice of CLs, we start our simulations from $9$ independent initial conditions.  
For example, in one of the initial conditions, the monomers of the ring polymer
are arranged along a circle of radius $30.73a$ such that one circle has $193$ monomers. 
The circle of monomers  are stacked up to form a cylinder.
Note that this will lead to monomer numbered as  $1$ and  the last monomer 
$N=4642$ to be at a distance much larger than $a$ though it is a ring polymer. 
Also,  the monomers which form CLs can be at distances much larger than $a$ as they get arranged along 
the cylinder. But these will come closer due to the presence of harmonic spring potentials acting between monomer-pairs
as the polymer is allowed to relax during the MC run.
In two other initial conditions, we arrange the monomers in circles of radius 
$40.92a$ and $36.94a$. For the next  three initial conditions,
 the monomers are arranged in squares of side $90a$, $80a$ and $70a$; these squares are stacked then up.
For the last three initial conditions, we arrange monomers in equilateral triangles of side $40a$,
$50a$, $60a$ and stack them to form a vertical column. By such initial conditions, we ensure that the monomer pairs which
constitute a CL are at arbitrary positions relative to each other in space.

The question we then ask is: as the chains relax from their initial conditions to their equilibrium conformations 
in different Monte Carlo runs, do all of them organize themselves in some particular 
set of conformations (in a statistical sense), though the initial configurations were very different? 
If they do, we can expect that the presence of CLs to play a significant role in the organization, since a normal
ring polymer is not expected to show structural organization.

We use  additional techniques to allow the chain to relax slowly over $10^5$ Monte Carlo iterations
to its equilibrium state without allowing the system to get stuck in some entangled and metastable 
state. We set spring constant of cross-links $\kappa^{initial}_c =0.01 \kappa_c $
at the start of the simulation and gradually ramp it in steps of $0.01 k_c$ every $1000 $ MC steps, 
as the CL monomers approach each other in the relaxation process. 
In a standard Metropolis step, a monomer attempts a displacement 
$\vec{dr} = \delta (r_1 \hat{i} + r_2 \hat{j} + r_3 \hat{k})$ in a random direction, where $\delta = 0.2 a$
and $r_1,r_2, r_3$ are random numbers. The attempt is accepted with Boltzmann probability. In addition, every 
$100$ iterations, we attempt displacements with $\delta = 1.2 \sigma$. This helps chains to cross each other at times
and overcome topological constraints which might arise as the chain relaxes from its initial condition.

\begin{figure}[!hbt]
\includegraphics[width=0.49\columnwidth]{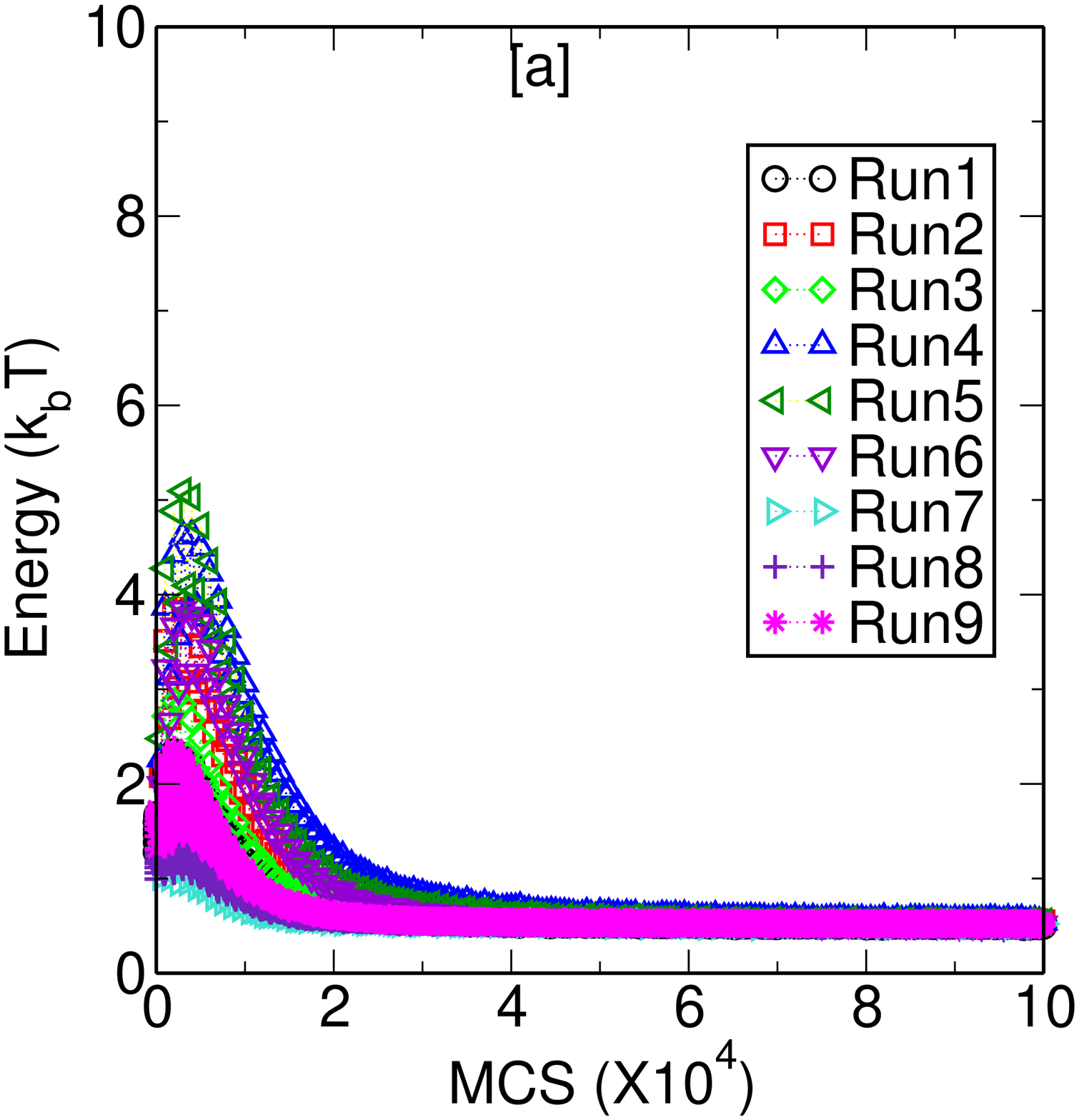}
\hfill
\includegraphics[width=0.49\columnwidth]{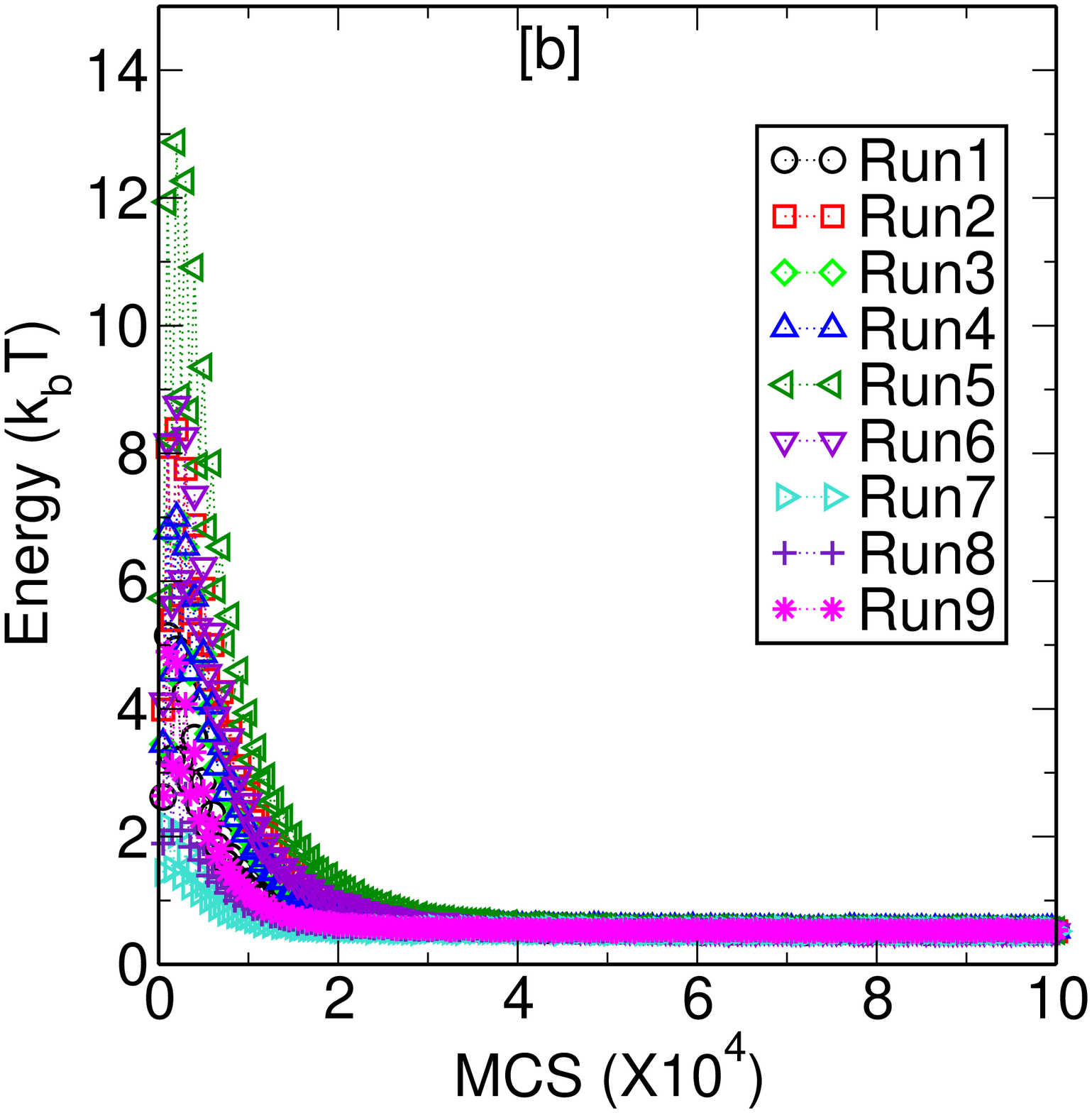}
\caption{\label{fig2}  The plot of the energy versus Monte Carlo steps (MCS) as the systems evolve to 
a relaxed state  from the nine different initial conditions.  The energies all converge and fluctuate around the same value 
establishing that the chains are unlikely to be stuck in some metastable states due to topological constraints arising from 
excluded volume interactions. Subplot (a) and (b) are for $47$ and $159$ CLs respectively.
}
\end{figure}

We monitor the potential energy as the chain relaxes. The value of energy relaxes to the same 
value at the end of $10^5$ iterations from the $9$ different runs, see Figure \ref{fig2}.  It gives us confidence
that the chain conformations are not stuck at metastable energy minima. From this initial state,
we evolve each of the $9$ different  chain conformations in independent simulation runs over the next 
$12 \times 10^6$ iterations and collect data to calculate and compare structural quantities. 
We carry out this comparison of statistical data from $9$ independent runs for 
each set of CLs, viz., chains with (a) $47$ (b)  $159$  CLs.

In addition, we also carry out similar calculations starting from $9$ independent initial configurations  
for  each of the $10$ distinct sets of the randomly chosen position of CLs (monomers which are cross-linked 
together are chosen randomly from the list of monomers). The effective number of CLs for RC-1 and RC-2 correspond
to the number of CLs in BC-1 and BC-2. In BC-1 and BC-2 set of CLs, 
there are some CLs which are not independent.  For example, monomer number $16$ and $17$ are 
cross-linked to monomer $2515$ and $2516$, respectively.  One cannot consider them as distinct CLs.
The list of cross-linked monomers is given in  Table-1 of Supplementary section TABLE $I$. 
Hence there are  fewer {\em effective} CLs than the number of CLs in BC-1 and BC-2.  Thus we compare the 
results of our simulation from bio-CLs (BC-1 and BC-2) with an equal number of  {\em effective} random CLs.
In each of these random set of CLs, we have the same number of {\em effective} CLs as the ones obtained
from biological data, which is less than the corresponding number of CLs in BC-1 and BC-2.  
Hence the list of randomly positioned CLs have just (a) $27$ effective number of CLs (we refer these as 
RC-1) and (b) $82$ effective CLs (referred as RC-2), corresponding to $47$ CLs in BC-1 and 
$159$ CLs in BC-2.  We can now  compare  structural data obtained from polymer simulations using BC-1 
and RC-1 on the one hand, and BC-2 and RC-2 on the other.

\section{Results}
Now we discuss the statistical quantities which we use to investigate the structure and conformation of 
the ring polymer.  We  aim to check if statistical quantities from $9$ different runs with the same set of CLs 
give similar results to  infer that the polymer has similar shape and conformation across runs.
We further compare data from $10$ different RC-1 and RC-2 CL sets with data from {\em E.Coli} CL set BC-1 and BC-2, though
in this manuscript we show data primarily from one representative RC set.

The first quantity we want to estimate is the size and extent of the polymer with CLs. 
To that end, we calculate the moment of inertia tensor ${\mathbf I}$ 
with respect to the center of mass (CM) of the polymer coil and diagonalize the matrix to get its principal 
moments for each microstate. We then calculate the average principal moments $I_1,I_2,I_3$. where $I_1$ is 
the largest eigenvalue and $I_3$ the smallest. 

\begin{figure}[!tbh]
\includegraphics[width=0.49\columnwidth]{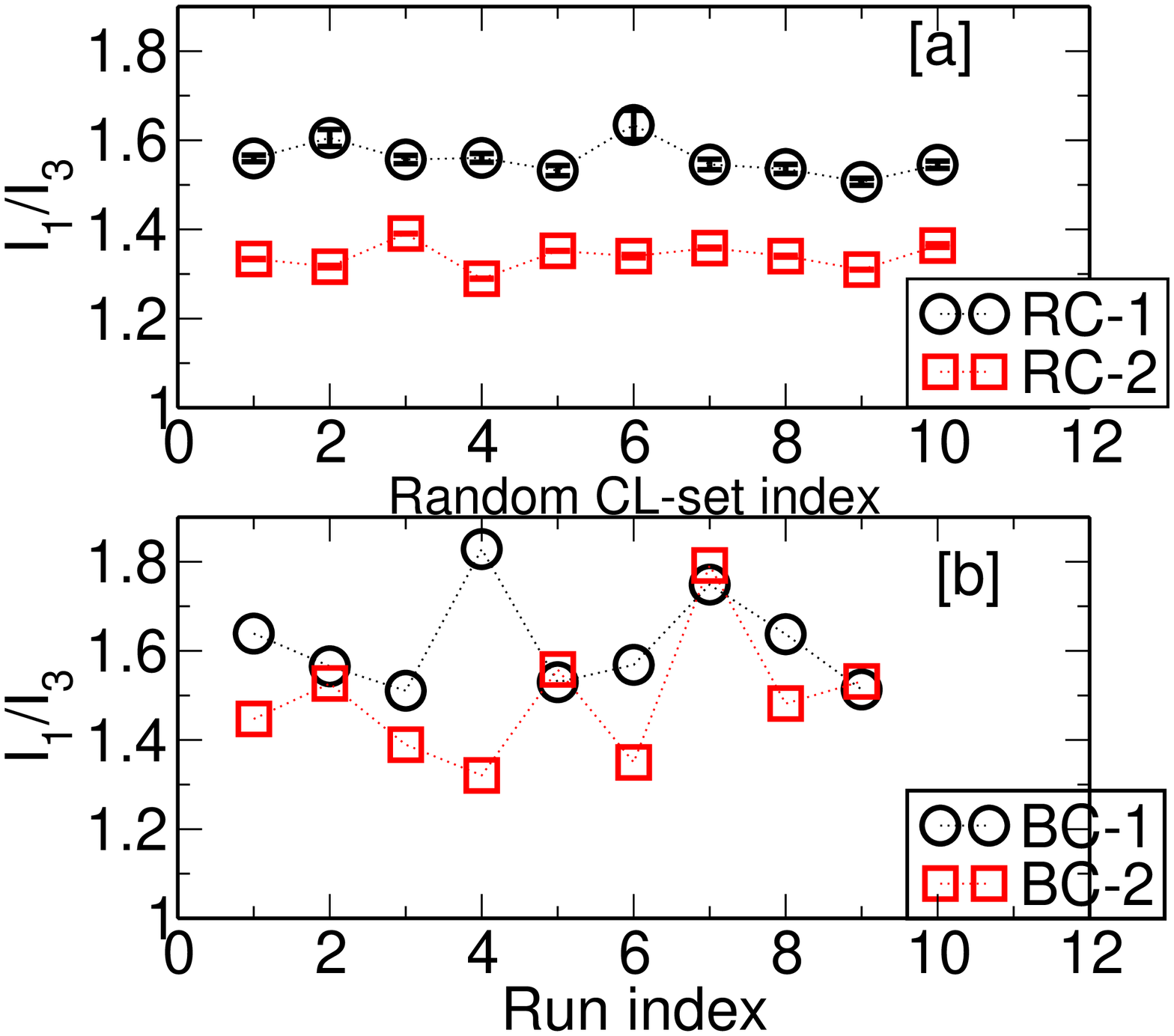}
\hfill
\includegraphics[width=0.49\columnwidth]{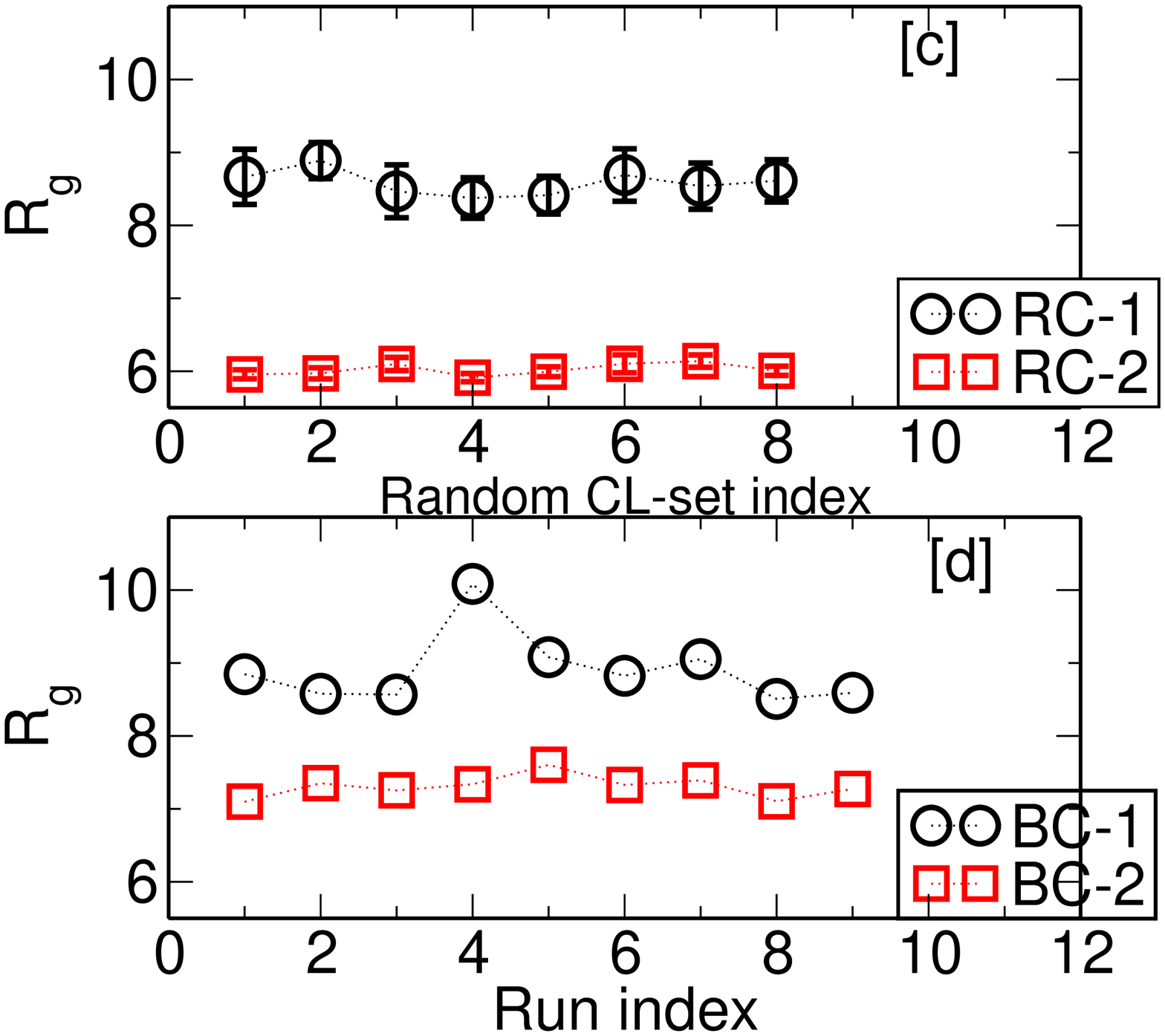} \\

\caption{\label{figI1}  (a) The value of $I_1/I_3$, the ratio of the largest and lowest eigenvalues of the  diagonalized 
moment of inertia matrix for distinct random CL sets (no. of CLs for each RC-set is equal to that for RC-1 and RC-2,
 respectively). For each random CL-set index, the plot shows the average and s.d. of $I_1/I_3$ over $9$ 
independent initial conditions; the s.d. is smaller than the size of the symbol.
(b) The value of $I_1/I_3$ versus run index for runs starting from $9$ independent initial conditions
 for biologically determined CLs: BC-1 and BC-2.
(c) The plot of Radius of gyration $R_g =\sqrt{(I_1 + I_2 + I_3)/3M)}$)
versus the random CL-set index. (d) Plot of $R_g$ and run index from 9 independent initial conditions 
for biologically determined CLs: BC-1 and BC-2.     
}
\end{figure}

In Fig \ref{figI1}(a) we show the values of $ (I_1/I_3)$ for distinct random CL-sets 
but having the same number of CLs as RC-1 and RC-2. 
For each random CL-set, the average is taken over $9$ independent initial configurations. 
In subplot (b) we show $I_1/I_3$ for Biologically determined CLs: BC-1, BC-2 for 9 independent initial conditions. 
In plot (c) $R_g = \sqrt{(I_1 + I_2 + I_3)/3M}$ for different random CL sets having same number of CLs as RC-1 and RC-2 is 
shown and in subplot (d) we show $R_g$ for $9$ independent initial conditions for BC-1 and BC-2 respectively. 
Here $M$ is the sum of masses of the individual monomers $M = \sum m_i$, $m_i =1$ is the mass of each monomer.    
The value of $I_1/I_3$ is the ratio of major and minor axes and gives a measure of shape asymmetry of the coil.
Comparing the value of $I_1/I_3$ in \ref{figI1}(a), (b) we see that  $I_1/I_3$ has a lower value for all 
ten RC-2 sets compared to BC-2 set.
A plausible explanation for this difference is given later in this paragraph  and confirmed by the end of this paper.
Subplots Fig \ref{figI1} (b) and (d)
show the values of $R_g$ obtained from randomly determined CLs and Biologically determined CLs. The calculated  value of $R_g$ 
for ring polymer without CLs and the average value is $\approx 12$. 
The value of $R_g$ decreases as we increase the number of CLs from  BC-1/RC-1 set to BC-2/RC-2 sets;
this decrease in the value  with increase in the number of {\em effective} constraints is expected. 
But interestingly, the change in the value of $R_g$ as we go from BC-1 to BC-2 is distinctly less
than the decrease in $R_g$ as we go from RC-1 to RC-2. 
We interpret the difference  between the two cases as follows: 
the {\em effective} CLs in BC-1 are already at critical positions along the contour which give partial  organization in the DNA. 
On increasing the number of CLs (BC-2), the organization of the molecule improves along the already established 
framework. On the other hand, an increase in the number of random CLs  leads to an overall shrinkage in the size of the coil
and not necessarily to accentuate a preferred set of conformations. 
The lower values of $I_1/I_3$ for all 10 sets of RC-2 compared to that in BC-2 also point towards such an understanding.
This idea will get  further substantiated in the rest of the paper.

\begin{figure}[!htb]
\includegraphics[width=0.4\columnwidth]{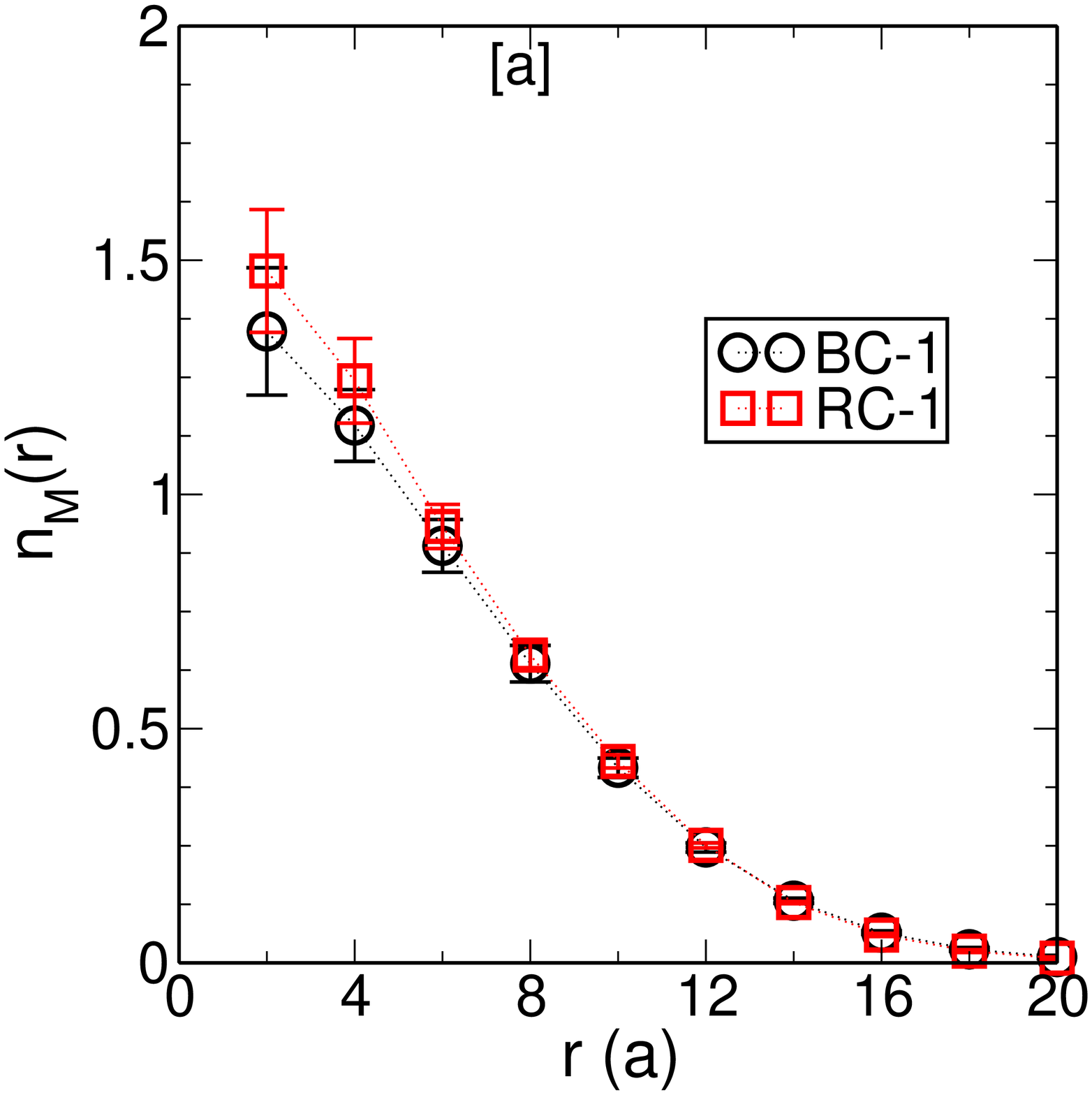} 
\hfill
\includegraphics[width=0.4\columnwidth]{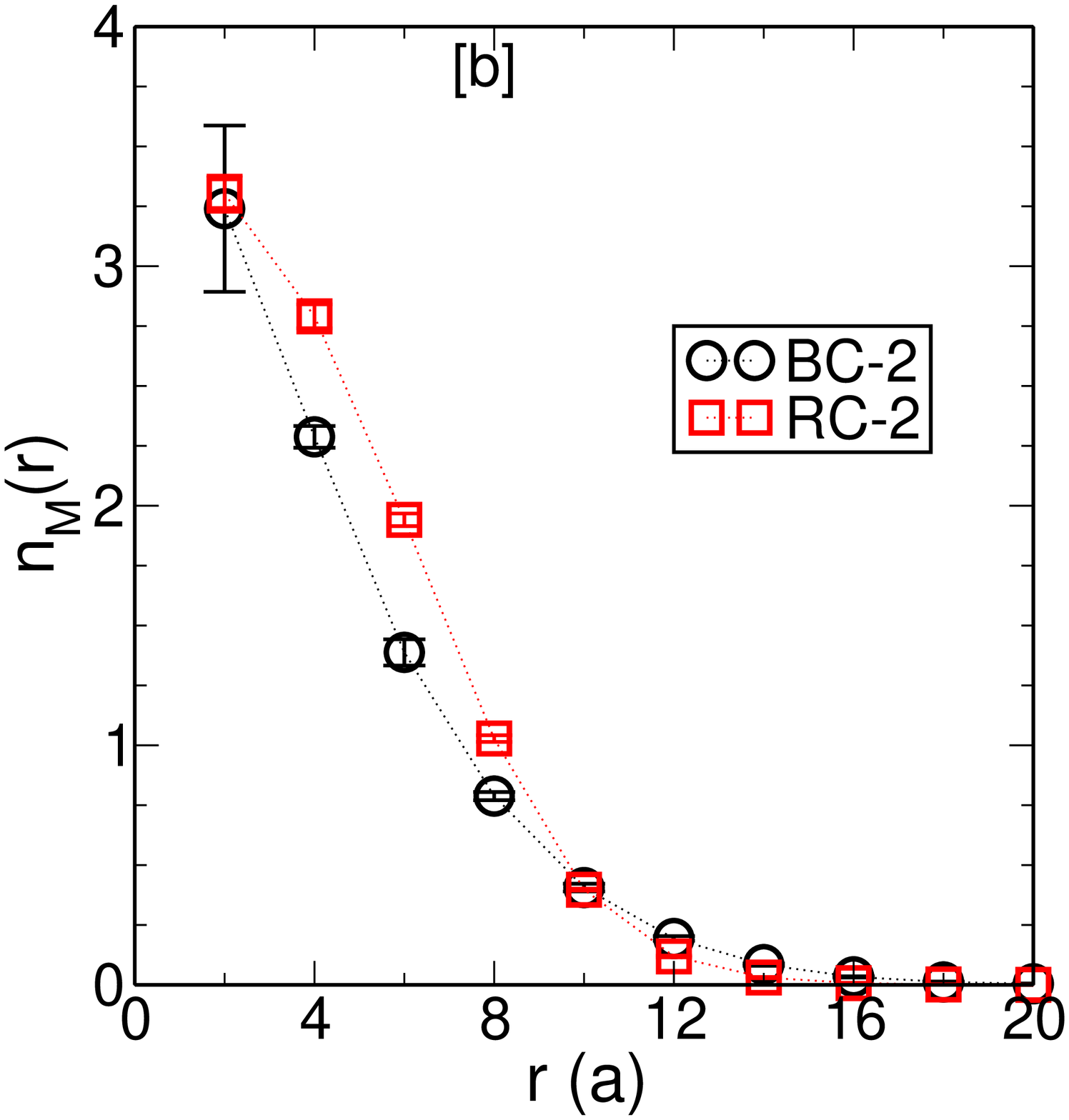} 
\caption{\label{fignm} 
The monomer number density of $n_M(r)$ is plotted versus $r$, where $r$ is the distance of the position
of the monomers from the center of mass of the DNA-polymer coil. Plot (a) is for BC-1, RC-1, whereas 
plot (b) is for BC-2,RC-2. $n_M(r)$ is averaged over $9$ independent initial conditions and error 
bar shows the standard deviation from the average.
}
\end{figure}

\begin{figure}[!htb]
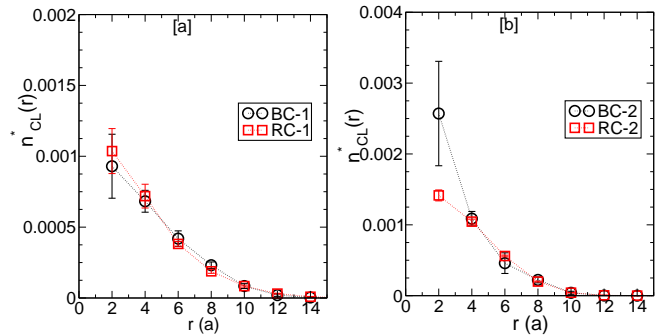

\includegraphics[width=0.49\columnwidth]{contact1_c_no_density.eps}
\includegraphics[width=0.49\columnwidth]{contact3_c_no_density.eps} \\
\caption{\label{figncl} 
The normalized CL number density of $n_{CL}^*(r)$ is plotted versus $r$, where $r$ is the distance of the position
of the CLs from the center of mass of the DNA-polymer coil. 
Subplot (a) is for BC-1 and RC-1, whereas subplot (b) is for BC-2 and RC-2.                   
The number of CLs in each case is further normalized by the total number of CLs in each case. 
Further $n_{CL}^*(r)$ is averaged over 9 independent runs starting from initial conditions and 
error bar shows the standard deviation from the average.
}
\end{figure}

To get some idea of how the monomers of the polymer are distributed in space. And if there is any difference in the 
radial arrangement of bio-CLs and random CLs, we investigate the radial distribution of monomer number densities  and the 
normalized CL number density with the distance $r$ from the center of mass (CM) of the polymer coil. 
The quantities $n_M(r)$ and $n_{CL}(r)$ are calculated by calculating the average number of monomers 
and CLs in radial shells of width $2a$ from the CM of the coil, divided by the volume of each shell. The CL-density is further 
normalized by the total number of CLs for the particular case under consideration to obtain $n_{CL}^*(r)$.
Data for $n_M(r)$ and $n_{CL}^*(r)$ from $9$ independent runs are plotted for each of set of CLs: BC-1, BC-2, 
and one set of RC-1, RC-2
in Figs.\ref{fignm} and \ref{figncl}, respectively. Small standard deviation from average for monomer number densities  and the 
normalized CL number density is an indication that the arrangement of monomers and CLs have relaxed 
to similar distributions and is independent of starting configuration of monomers. 

Comparing subplots (a),(b) of Figs. \ref{fignm} and \ref{figncl} for BC-1 and BC-2, establishes that coils with a higher 
number of CLs lead to more number density of monomers and CLs at the center of the coil. As the coil gets into a more compact 
coil structure with increased number of CLs in set RC-2 as compared to RC-1, we again see an increase in the  
number density of monomers, CLs (suitably normalized) at the center. 
Comparing data for BC-1 and RC-1, respectively from Figs. \ref{fignm} and \ref{figncl}, we observe that 
the distribution  of monomers and CLs are similar at different  $r$ for the 2 cases. 
In contrast, the normalized density of CLs at the central region is more for BC-2 compared to that for RC-2.
Moreover, monomer density $n_M(r)$ for BC-2 is lower than that for RC-2 at the center of the coil,
whereas there are more monomers present at the periphery (for $r>12s$) for BC-2 as compared to RC-2. Since
the number of monomers in each shell is divided by the volume of the shell, the difference in the number of monomer at 
the periphery is just discernable from the number density $n_M(r)$ plots.
Lastly, the number density of monomers/CLs drops down significantly beyond a distance of $8a$ from the coil's center. 
Other nine sets of randomly chosen CLs also in tune with the above observations (data not shown). 

\begin{figure}[!tbh]
\includegraphics[width=0.45\columnwidth]{e_contact3_inner.eps}
\hfill
\includegraphics[width=0.45\columnwidth]{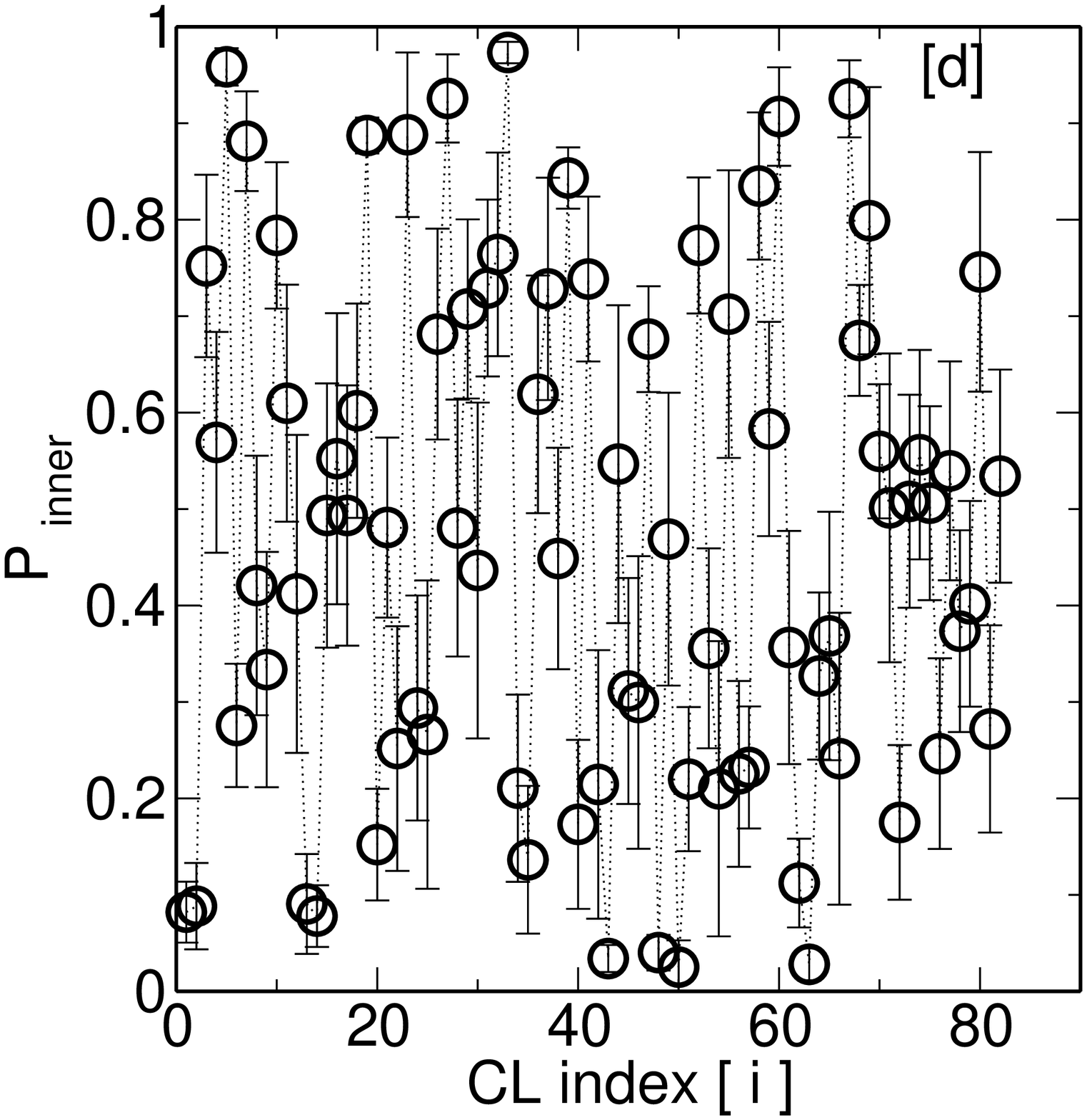}
\hfill
\vskip0.3cm
\includegraphics[width=0.45\columnwidth]{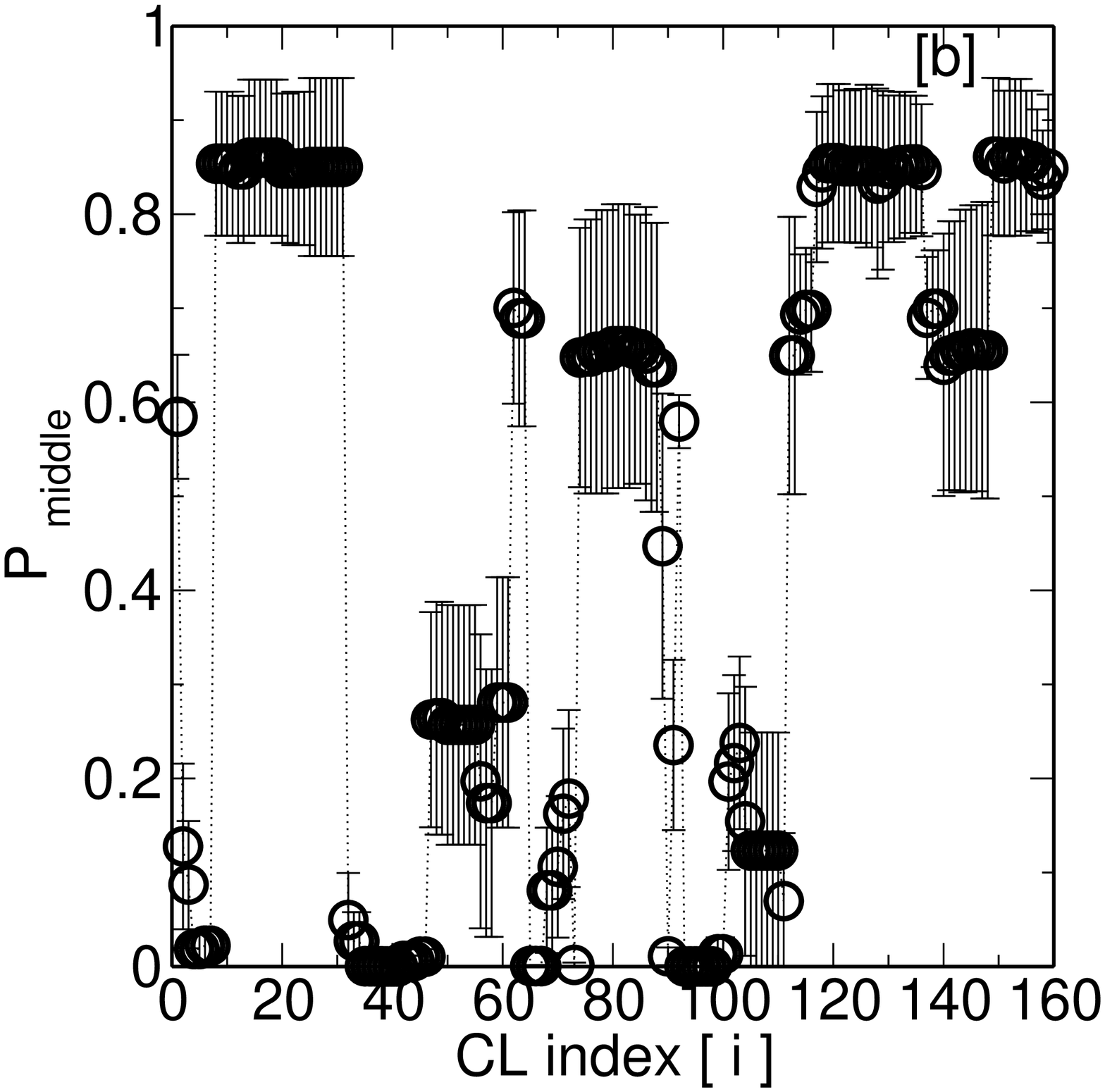}
\hfill
\includegraphics[width=0.45\columnwidth]{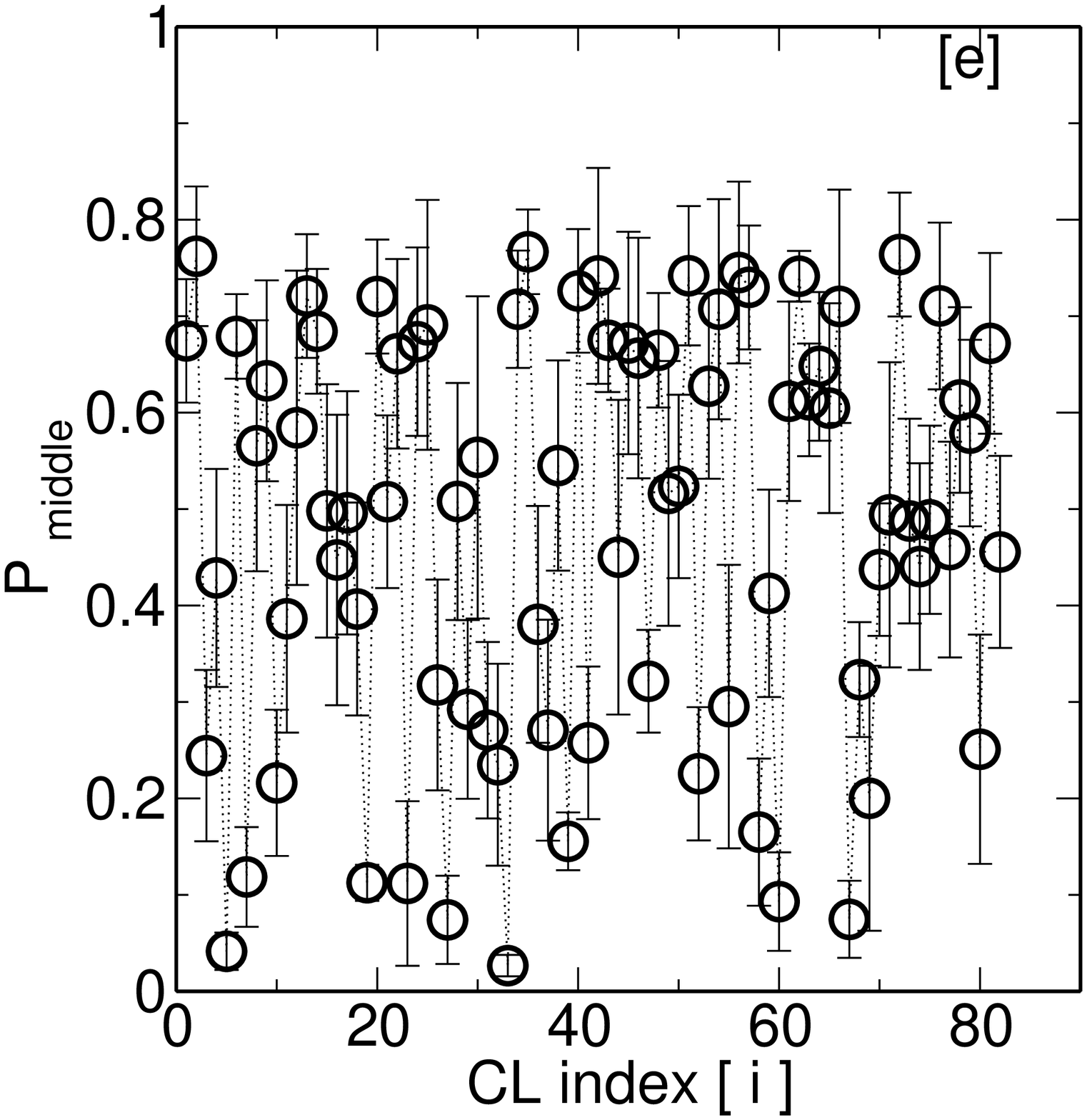}
\hfill
\vskip0.3cm
\includegraphics[width=0.45\columnwidth]{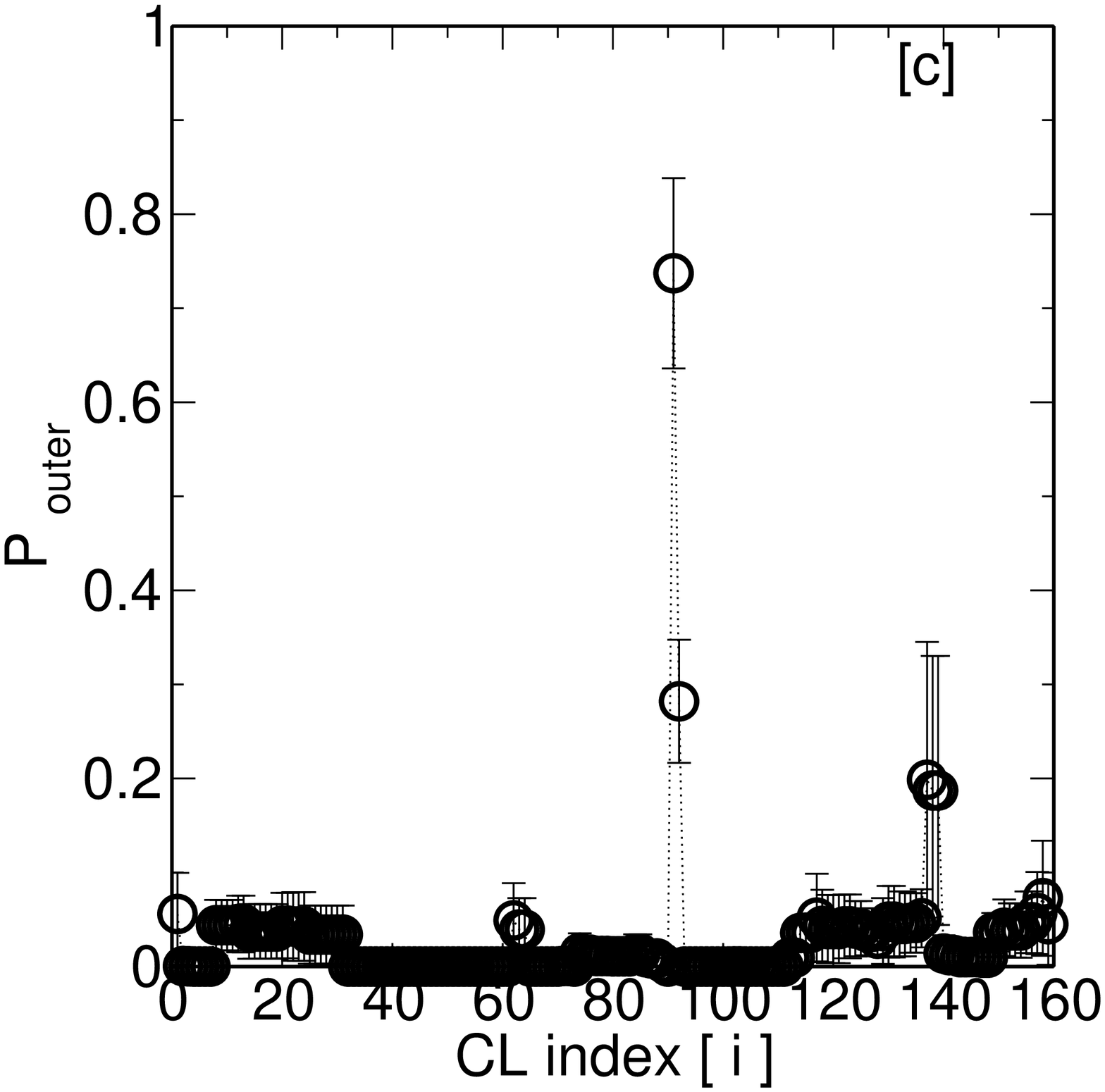}
\hfill
\includegraphics[width=0.45\columnwidth]{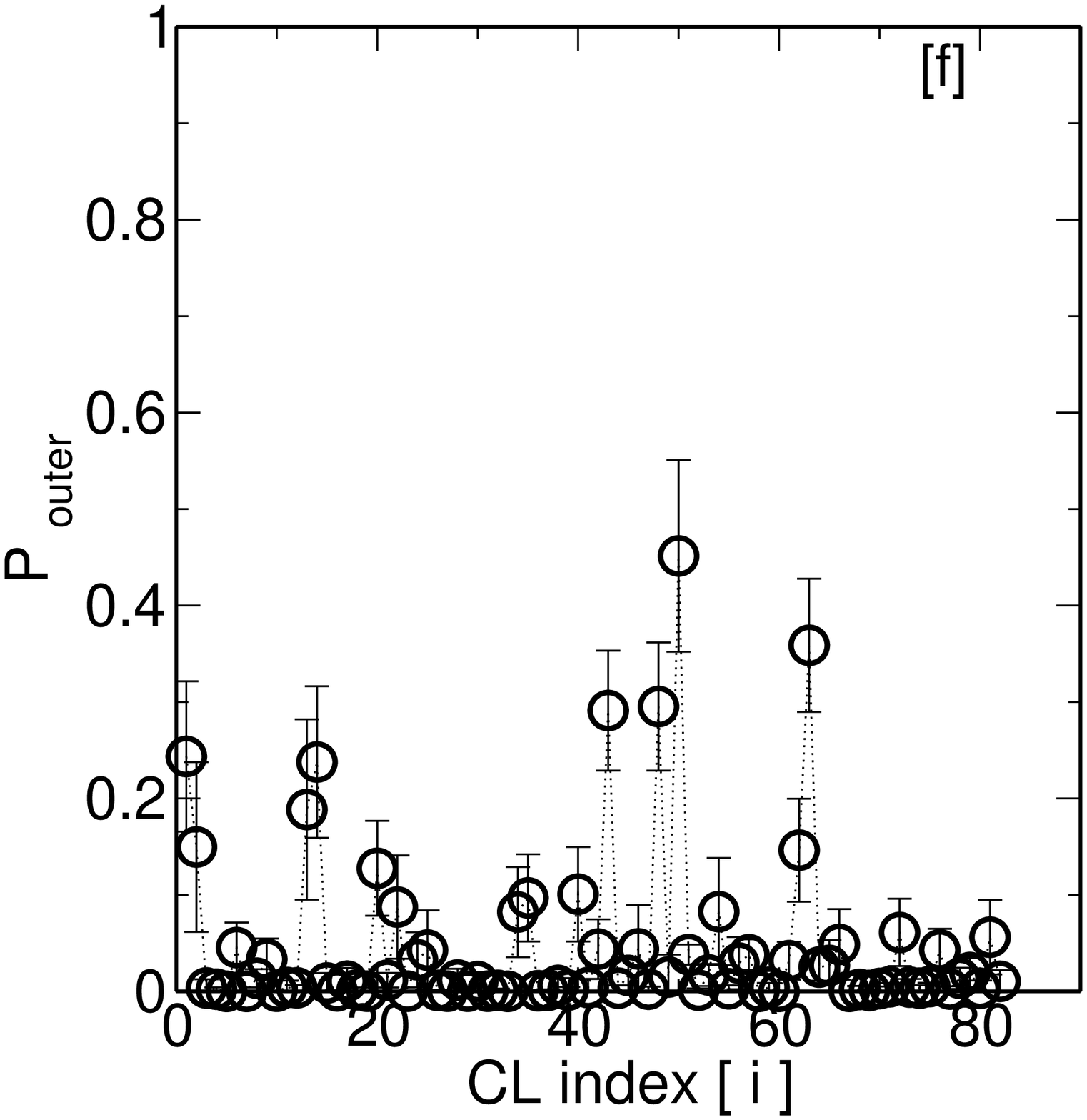}
\vskip0.3cm
\caption{\label{figprobcl2} 
Subplots (a), (b) and (c) (on left column) show the probabilities of individual CLs to be found in the inner, 
middle and outer region for CL-set BC-2. The x-axis is an index for CLs. The average is taken over 9 independent runs
starting from different initial conditions, and standard deviation from the average is shown as an error bar. 
Small error bars indicate 
that the probability of finding CL $i$ is the same across different runs. 
Data (d), (e), (f) on the right column is for the set of random CLs RC-2. The set of 159 CLs for {\em E. Coli} are referred
to as BC-2. Dataset with  fewer number of {\em E. Coli} and random CLs (referred as BC-1 and RC-1, respectively) are shown in Supplementary
Section Fig.16. 
}
\end{figure}

\begin{figure}[!tbh]
\includegraphics[width=0.49\columnwidth]{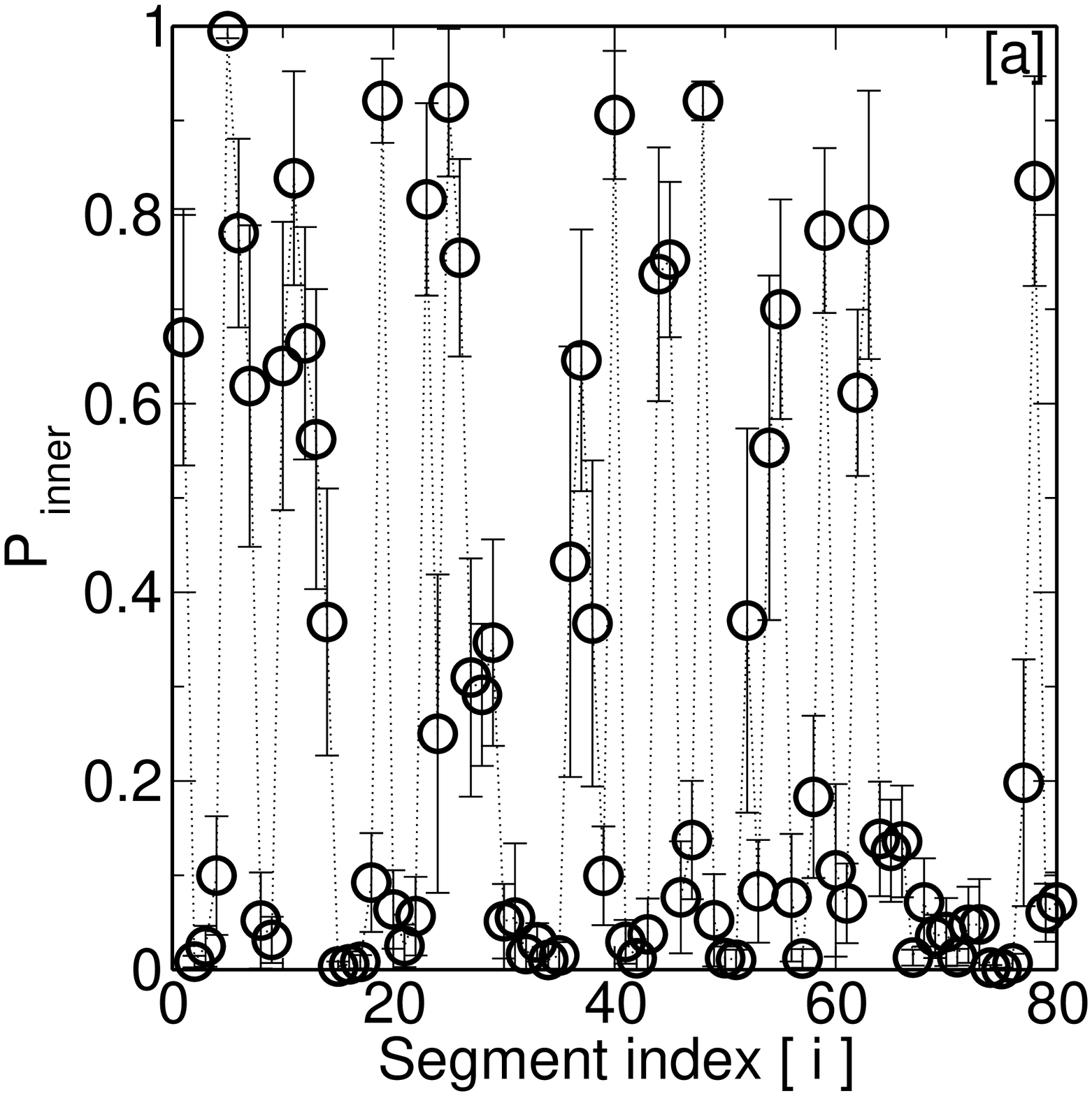}
\hfill
\includegraphics[width=0.49\columnwidth]{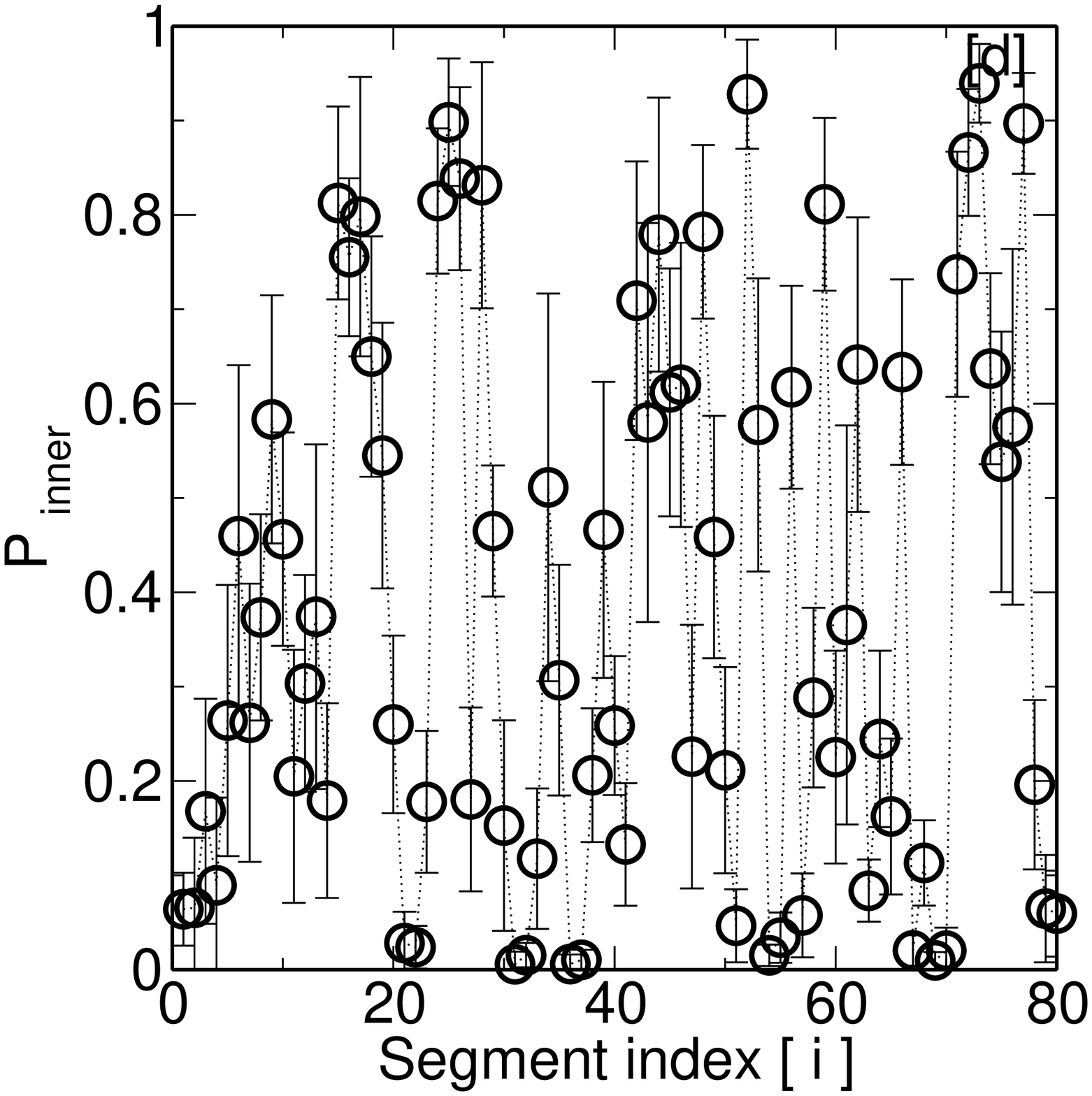} \\
\hfill
\vskip0.3cm
\includegraphics[width=0.49\columnwidth]{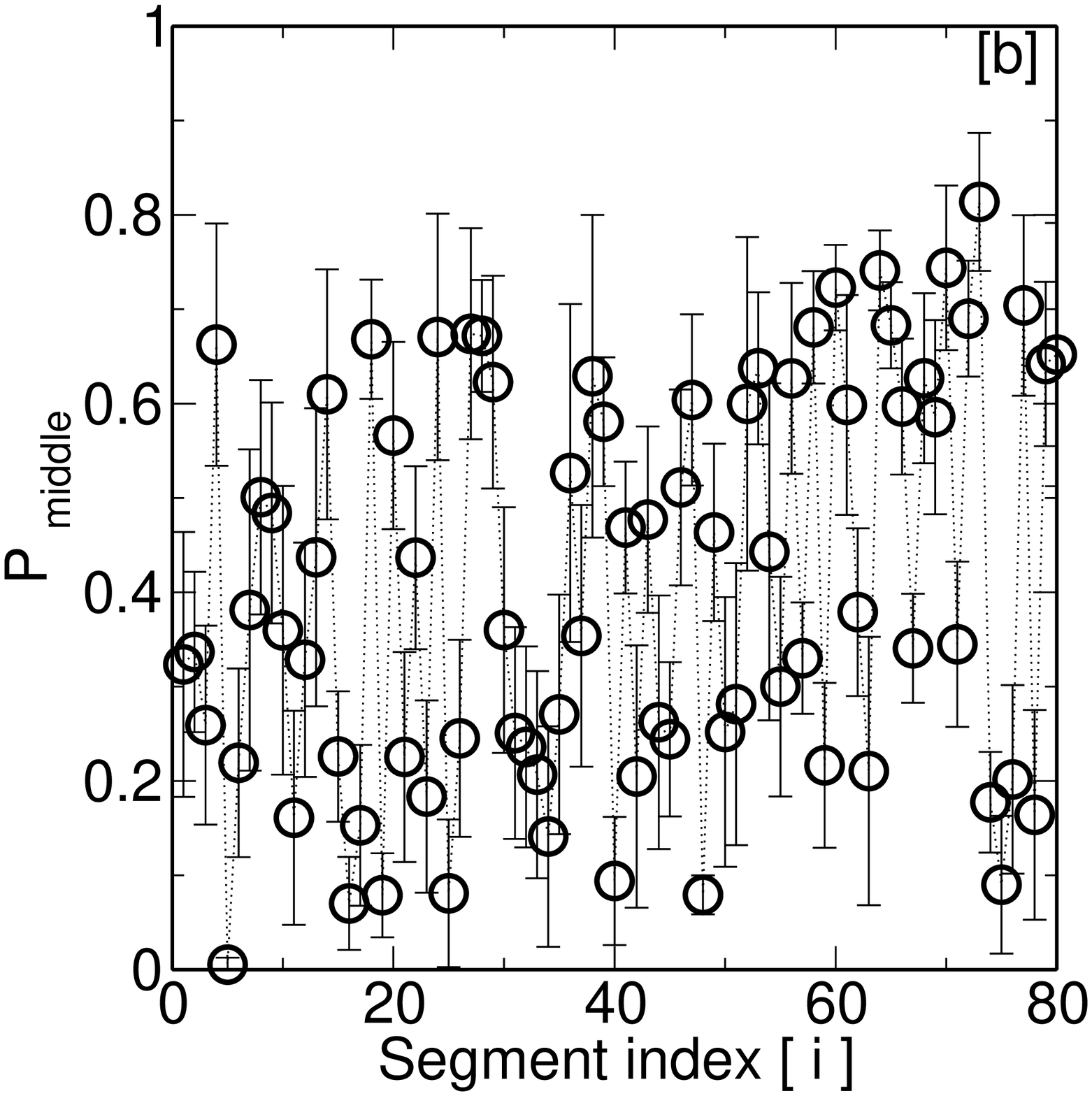}
\hfill
\includegraphics[width=0.49\columnwidth]{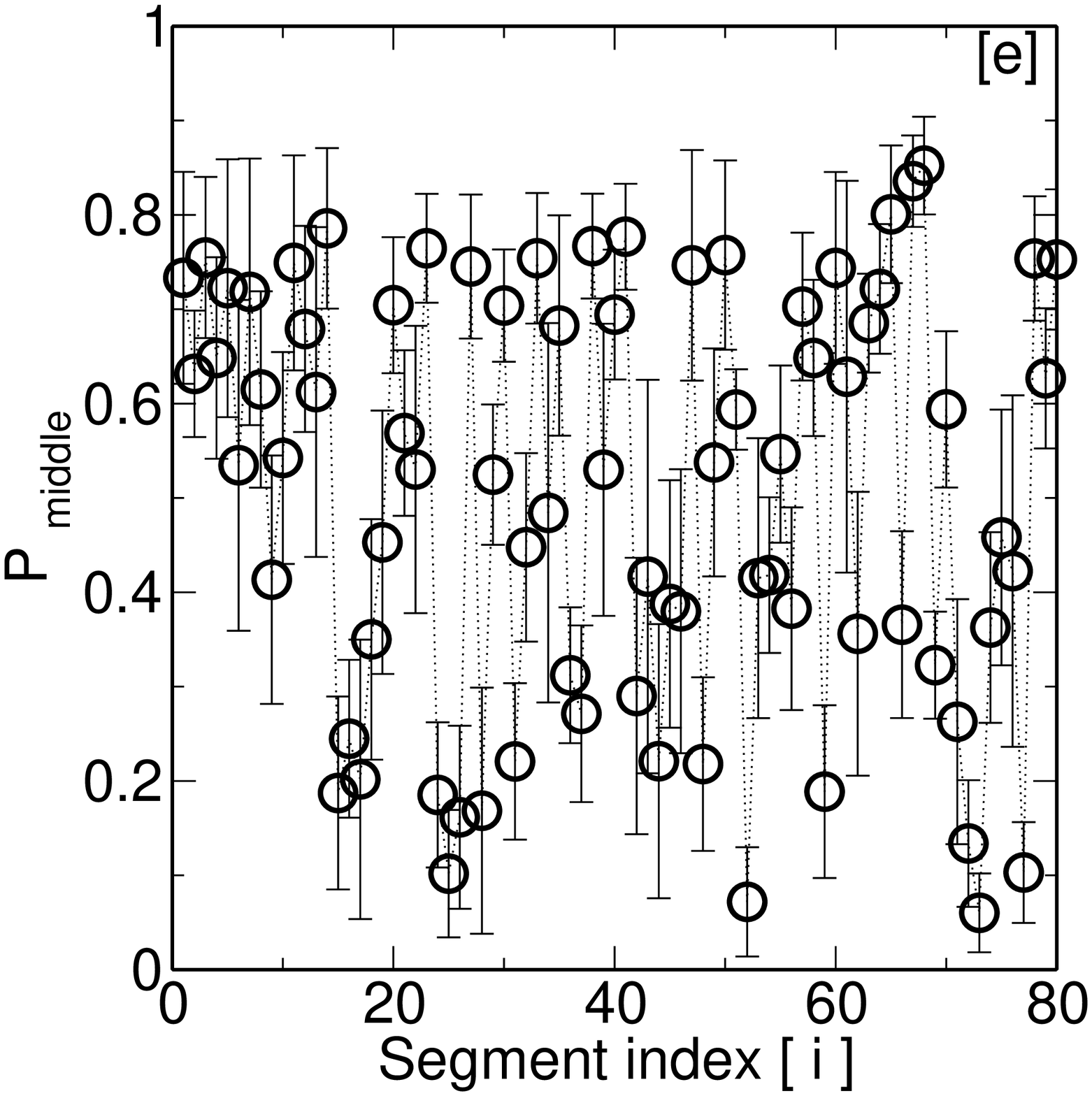} \\
\hfill
\vskip0.3cm
\includegraphics[width=0.49\columnwidth]{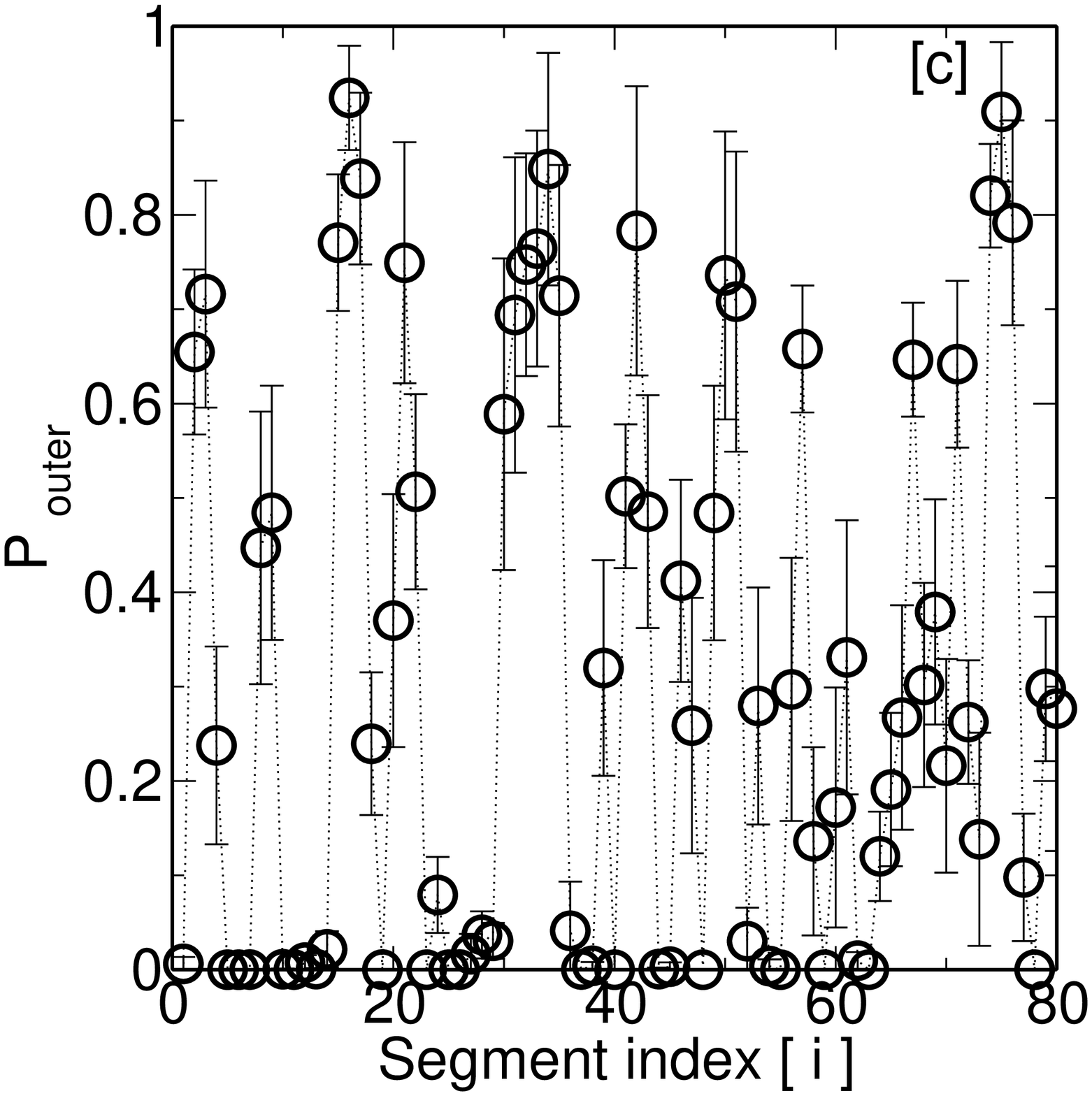}
\hfill
\includegraphics[width=0.49\columnwidth]{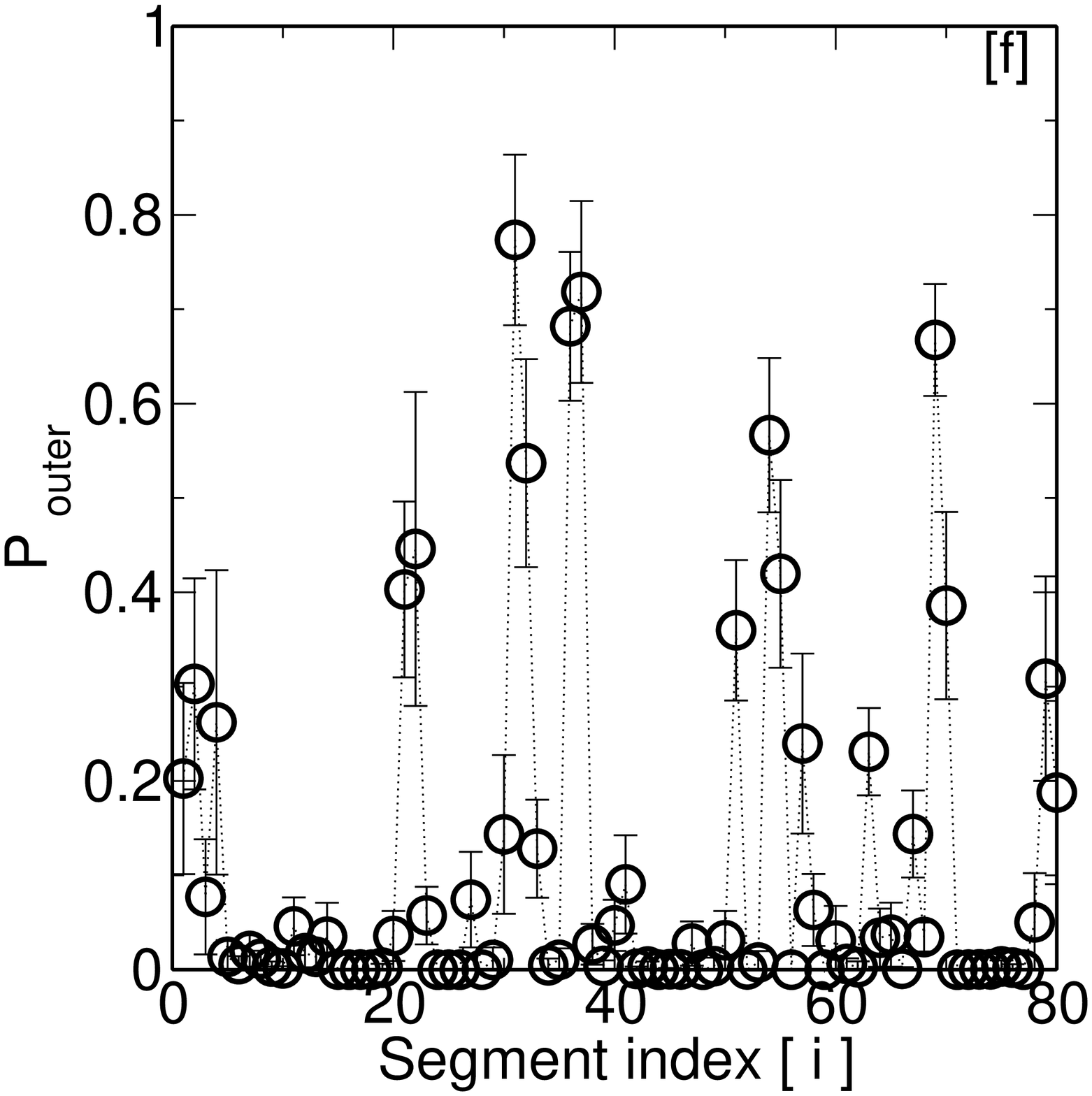} \\
\hfill
\vskip0.3cm
\caption{\label{probsegment2} 
Subplots (a), (b) and (c) (in the left column) shows the probabilities of  the center of mass (CM) of  $80$ DNA-polymer
segments  to be found in the inner, middle and outer region of {\em E. Coli} DNA coil. The x-axis is segment index.
In each case average values of $P_{inner},P_{middle}$ and $P_{outer}$ are taken over 9 runs starting 
from independent initial conditions, 
deviation from the average is shown as error bars. Small error bars 
indicate that the probability of finding CM of segment $i$ in a particular region is nearly the same across
 different runs.  Data on subplots (d), (e), (f) is for random choice of cross-link position (set RC-2) 
with $82$ CLs in a chain with 4642 monomers.  Each segment has 58 monomers.
The dataset with fewer  CLs (referred as BC-1 and RC-1, respectively) are shown in Supplementary Section Fig.17.
}
\end{figure}

To gain some more insight about the global structural organization of the DNA-coil, the simplest question to 
ask is whether a particular CL is always found near the center of the coil or near the periphery of the coil. 
To this end, we compute the probability of each of the CLs to be found in the {\em inner,middle}, and {\em outer} 
regions of the DNA-coil. We use the cutoff radii $R_{inner} = 5a$, 
$R_{middle} = 9a $ (chosen from the knowledge of the value of $R_g \approx 8a$) and calculate 
the probability $P_{inner}, P_{middle}, P_{outer}$ of finding the $i-$th CL within distance $ r< R_{inner}$ (inner region),
$R_{inner} < r < R_{middle}$ (middle region) and $r> R_{middle}$ (outer region), respectively, from the coil's center of mass. 
If the  values of  $P_{inner}, P_{middle}, P_{outer}$ for each CL has small deviation from the average value 
in each of the $9$ independent runs, 
it would  indicate that the presence of CLs leads to similar organization of the DNA across independent runs.
Also, we  compare the probability distribution of  CLs for runs with bio-CLs and random-CLs to investigate if bio-CLs lead to 
 organization distinct from that obtained with random-CLs. 

We carry out the same exercise for different segments of the polymer chain.
The {\em E. Coli} chain with $4642$ monomers is divided into $80$ segments with $58$ monomers in each of segment and the segments
are labeled from $i =1,2...N_s$ as we move along the contour. We can then calculate the location of the CMs
of each segment, and find out the probability of finding the CMs in the central, middle and outer region.
The segments in a  random-walk polymer model (without CLs) can take any conformation, and there is no reason to believe that 
certain segments will preferably be found in the inner or outer regions of the coil. 
If the segments were completely delocalized, we would expect the polymer in different microstates to contribute to 
all the  $P_{inner}, P_{middle}, P_{outer}$ quantities for each segment.
The question is to what extent will this basic behavior of polymer coils get modified by the presence of bio-CLs and random-CLs? 

Probability data about the location of CLs and segments for BC-2 and RC-2 is given in \ref{figprobcl2}
and \ref{probsegment2}, respectively. Data for BC-1 and RC-1 is given in the Supplementary data section Fig.16, 17.   
Furthermore, from Figs \ref{figprobcl2} and \ref{probsegment2}(a),(b) and (c) we see that  
some CLs  (e.g. the CL with index $60$) has the nearly equal 
probability of being in the inner or middle region of the coil, but very low probability to be found in the outer region.
For BC-2, most CLs are found in the inner and middle regions of the coil whereas for RC-2 CL set there are some CLs 
at the periphery; refer Fig.\ref{figprobcl2}. On the other hand, from Fig.\ref{probsegment2} we see a larger number 
of segments have a finite probability to be in the outer regions for BC-2 as compared to data for RC-2. The data
consistently shows that the position of CL, as well as segments are localized in space across different runs.

Having established that the CLs and segments of DNA-polymer coil have some degree of radial organization, 
we try to extract more detailed structural information about the position 
of segments relative to each other within the coil.  We calculate the probability of  each CL 
(alternatively, each  segment) to be in proximity to other CLs (alternatively, other segments).
If there are no particular well-defined relative positions of CLs/segments within the chain-coil, 
there is no reason to expect CM of some segments (or independent CLs) to be found spatially close to 
each other with high probability, especially when the segments/CLs are separated along the chain 
contour. We define two CLs/segment's CM to be close to each other if the 
distance $r$ between the CLs/segment-CMs is $<5a$, which is just more than $0.5 R_g$. 
We emphasize that we have cross-linked monomers, these constraints are at the monomer ($1000$ BP) 
length scale, whereas we are investigating the organization of polymer segments at a much larger
length-scale. The position of CLs, position of CM of segments are just 2 different markers of 
different sections of the chain and we use relative position of both to identify spatial correlations 
between different sections of the chain. 
	
\begin{figure}[H]
\begin{center}
\includegraphics[width=0.49\columnwidth]{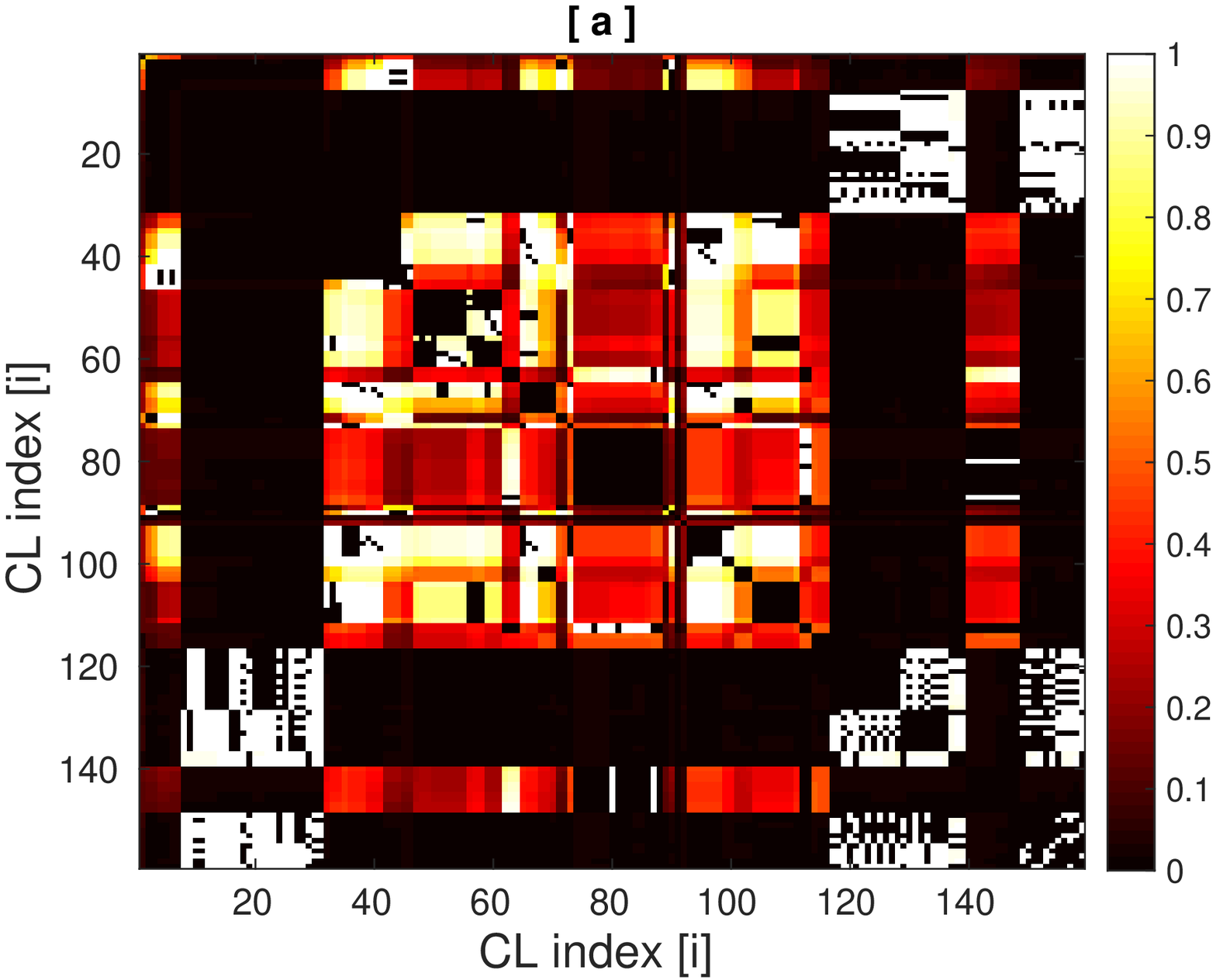}
\hfill 
\includegraphics[width=0.49\columnwidth]{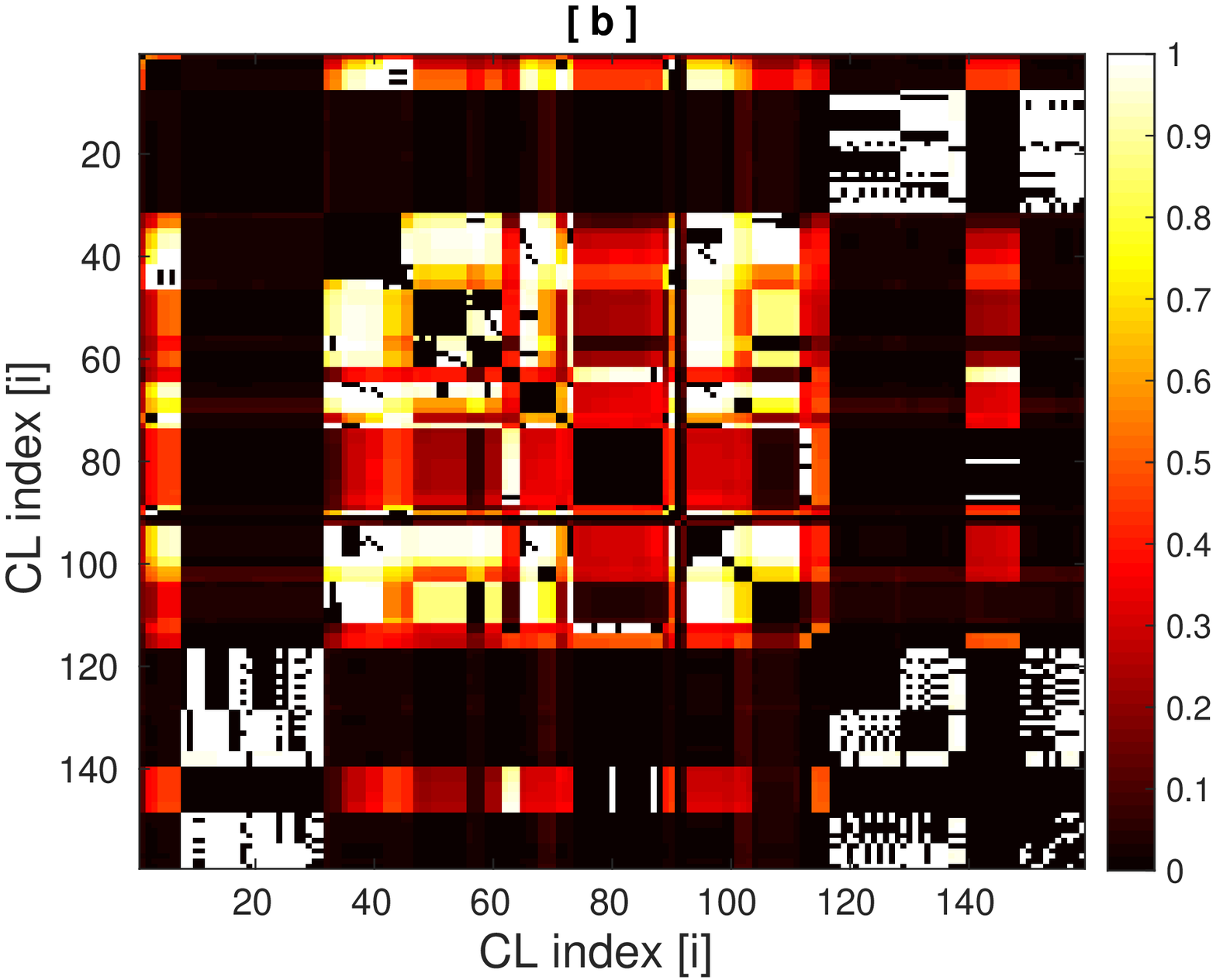} \\
\vskip0.05cm
\includegraphics[width=0.49\columnwidth]{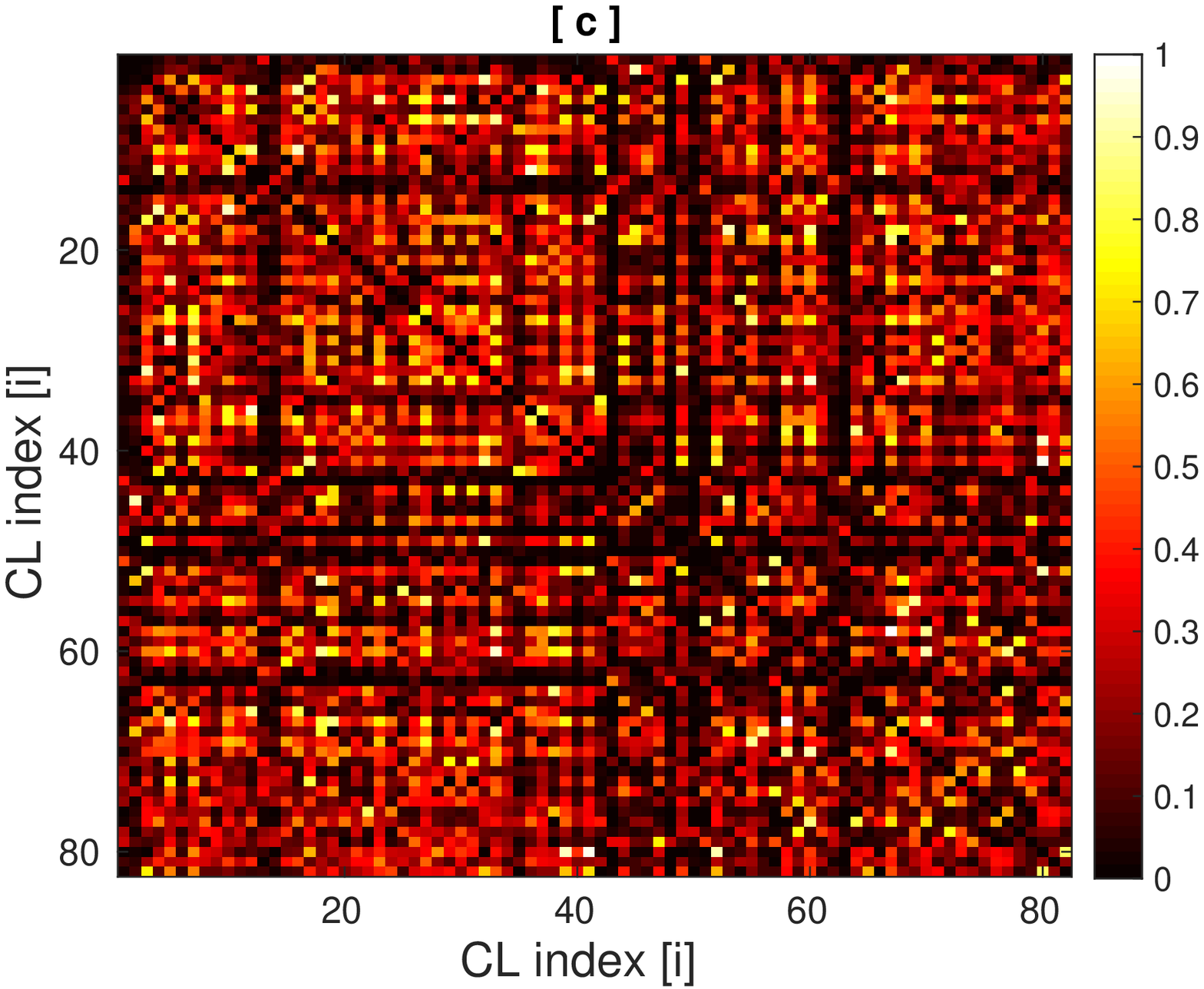} 
\hfill
\includegraphics[width=0.49\columnwidth]{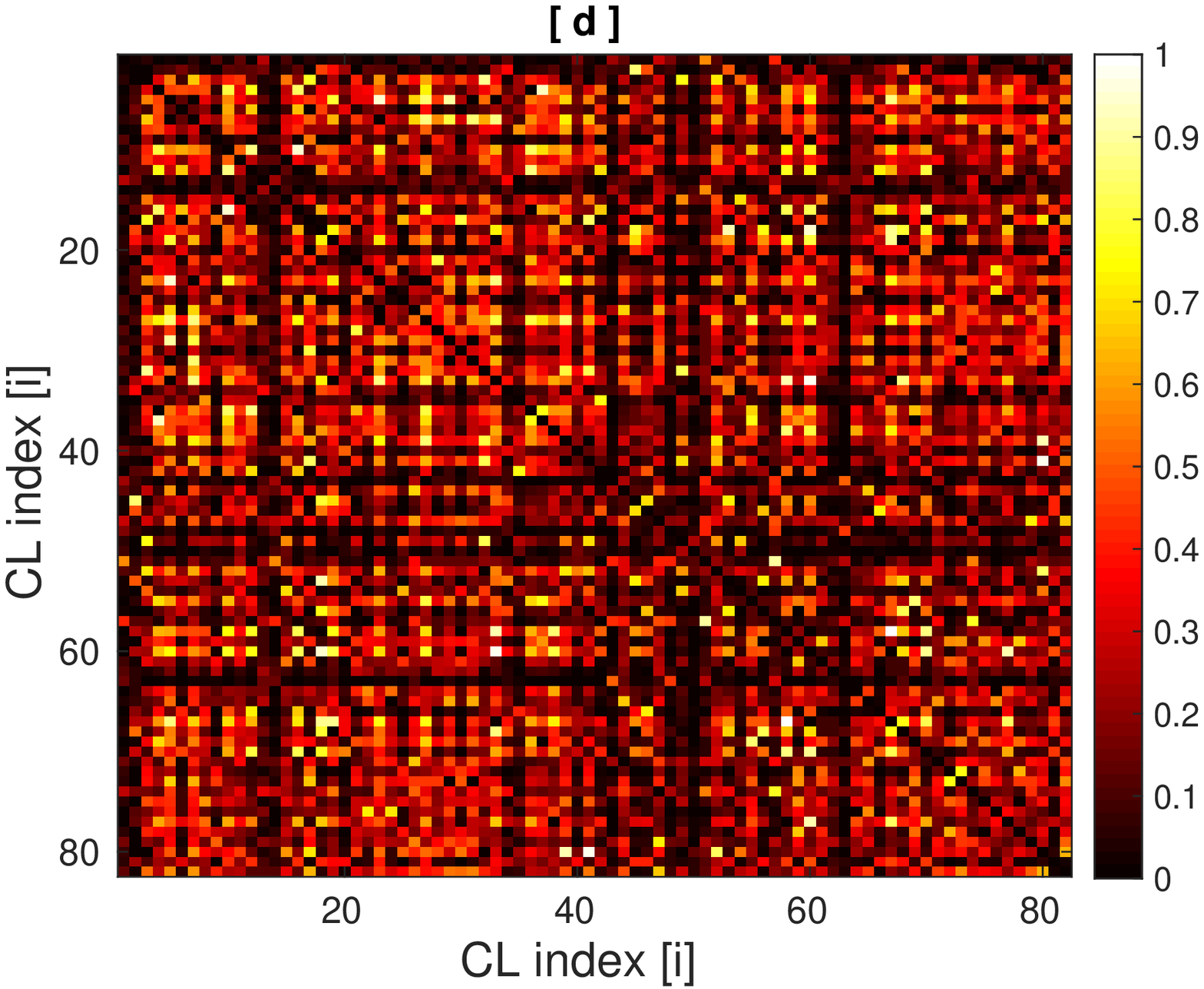} \\
\hfill
\end{center}
\caption{\label{posclbc2} 
Colormaps to represent probability $p(i,j)$ to find CLs $i$ spatially close CLs $j$. 
The top 2 figures are data from 2 independent runs with BC-2 with 159 CLs ($i,j =1...159$) 
and the bottom two subplots are from 2 independent runs with RC-2
with $i,j=1,....82$.  More colormaps from independent runs are given in Supplem. Section:  Fig.19. 
}
\end{figure}

\begin{figure}[H]
\includegraphics[width=0.7\columnwidth]{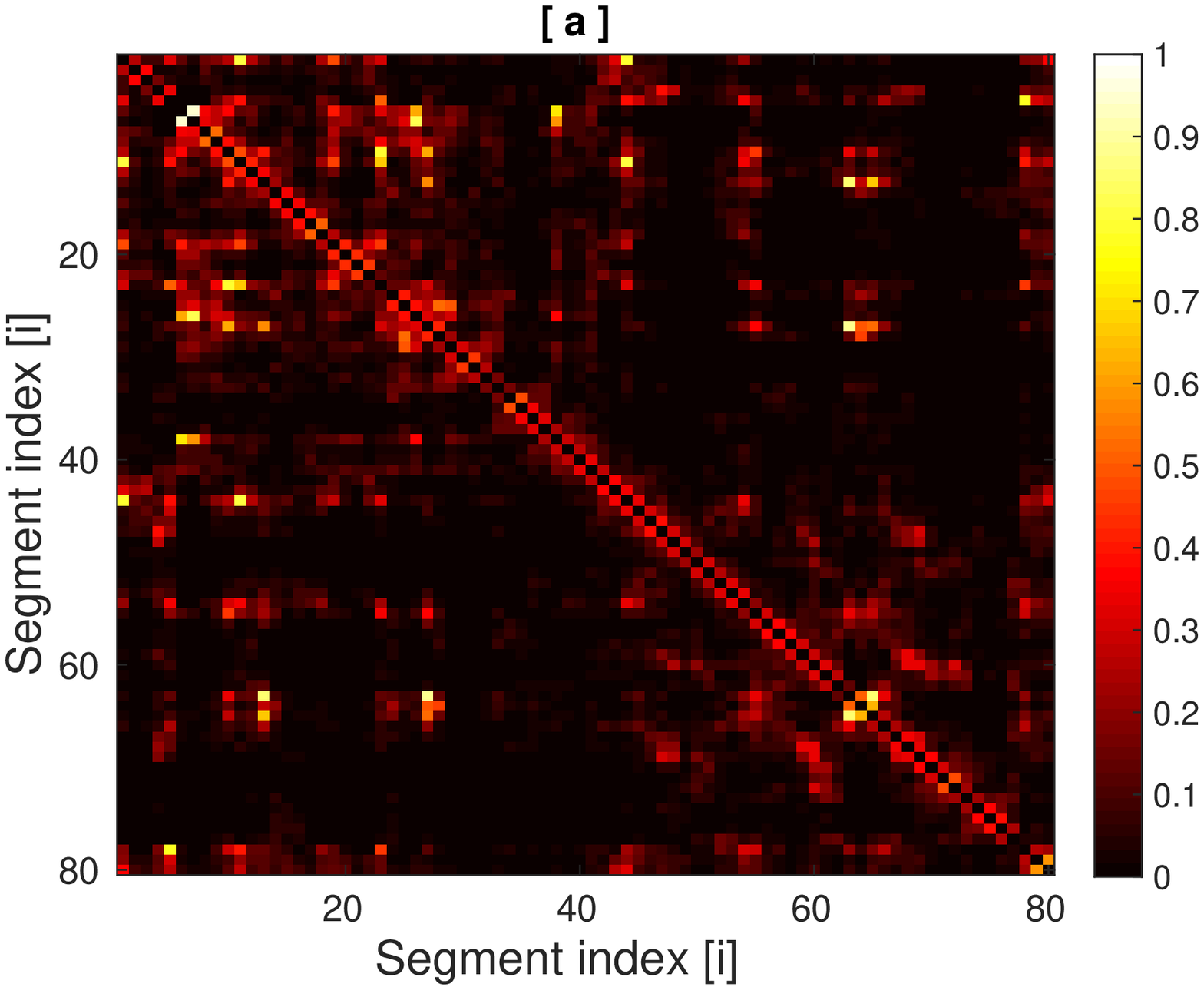} \\
\vskip0.05cm
\includegraphics[width=0.7\columnwidth]{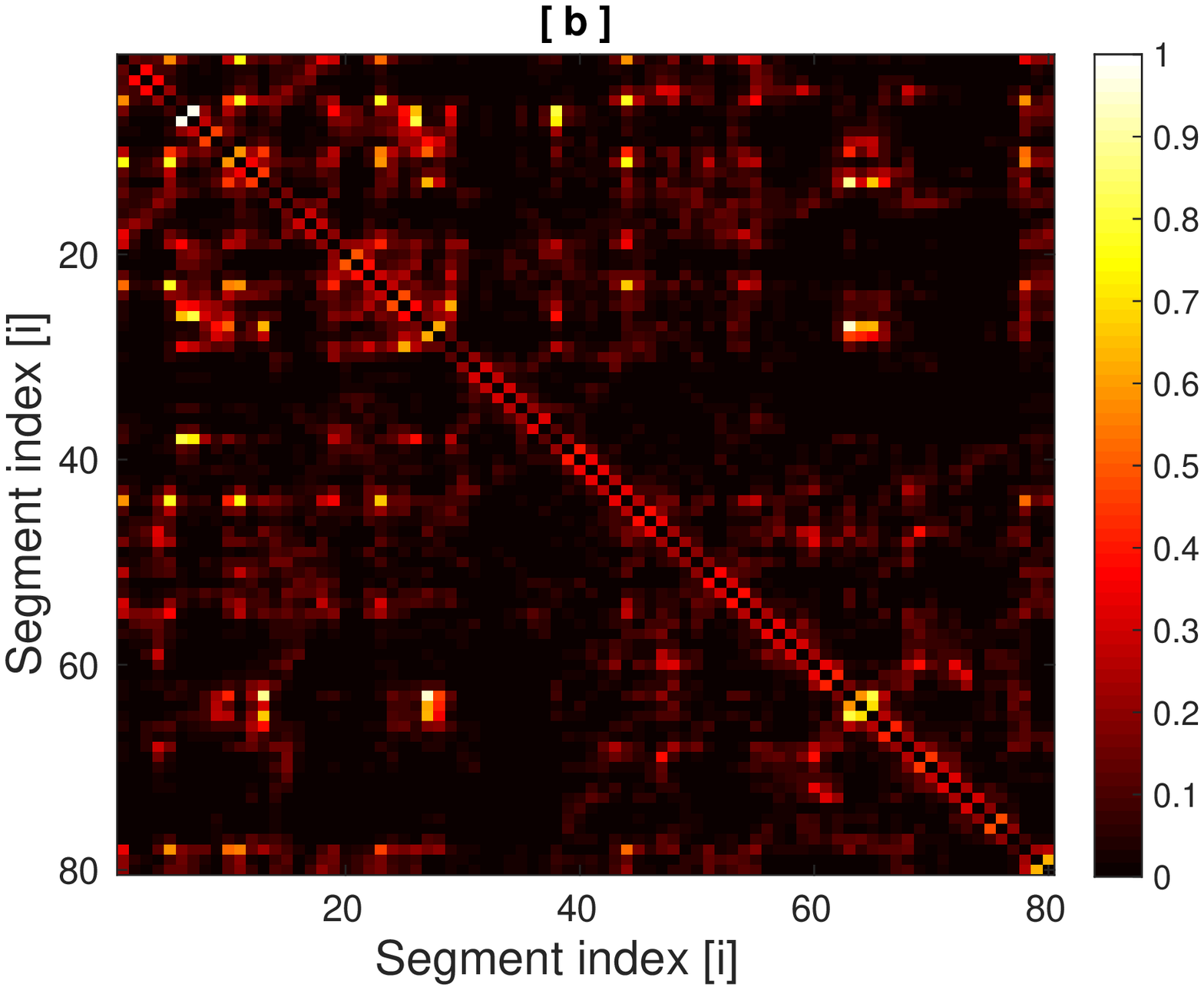} \\
\vskip0.05cm
\includegraphics[width=0.7\columnwidth]{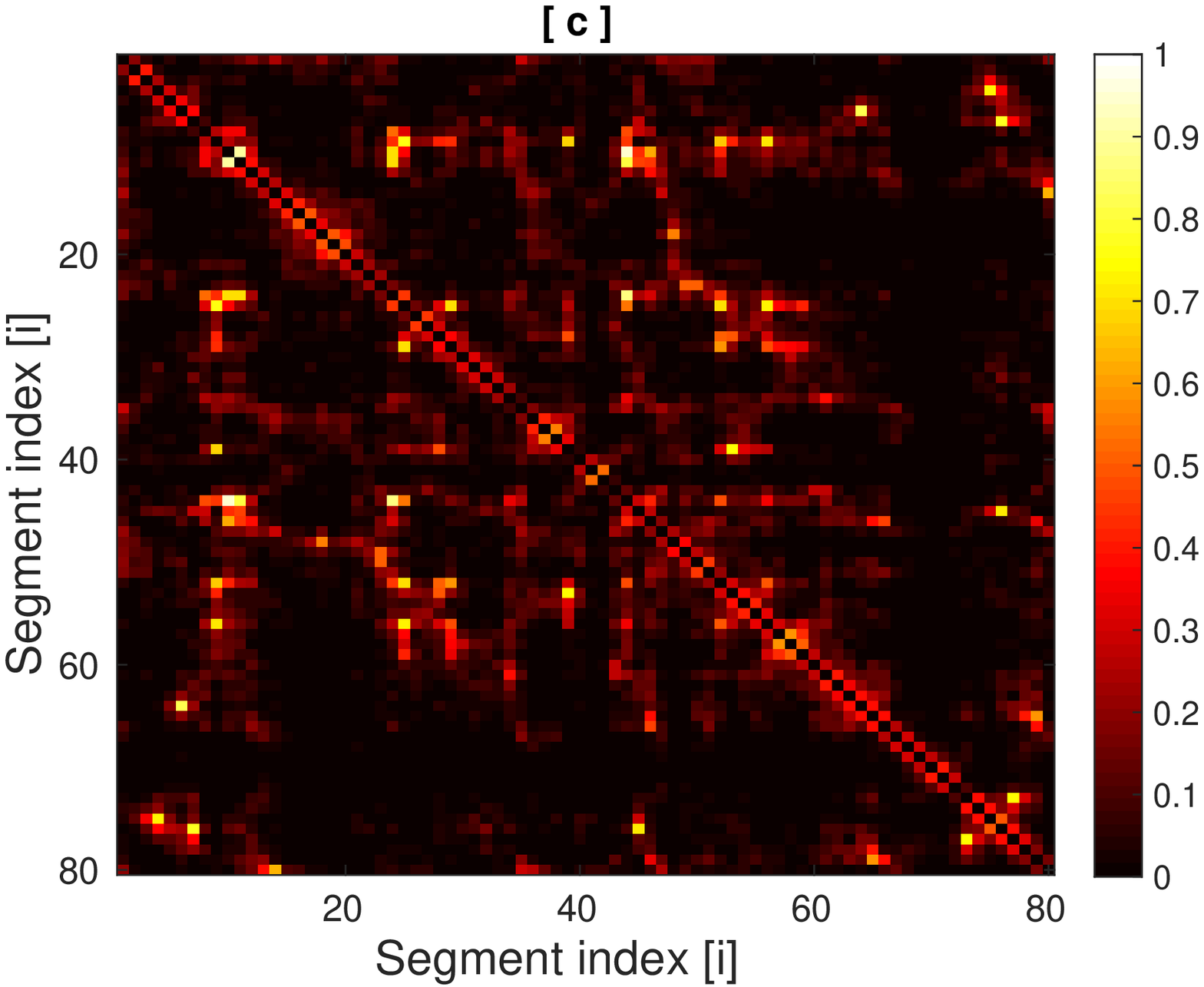} \\
\vskip0.05cm
\includegraphics[width=0.7\columnwidth]{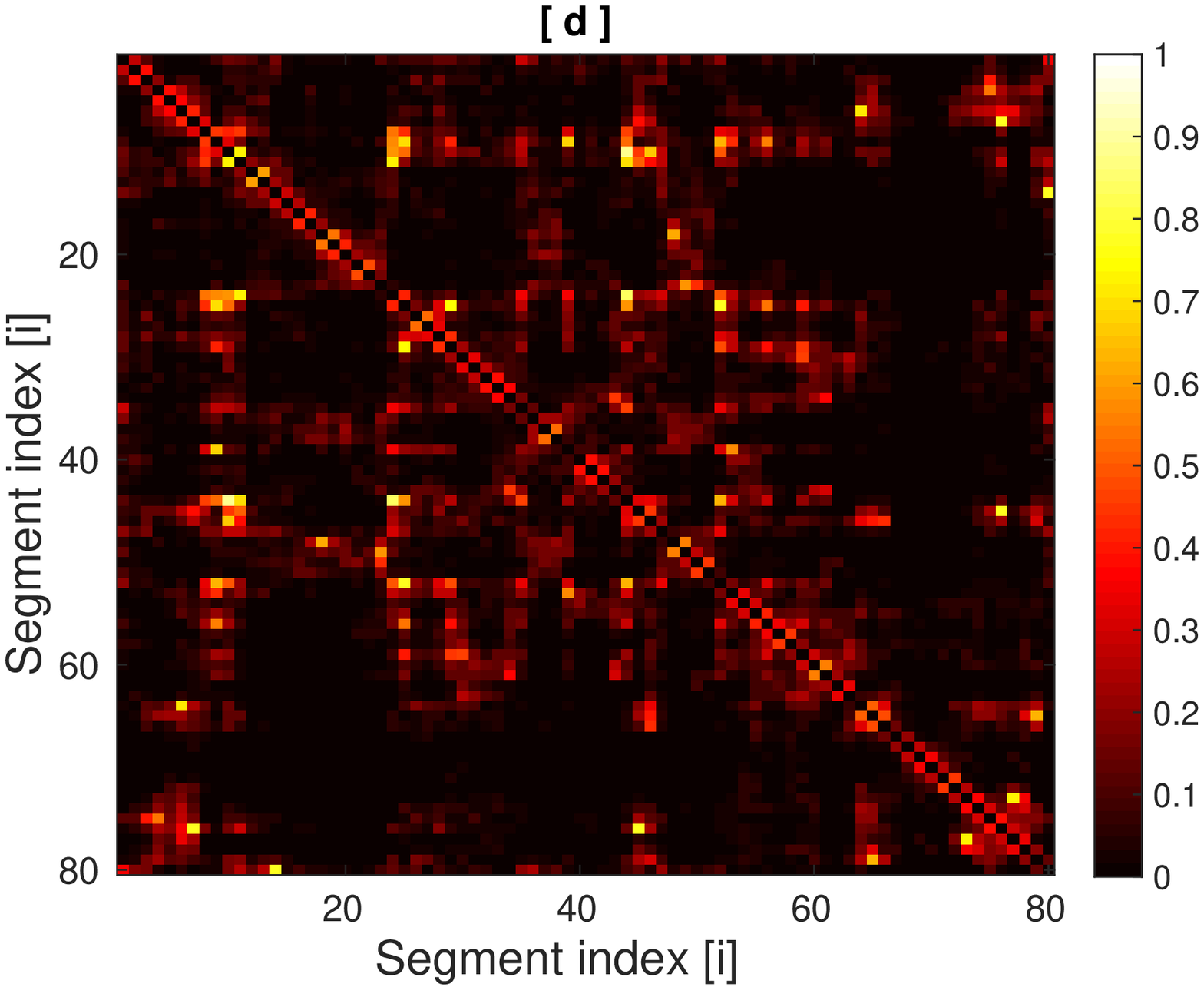} \\
\hfill
\caption{\label{possegbc1} 
Colormaps to represent probability $p(i,j)$ to find CM of segment $i$ spatially close to  CMs of other chain segments $j$. 
There are $80$ segments in the {\em E. Coli} polymer  with $58$ monomers per segment. 
The top 2 figures are runs with BC-1 and the bottom two for RC-1.
More colormaps from independent runs are given in Supplem. Section:  Fig.18. 
}
\end{figure}

\begin{figure}[H]
\includegraphics[width=0.7\columnwidth]{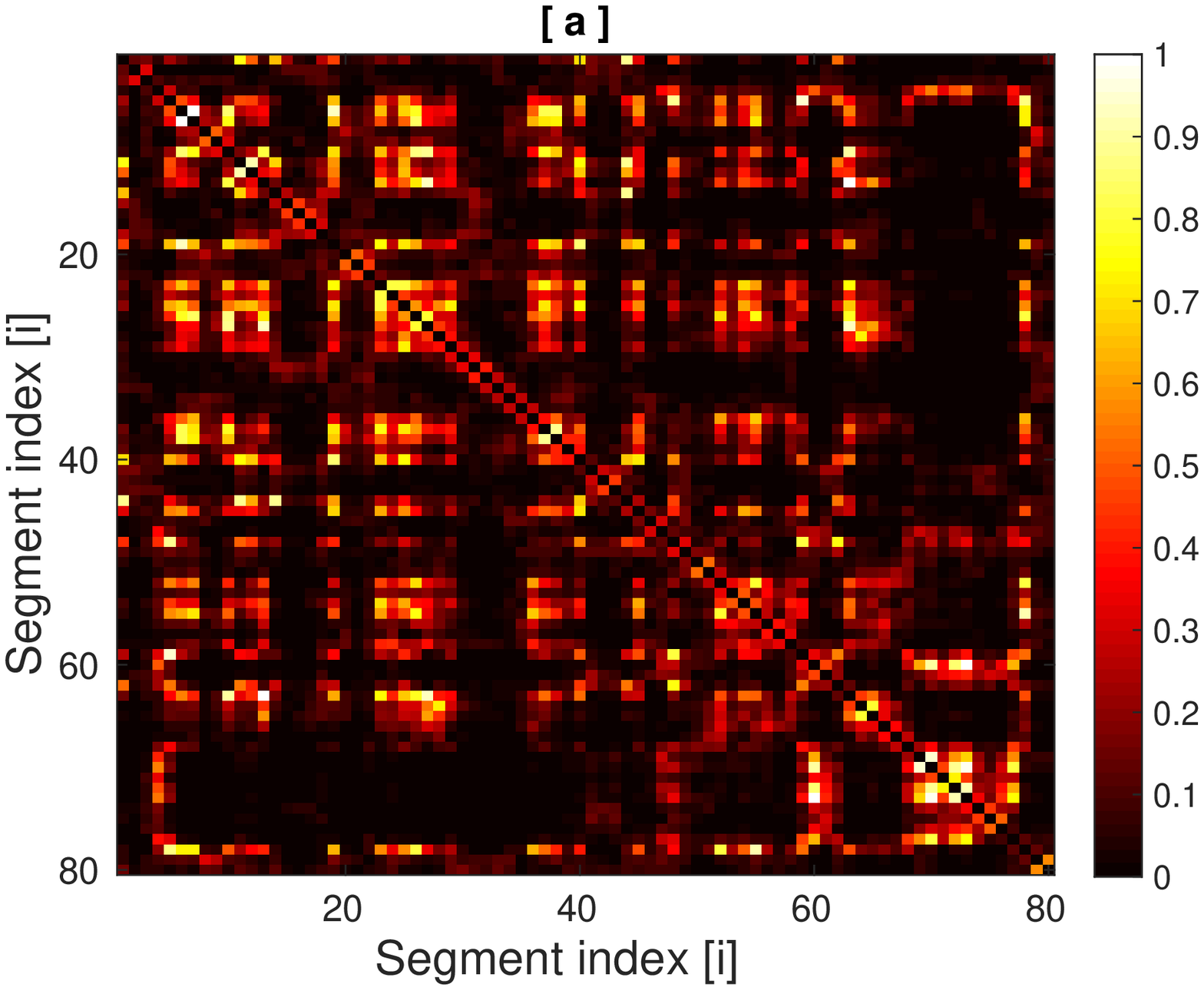} \\
\vskip0.05cm
\includegraphics[width=0.7\columnwidth]{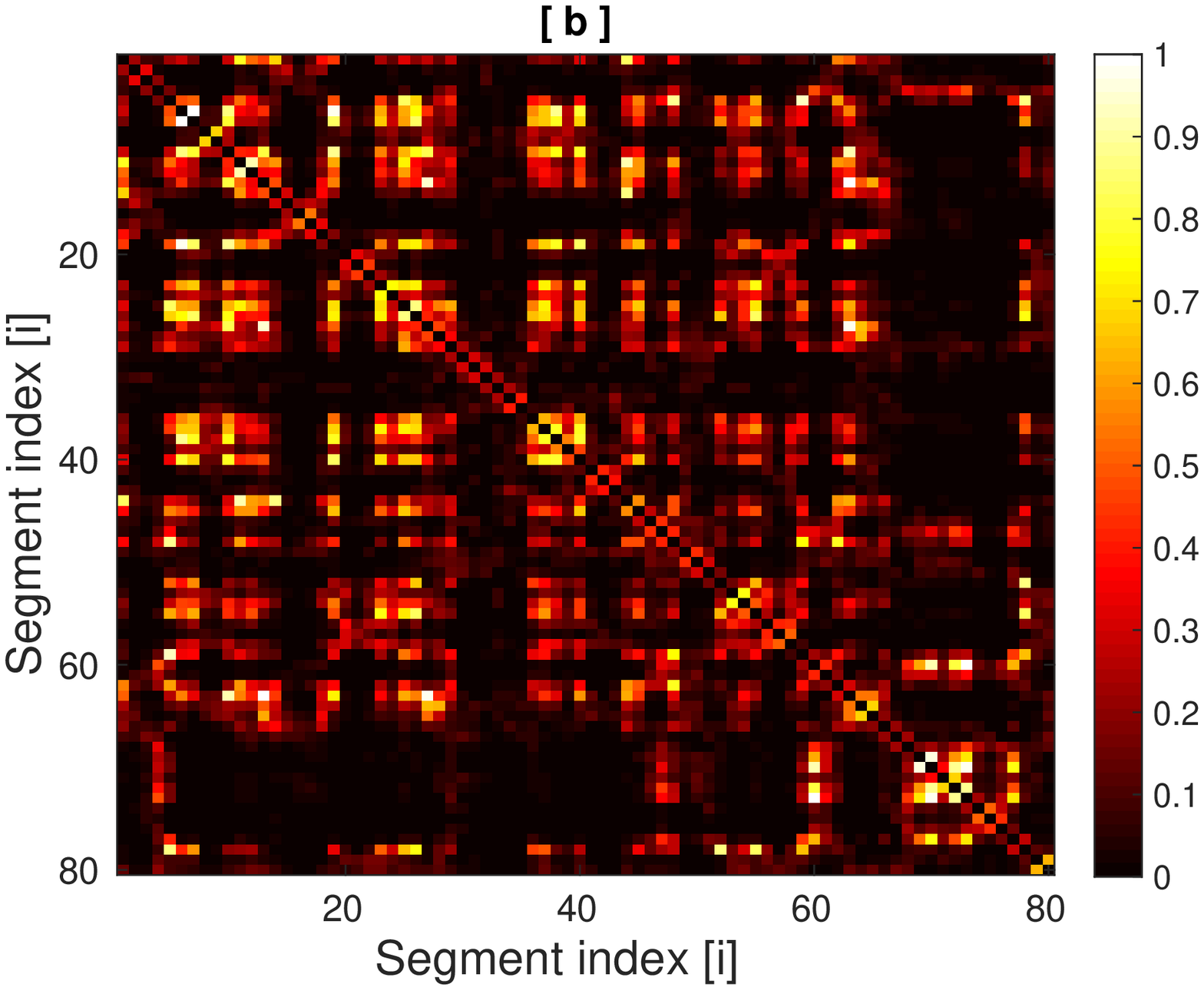} \\
\vskip0.05cm
\includegraphics[width=0.7\columnwidth]{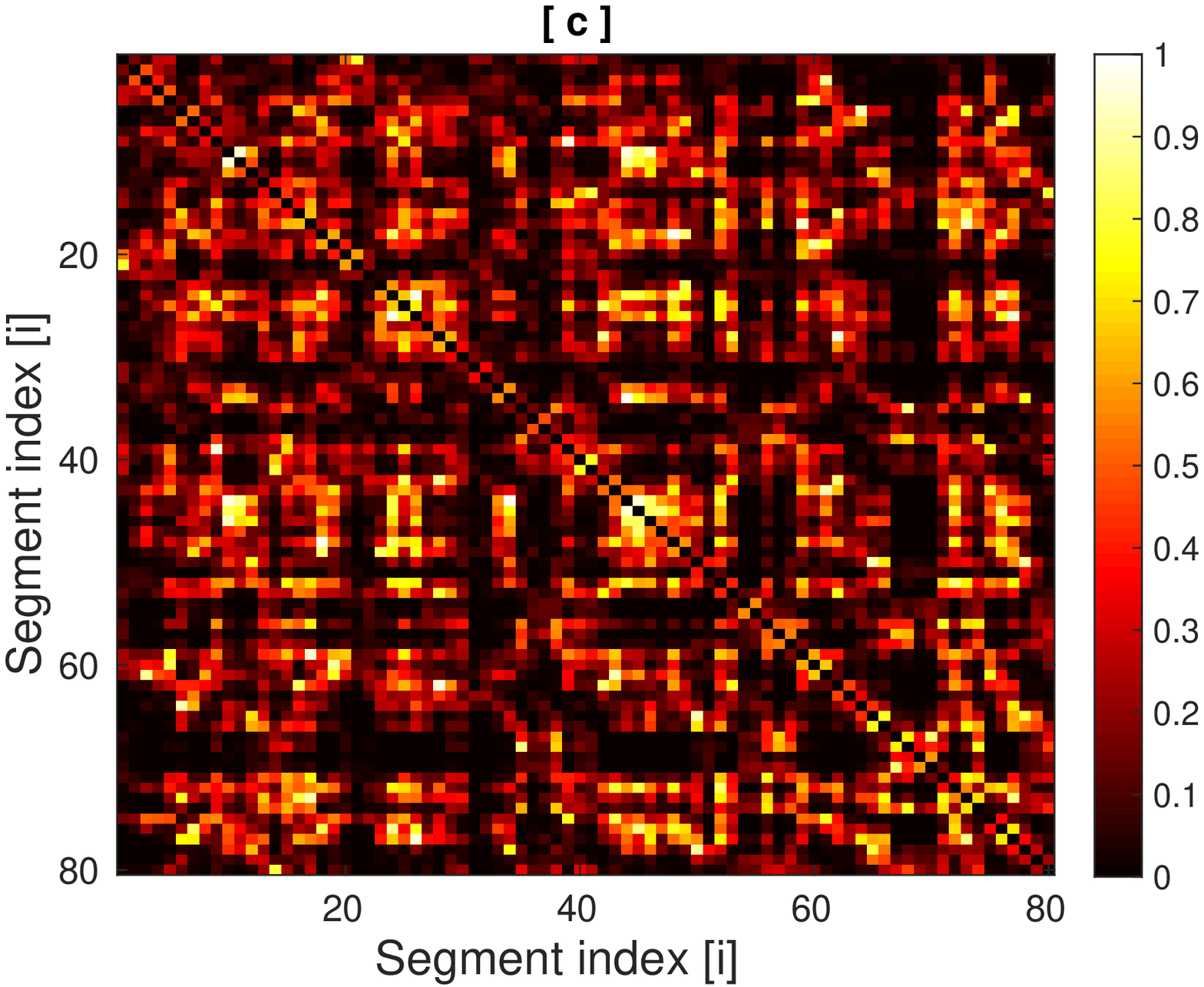} \\
\vskip0.05cm
\includegraphics[width=0.7\columnwidth]{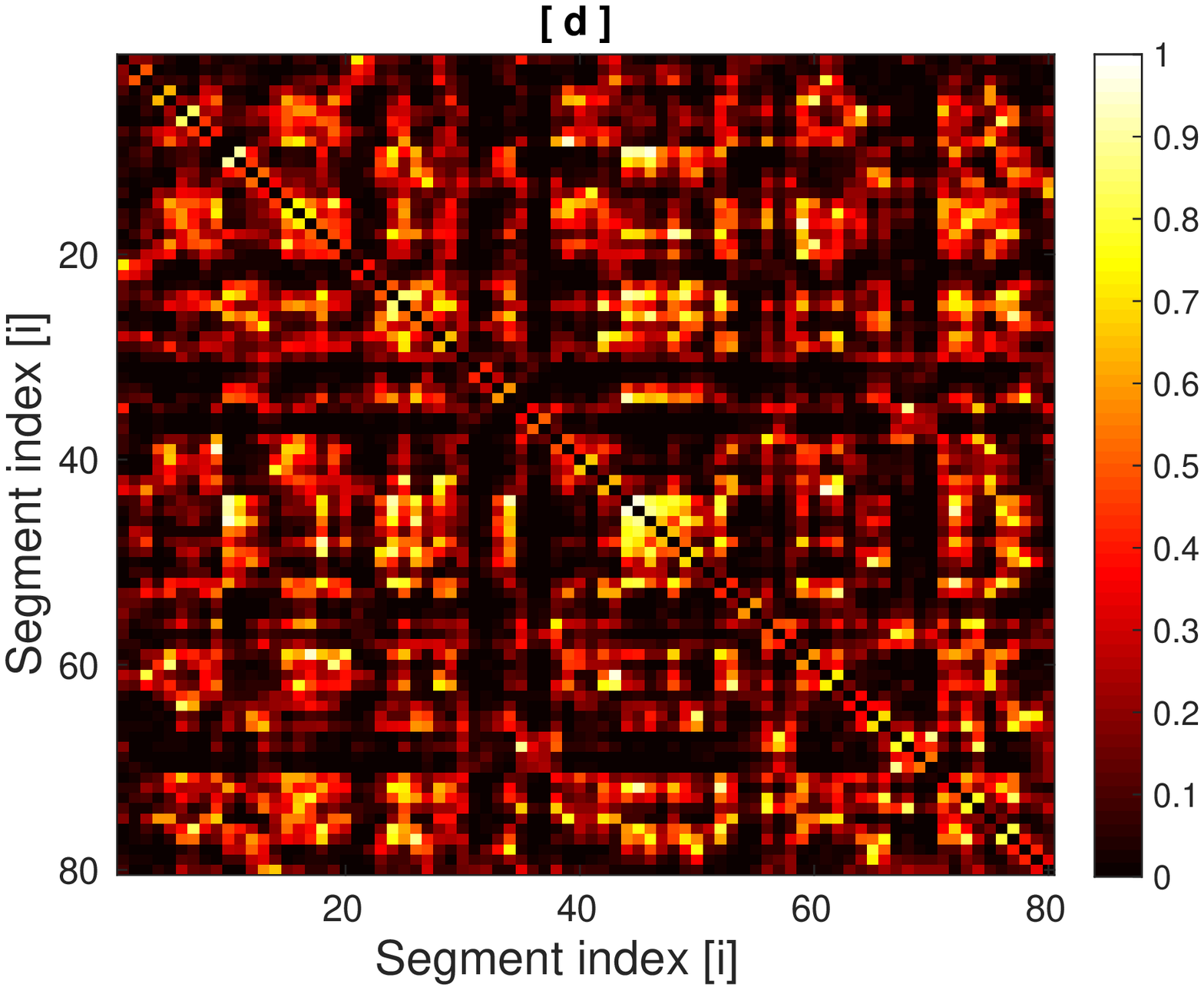} \\
\hfill
\caption{\label{possegbc2} 
Colormaps to represent probability $p(i,j)$ to find CM of segment $i$ spatially close to  CMs of other chain segments $j$. 
There are $80$ segments in the {\em E. Coli} polymer  with $58$ monomers per segment. 
The top 2 figures are runs with BC-2 and the bottom two subplots RC-2.
More colormaps from independent runs are given in Supplem. Section:  Fig.20. 
}
\end{figure}

In Fig. 
\ref{posclbc2} we show colormaps showing the average probability $p(i,j)$ 
of finding each pair CLs $i,j$ at distances of $< 5a$ for 
 BC-2, RC-2 for two independent runs.
As the Monte Carlo simulation evolves, at each microstate if the distance $d$ 
between a pair of CLs is such that $d<5 a$, a counter $c(i,j)$ for pair $i,j$ is incremented. 
The probability $p(i,j)$ at the end of the MC-run is the value of $p(i,j) = c(i,j)/N_{micro}$, 
where $N_{micro}$ is the number of microstates over which data is calculated for calculation.   
The x-axis and the y-axis represent CL indices $i,j$, and the colored pixel indicates the value of $p(i,j)$. 
The top two colormaps of Fig.\ref{posclbc2}) represent data obtained for BC-2,
and  the bottom two colormaps  Fig.\ref{posclbc2} show corresponding data  from two independent runs with 
RC-2 set of CLs.  A pair of CLs which are near each other along the contour of the chain will have the 
distance $d < 5a$ between them by default, and will show up as high probabilities in the colormap. 
We set these $p(i,j)=0$ in the calculation if the monomers constituting pair of nearby CLs are separated 
by less than $6$ monomers along the contour.
We do this because we want to see only non-trivial correlations between different CLs. 
Following  Fig. 
\ref{posclbc2}, the colormaps show probability of finding a pair of segment-CMs
within distance of $5 a$ for BC-1/RC-1 and BC-2/RC-2 is shown in Figs.\ref{possegbc1} and \ref{possegbc2}, respectively. 
Note that these probability colormaps give much more detailed information than a pair correlation function $g(r)$, which 
would just give the average distance between CLs or segment-CMs.

Data showing probabilities $p(i,j)$ to find segments CM within a distance of $5a$ is shown in 
Figs.\ref{possegbc1} and \ref{possegbc2} for BC-1,RC-1 and BC-2,RC-2, respectively. 
We arrive at some conclusions  by comparing different pairs of probability-colormaps in
Figs. \ref{posclbc2},\ref{possegbc1} and \ref{possegbc2}. Firstly, comparing 
colormaps for data from different initial conditions, e.g. compare the top two colormaps  in
each of the figures which are for BC-1/BC-2 (or equivalently compare the bottom two 
colormaps which are for RC-1/RC-2), shows bright and dark patches at equivalent positions in the map. 
Thus the same set of CLs and segments are spatially near
each other in both the runs, i.e. the polymer organization is similar in both the runs. Additional colormaps 
from two more independent runs for each set of CLs are also given in the Supplementary section for further comparison.
The reference to relevant colormaps in the Supplementary section is given in the figure caption of each figure,
and these further reiterates our conclusion that the structural organization of DNA-polymer is similar across
different runs for the same set of CLs. Thus we find further evidence of our hypothesis that 
the set of CLs decides the large scale structure of the polymer.

\clearpage

Secondly, the number  of  the bright pixels are much more in colormaps obtained using CL sets BC-2 and RC-2 
(Figs.\ref{possegbc2}) as compared to colormaps for BC-1 (Fig. \ref{possegbc1}). 
It is not surprising as more constraints due to the presence of higher number of CLs 
lead to relatively more compact well-defined 
structure and a large number of CLs (or segments) near one another. With the few bright 
patches for BC-1, RC-1 CL set with $27$ {\em effective} CLs, one cannot clearly define the mesoscale conformation 
of the whole chain, though there are indications of the emergence of structure. However, a set of  $82$ effective CLs 
for BC-2, RC-2 might be enough to deduce  and define the large-scale organization of DNA-polymer as we now know
which segments are neighbors of a particular segment.

Thirdly, comparison of colormaps for  BC-2 and RC-2, especially in Figs. \ref{possegbc2} show a different 
nature of the organization of the DNA polymer. For BC-2 adjacent segments show higher propensity 
to be together, which can be deduced by observing that there are clusters of adjacent bright pixels.
Comparatively, bright pixels are  scattered more randomly in the colormaps for RC-2. From the colormaps, we can 
clearly, observe that there is a difference in the nature of patterns for BC-2 and RC-2.

\begin{figure}[!ht]
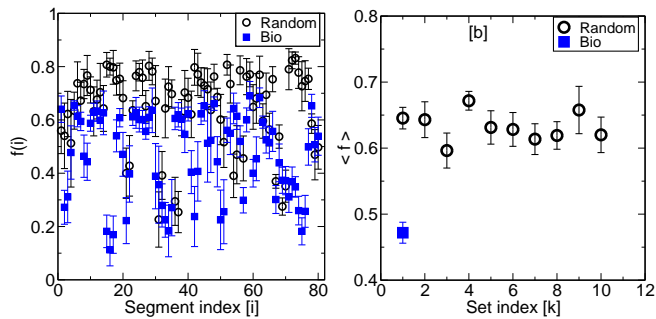

\includegraphics[width=0.49\columnwidth]{diff_cm_dom_corr1.eps}
\includegraphics[width=0.49\columnwidth]{diff_cm_dom_corr1_avg.eps}
\caption{\label{diff_colormap}
Subplot (a) shows number of pixels $f(i)$ with probability $p(i,j) > 0.05$ for a particular segment $i$ 
in the Fig. \ref{possegbc2}, normalized by the total number of segments versus the segment index.
Subplot (b) shows $<f>$ for $10$ distinct random CL sets, labelled $K=1..10$ and  one biologically determined CL set. 
 The error bars show the standard deviation in $f_{av}$ (see text) calculated for the $9$ independent runs for  each 
CL-set. Eighty-two pairs of monomers have been chosen randomly and then cross-linked for each CL set.   
}
\end{figure} 

Fourthly and importantly, the reasons for the  formation of clusters of bright pixels seen in the top two colormaps of Figs. 
\ref{posclbc2} (for CLs) is not the same as that of Fig.\ref{possegbc2} (for segment-CMs). 
To understand the bright patches of Fig. \ref{posclbc2}, we remind the reader
that the  CLs are often found adjacent to each other  along the chain contour for BC-1 and BC-2. 
Suppose, CL-$i$, CL-$j$ and CL-$k$ are next to each other along the chain. Note that then $p(i,j)$, $p(i,k)$,$p(j,k)$ 
has been explicitly put to zero. But if CL-$m$, which is far from $i$ and $j$ along the contour, 
comes within a distance of $5a$ from 
CL-$i$, then CL-$m$ is also automatically close to CL-$j,k$ and three adjacent pixels will appear in the colormap, viz., 
$p(i,m)$, $p(j,m)$,$p(k,m)$.  Thus, the bigger bright patches for BC-2 in Fig. \ref{posclbc2} 
should not necessarily be interpreted  as evidence for a  more organized polymer. 
A similar arrangement of bright/dark pixels across runs is just evidence of similar organization across different runs.
 
To quantify the differences in the colormaps of BC-2 and RC-2 in Fig. \ref{possegbc2},
we calculate the number of segments, $n_{seg}(i)$, 
which are near (i.e. within distance $d<5a$) to the CM of the $i$-th segment 
with probability $p(i,j) >0.05$. That is we count the number of non-black pixels in the colormaps of fig.\ref{possegbc2})
for a particular segment with index $i$. 
Then we divide $n_{seg}(i)$ by the total number of segments to get $f(i)$ to get an estimate of the 
fraction of a total number of segments which approach segment $i$ with any finite probability.
It is shown in the Fig.\ref{diff_colormap} for RC-2 and BC-2. A cutoff of $0.05$ for the value of $p(i,j)$
is appropriate as anyways most of the colormap is black and deep red going upto yellow for very few pixels.
From the figure,  we observe that  the value of $f(i)$ is relatively high  for RC-2 set of CLs as compared 
to $f(i)$ for bio BC-2, this suggests for random CL-set many more segments can approach
a particular segment for RC-2 compared to that for BC-2. We interpret this as a more  spatially organized
structure with BC-2 cross-links, as it has fewer but well-defined neighbors as can also be checked from the 
colormap of Fig.\ref{possegbc2}.   
As an example,  segments with indices 70-78 for BC-2 are only close to their adjacent segments 
(bright diagonal patch in the colormap) giving relatively very low value of $f(i)$ in Fig.\ref{diff_colormap}(a).

We have also obtained colormaps for the $10$ different sets of random CLs (data not shown),
and for each CL-set we can calculate $f(i)$ for each segment index $i$. Moreover, we can calculate 
$f_{av} $, that is the average value of $f(i)$ summed over all the segment indices, i.e.  
$f_{av} = (\sum_i f(i))/80$.  Furthermore, we can calculate the mean of $f_{av}$ over $9$ independent 
runs for each CL set, and thereby obtain $\langle f \rangle$. In Fig.\ref{diff_colormap}(b), we plot 
$\langle f \rangle$ versus the random CL-set index, each set has the same number of CLs as in RC-2. 
We compare this data with the $\langle f \rangle$ for the one set of biologically obtained CLs: BC-2.
We clearly see that for each random CL-sets the quantity $\langle f \rangle$ has relatively higher value 
than $\langle$f$\rangle$ for  BC-2. Observing the differences in colormaps for BC-2 and RC-2,
we claim that the position of CLs along the chain for DNA are not completely random. 
An equivalent number of CLs in random positions also give an organized structure in that the colormaps
from $9$ independent runs look similar, but the nature of organization 
is very different from the case where biological position of CLs are chosen. 

To extract further insight into the structural organization of the DNA-polymer, we would next probe whether the
segments are at geometrically fixed positions with respect to each other, of course
accounting for thermal fluctuations. Thereby, we next calculate  the angular correlations between CLs
and equivalently between segment's CMs.

\begin{figure}[!ht]
\includegraphics[width=0.7\columnwidth]{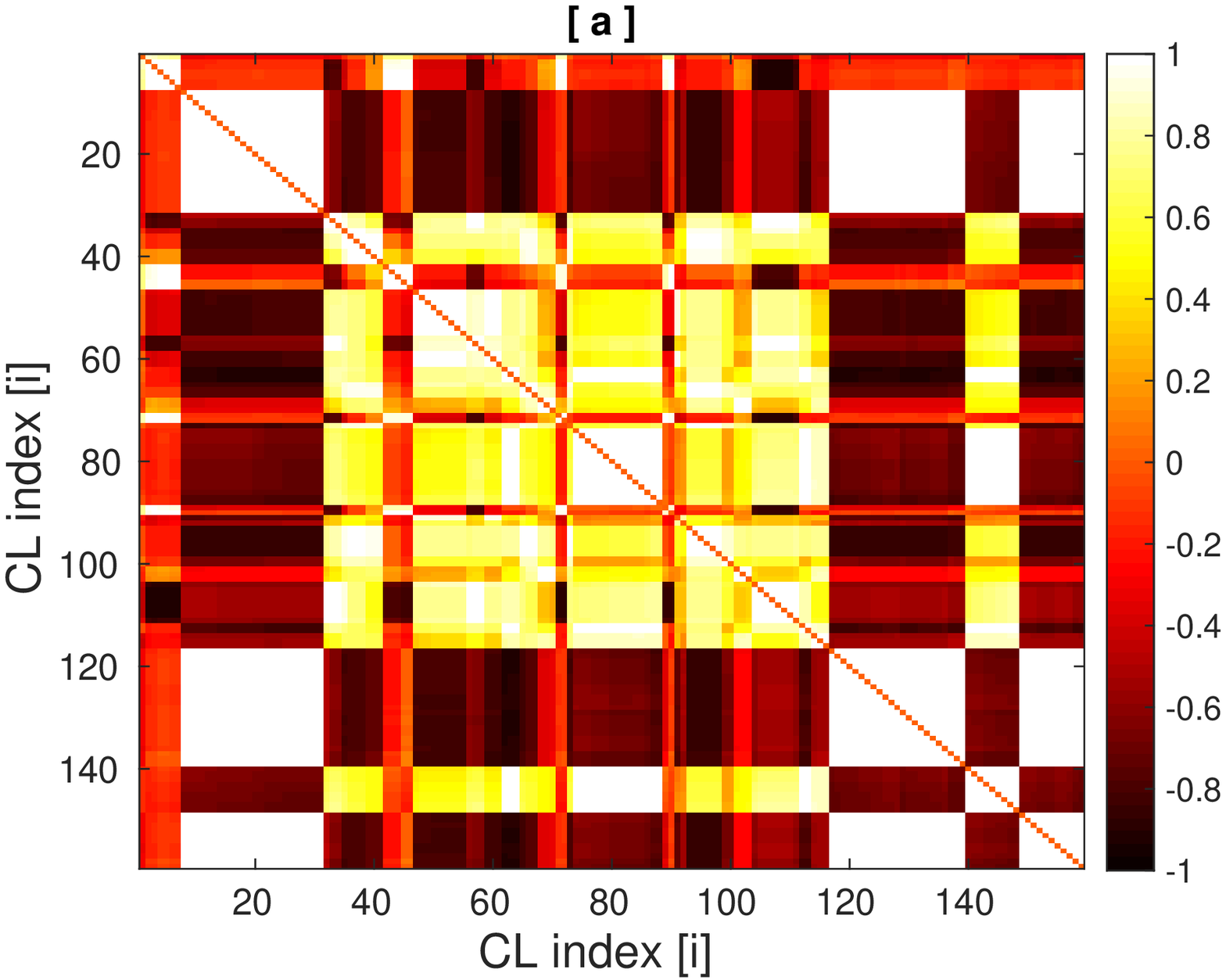} \\
\vskip0.05cm
\includegraphics[width=0.7\columnwidth]{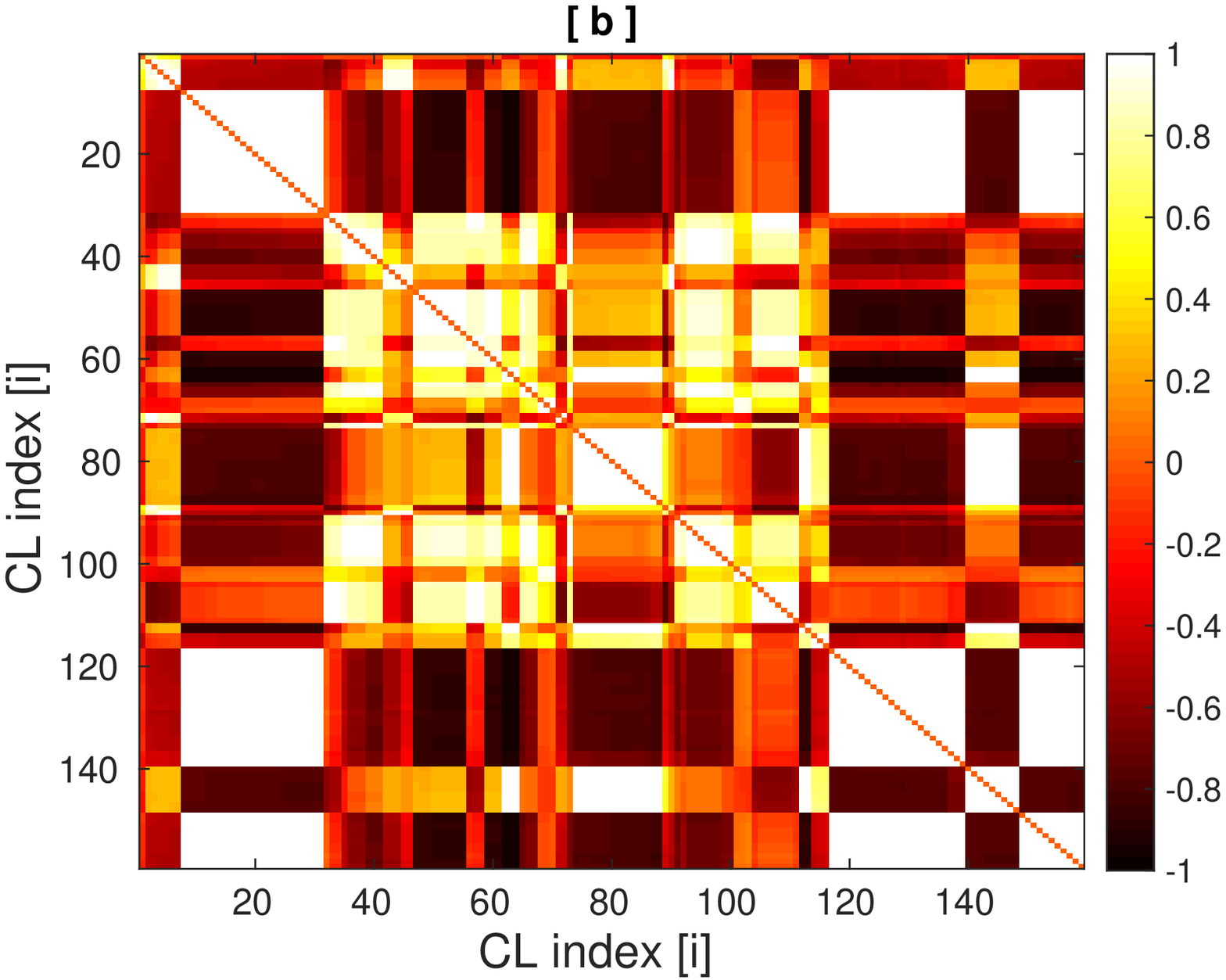} \\
\vskip0.05cm
\includegraphics[width=0.7\columnwidth]{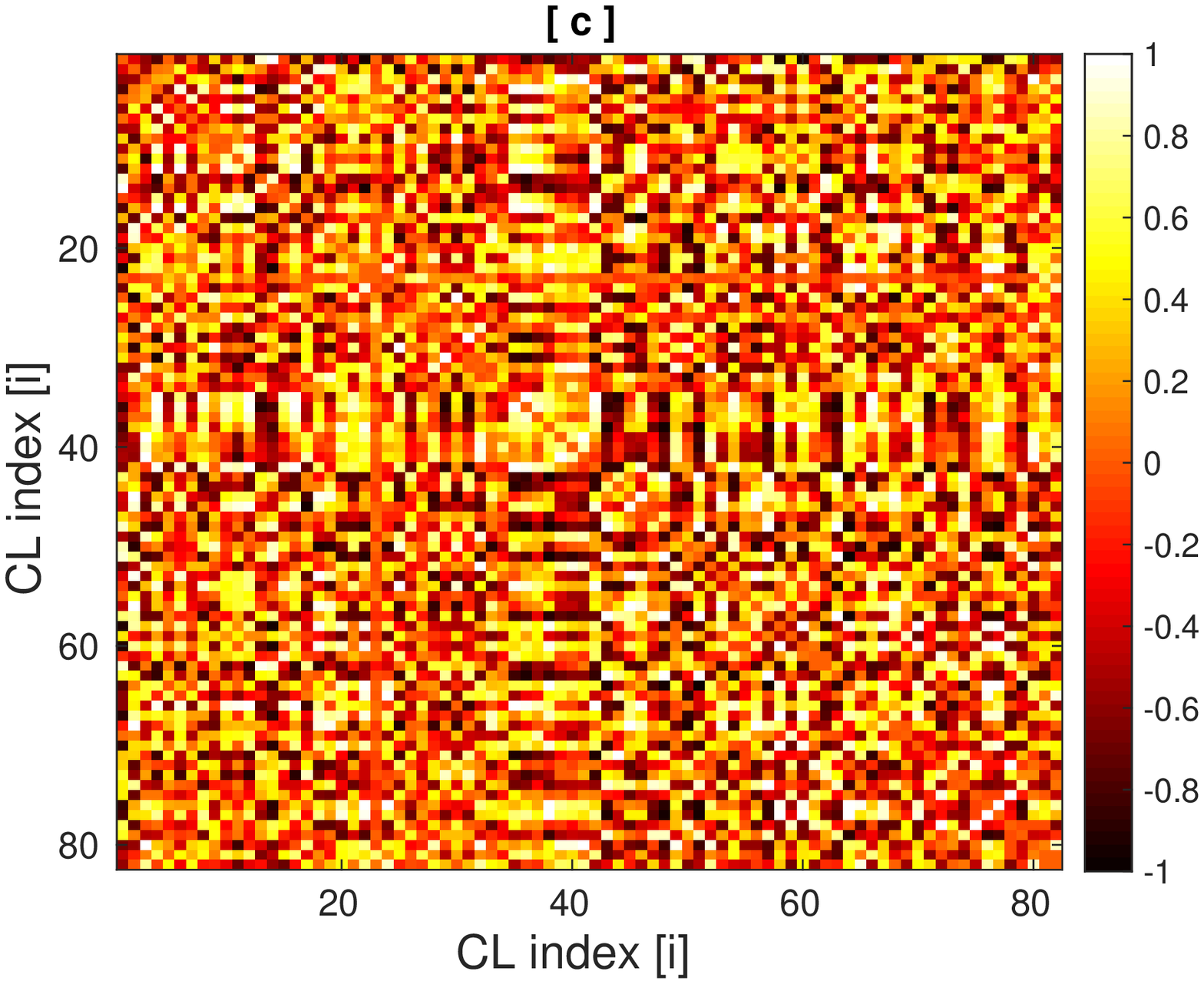} \\
\vskip0.05cm
\includegraphics[width=0.7\columnwidth]{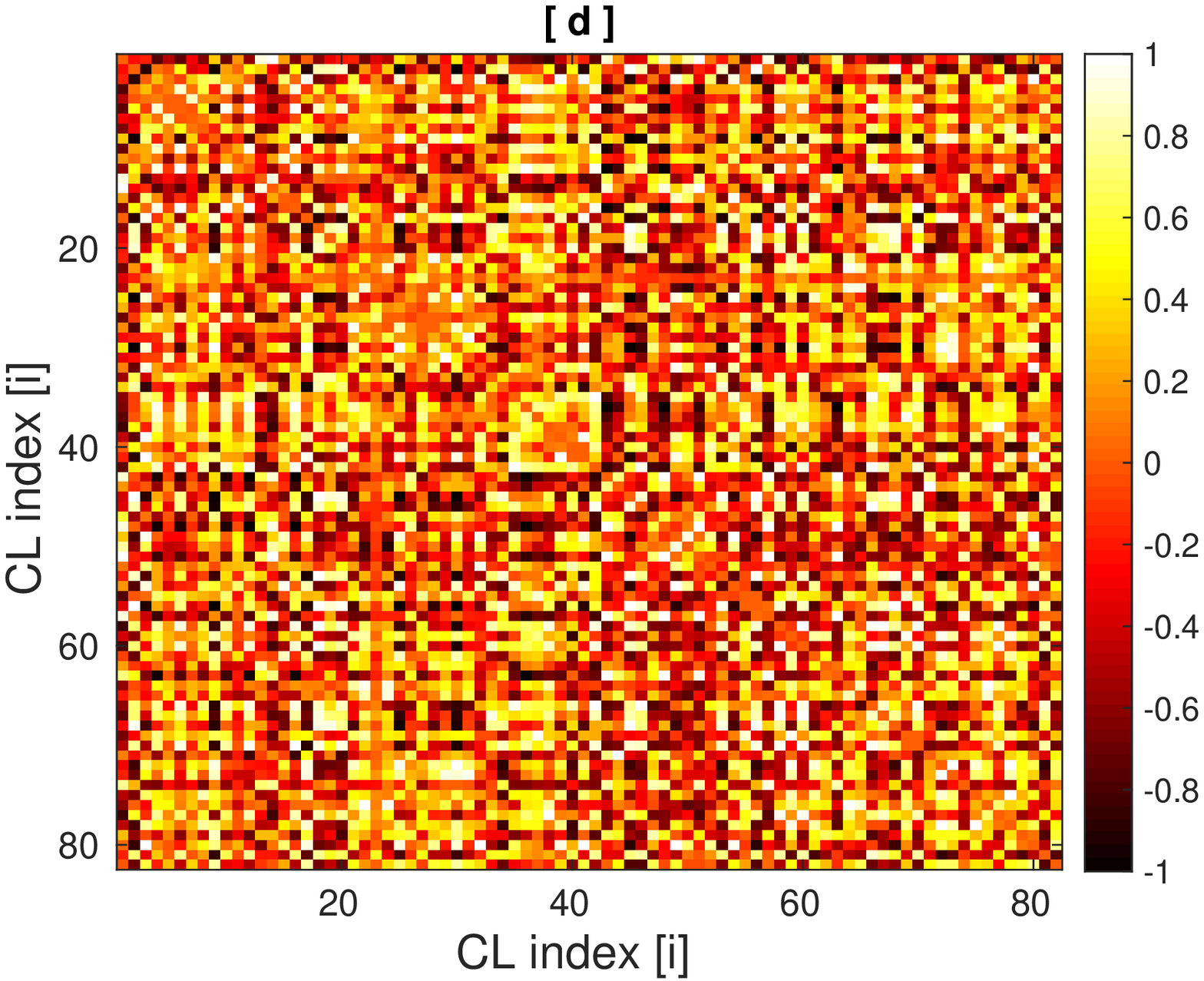} \\
\caption{\label{angclbc2}
Colormaps to investigate the angular location of different CLs with respect to each other. Subplots (a),(b)
are for BC-2 and  (c),(d) for RC-2 with different initial conditions, respectively.
Refer Supplem. Section Fig.21 to compare with more colormaps from independent runs.
}
\end{figure}

\begin{figure}[H]
\includegraphics[width=0.7\columnwidth]{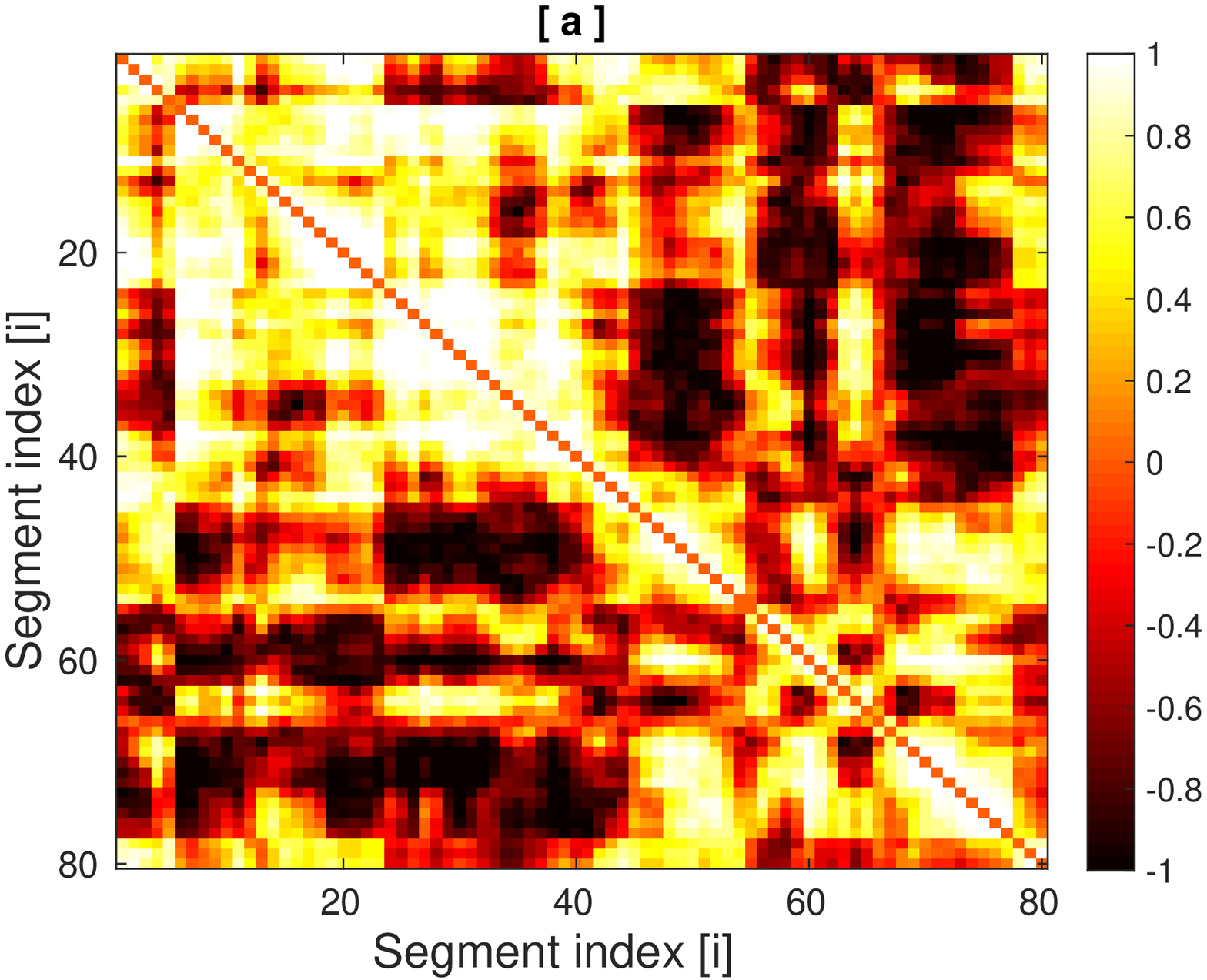} \\
\vskip0.05cm
\includegraphics[width=0.7\columnwidth]{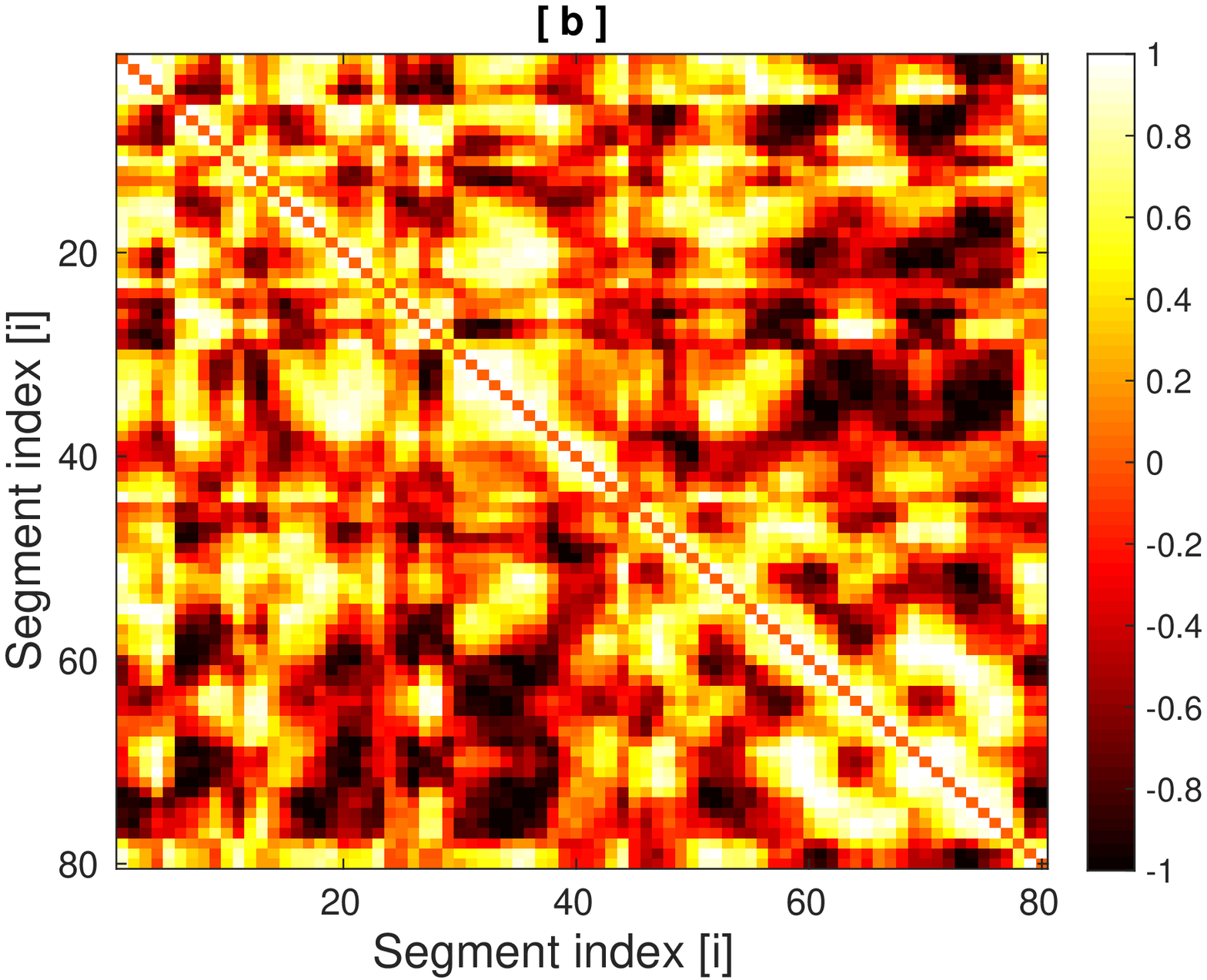} \\
\vskip0.05cm
\includegraphics[width=0.7\columnwidth]{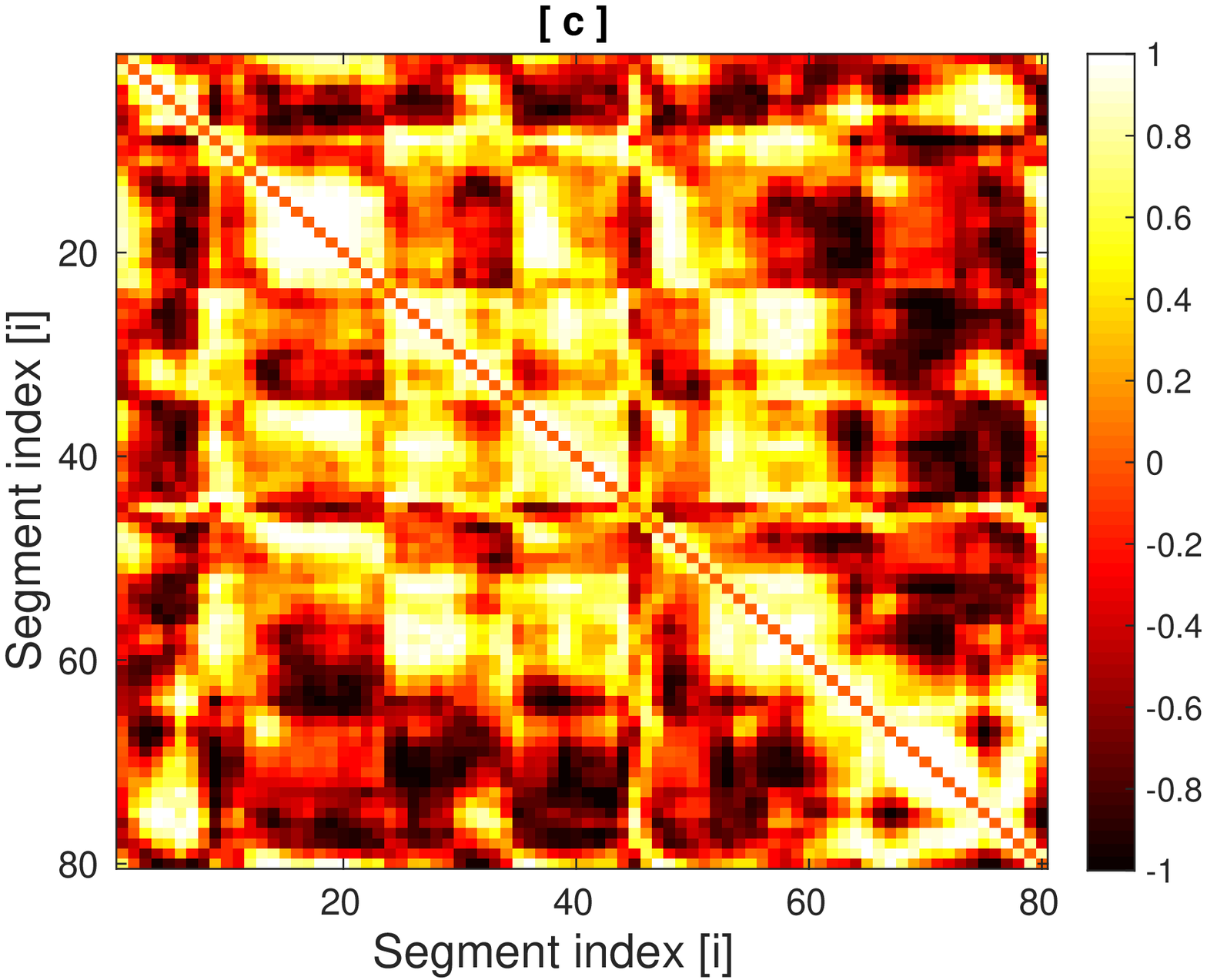} \\
\vskip0.05cm
\includegraphics[width=0.7\columnwidth]{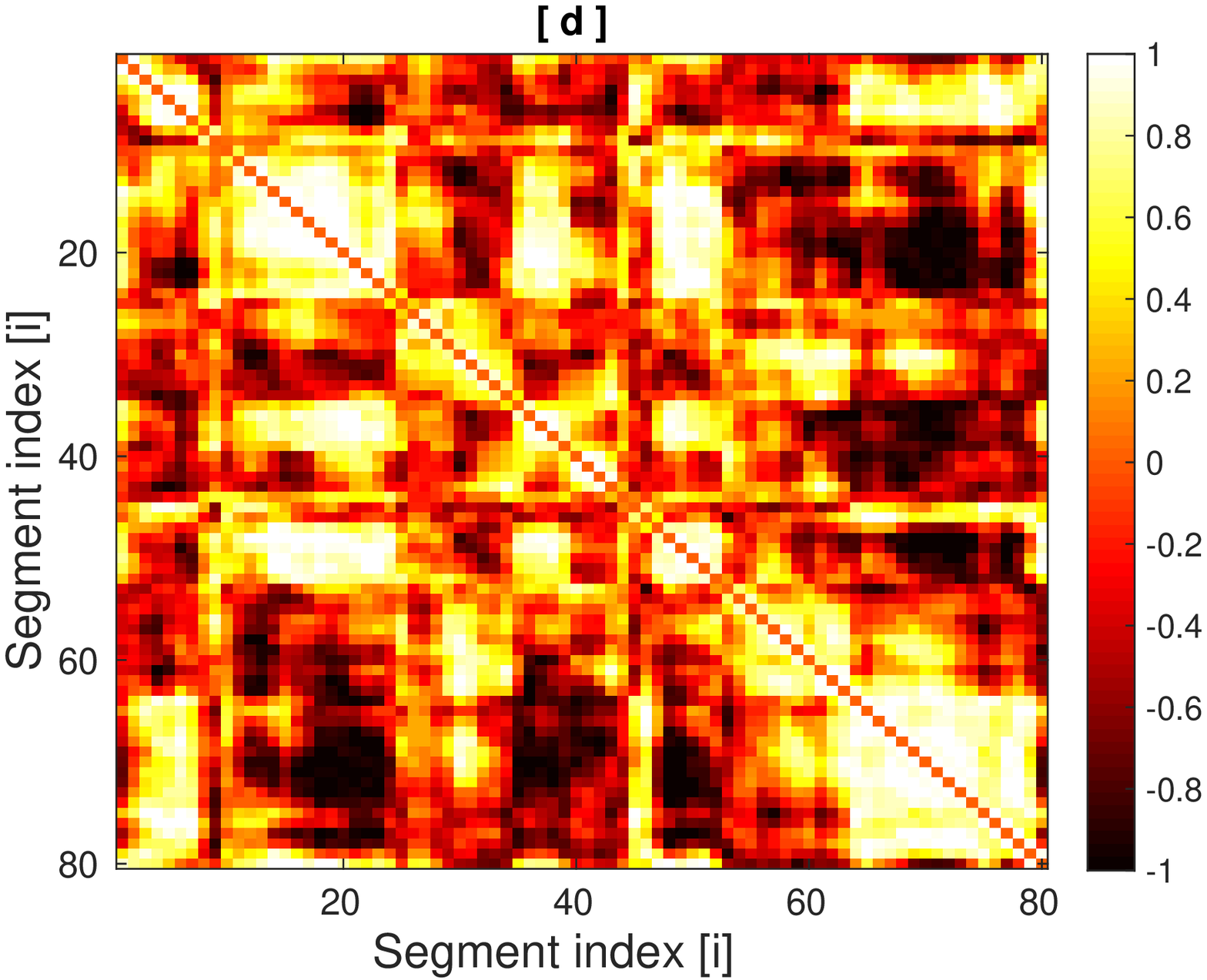} \\
\hfill
\caption{\label{angsegbc1}
Colormaps to investigate the angular location of different DNA-polymer segments respect to each other. Subplots (a),(b)
are for BC-1 and  (c),(d) for RC-1 with different initial conditions, respectively.
Refer Supplem. Section Fig.22 to compare with more colormaps from independent runs.
}
\end{figure}

\begin{figure}[H]
\includegraphics[width=0.7\columnwidth]{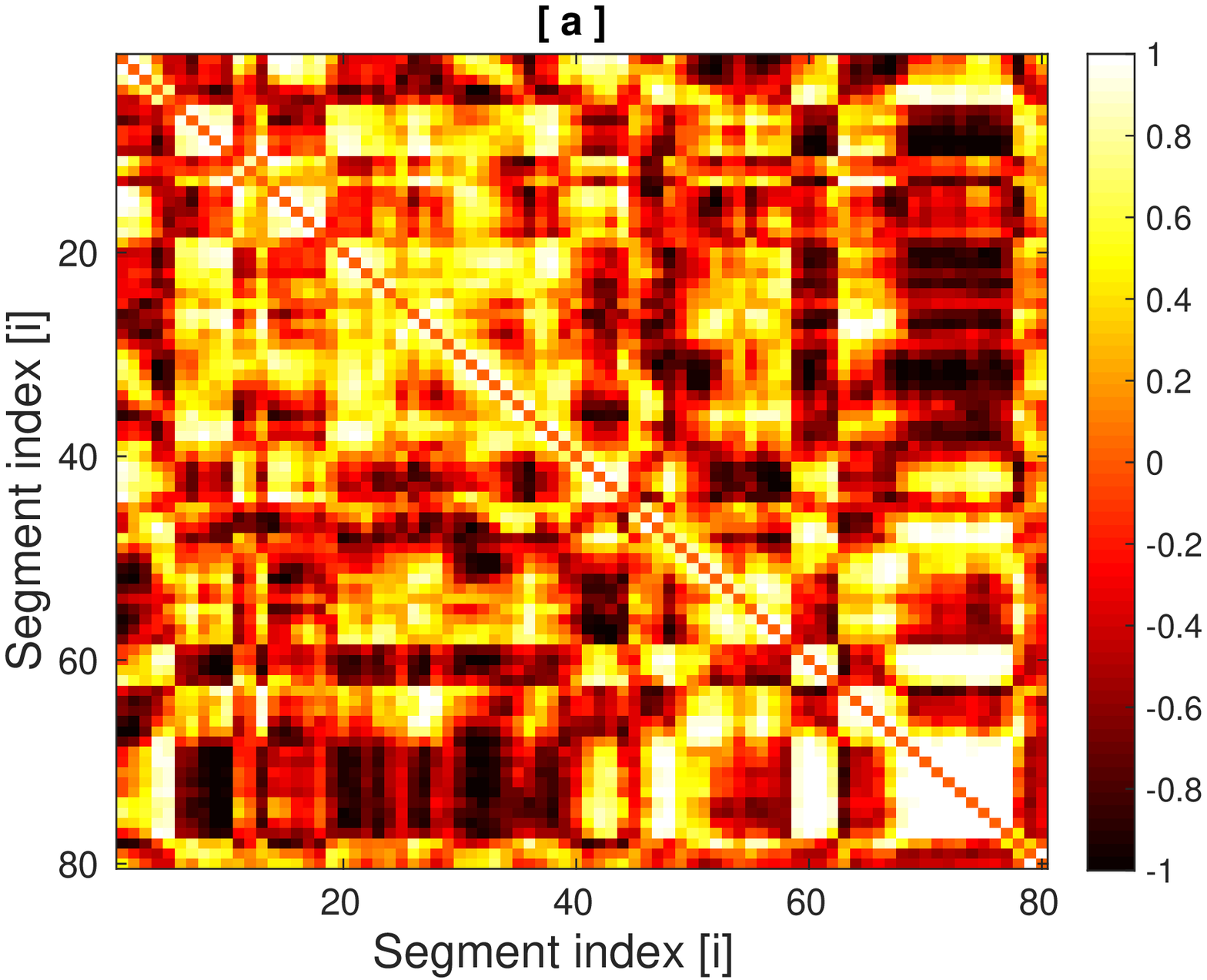} \\
\vskip0.05cm
\includegraphics[width=0.7\columnwidth]{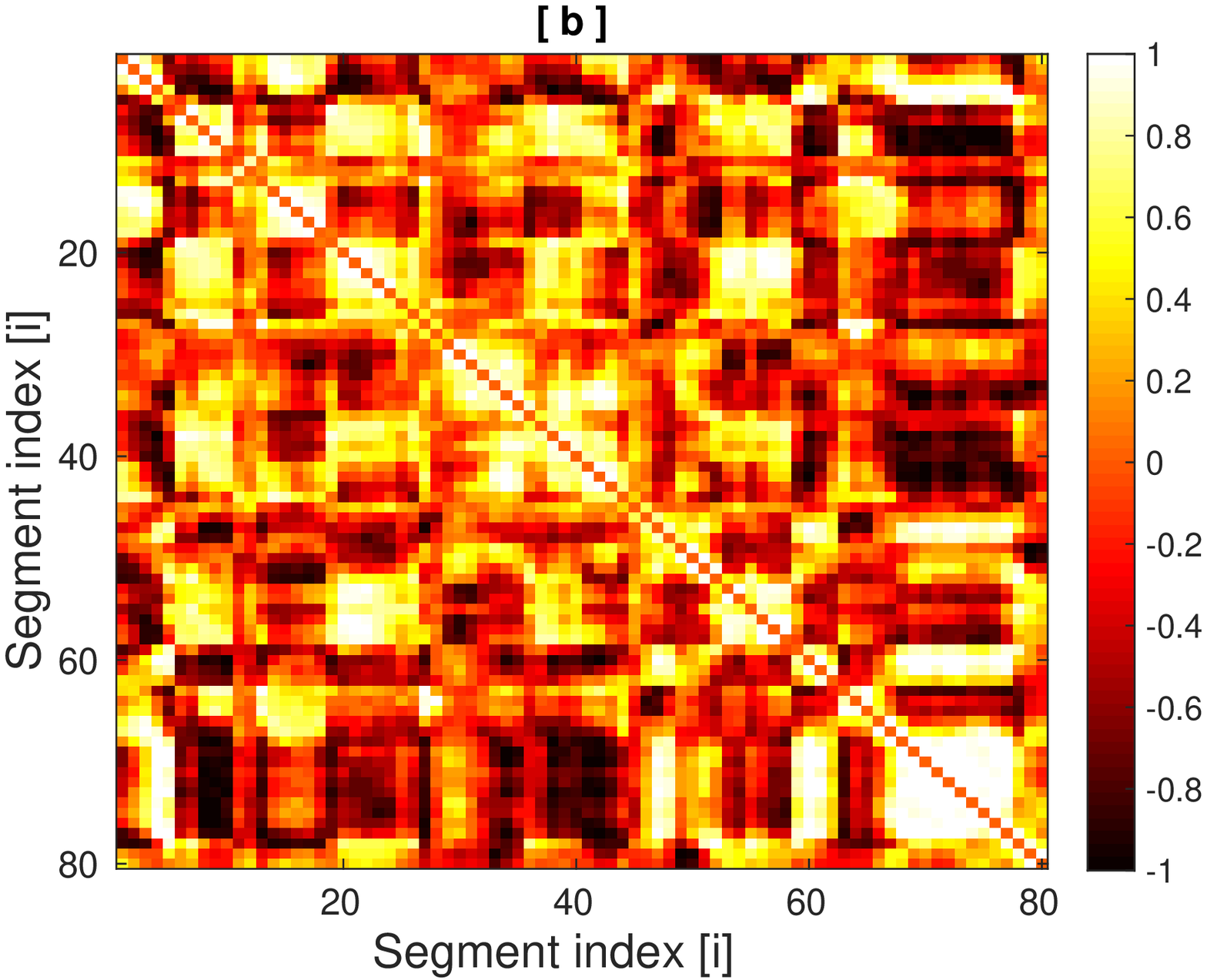} \\
\vskip0.05cm
\includegraphics[width=0.7\columnwidth]{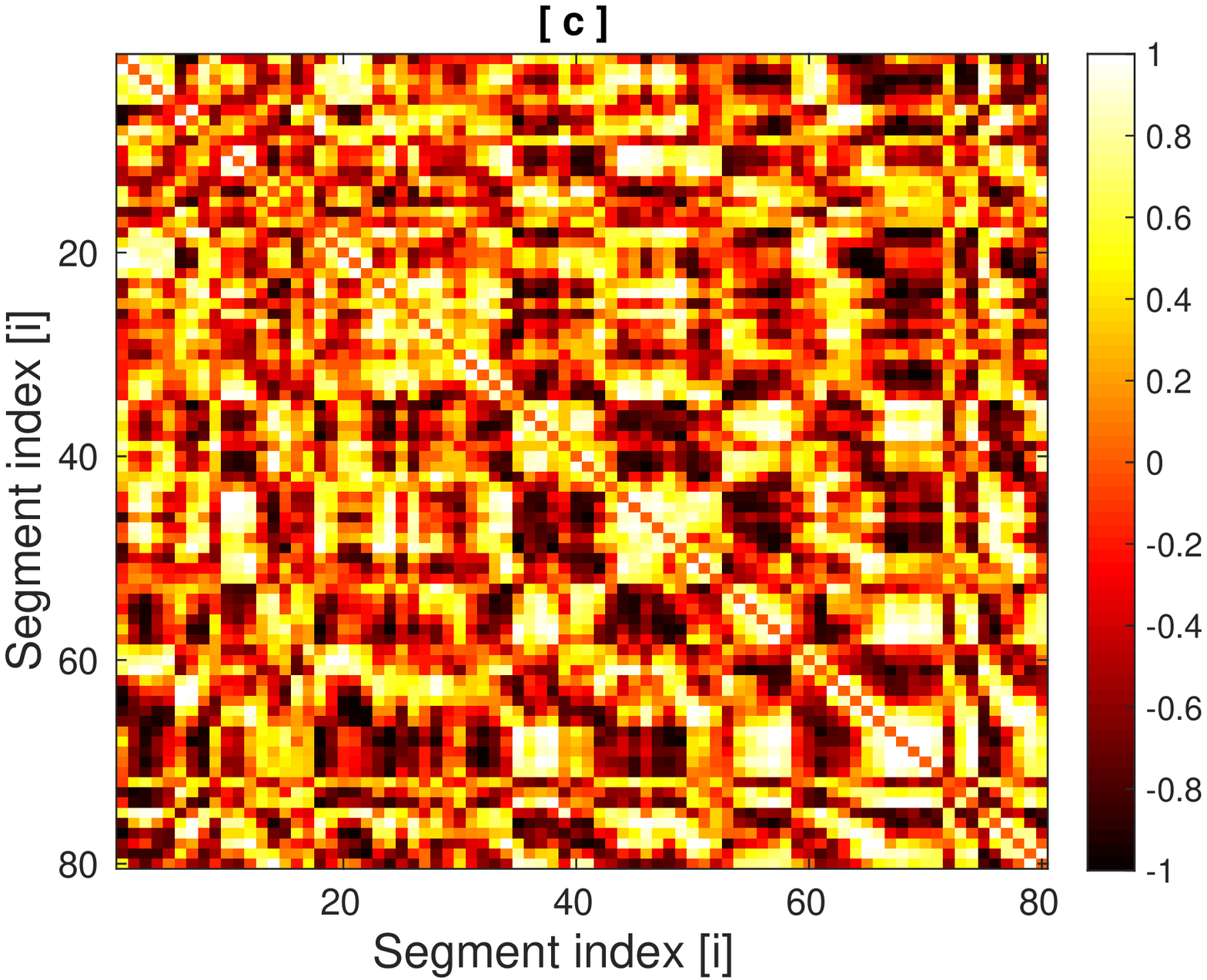} \\
\vskip0.05cm
\includegraphics[width=0.7\columnwidth]{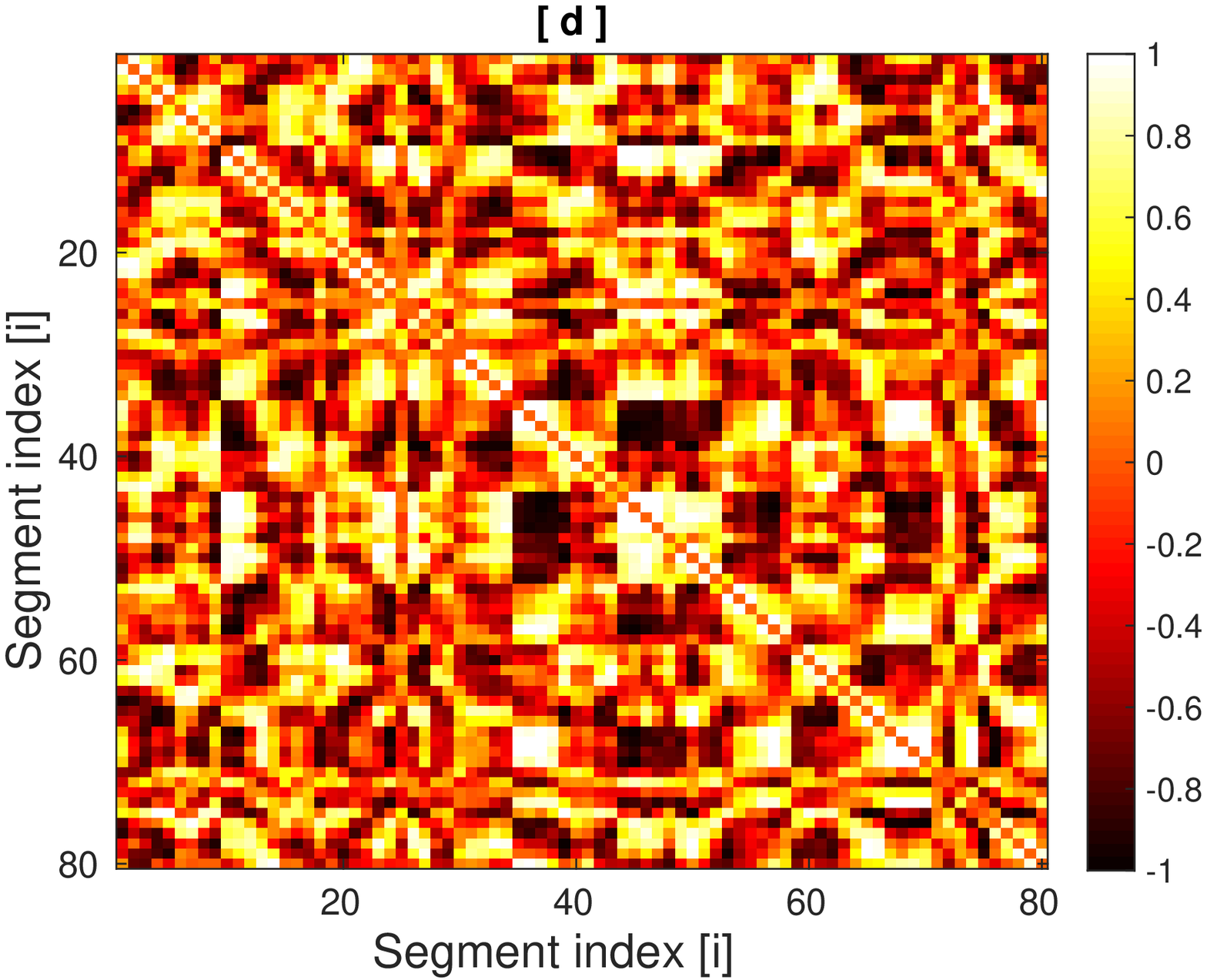} \\
\caption{\label{angsegbc2}
Colormaps to investigate the angular location of different DNA-polymer segments with respect to each other. Subplots (a),(b)
are for BC-2 and  (c),(d) for RC-2 with different initial conditions, respectively.
Refer Supplem. Section Fig.23 to compare with more colormaps from independent runs.
}
\end{figure}

To that end,
we calculate the dot product of the radial vectors from the CM of the polymer coil to the respective positions of a pair
of CLs ($i,j$) and check if the value of $\cos(\theta_{ij}) >0$ or $<0$, where $\theta_{ij}$ is the 
angle between the two  vectors. If the value of $\cos(\theta_{ij})>0$,
we can say that the two CLs are on the same side/hemisphere of the coil, and increment counter $c^{opp}(i,j)$ by
$1$. If $\cos(\theta_{ij})<0$ we decrement $c^{opp}(i,j)$ by $1$. For all possible pairs of CLs,
we calculate the average value of $\langle c^{opp}(i,j) \rangle$ suitably normalized by the number of snapshots used to calculate
the average.  The value of $\langle c^{opp}(i,j) \rangle \approx -1$ would indicate that the pair of CLs $i,j$
are always on two opposite hemispheres. A value of $\langle c^{opp}(i,j) \rangle \approx 1$ means that the two
CLs remain on the same hemisphere. 
 We should not interpret $\langle c^{opp}(i,j) \rangle 
\approx 0$ as we cannot claim that the average angle between the radial vectors
is nearly a right angle. The reason is that if the CLs are closer to the center of
the DNA-coil, small positional displacements could cause the quantity $c^{opp}(i,j)$ to fluctuate
between $1$ and $-1$ and cause $\langle c^{opp}(i,j) \rangle$ to average
out to zero. The $\langle c^{opp}(i,j) \rangle$ data for all pairs of CLs are given in
Fig.
\ref{angclbc2} for 
BC-2/RC-2 respectively,
the corresponding data for relative angular positions for the segment's
CMs are given in  Figs.\ref{angsegbc1} \& \ref{angsegbc2} for BC-1/RC-1 and BC-2/RC-2. As before, the top two colormaps in
all the four figures are from two independent initial conditions with BC-1/BC-2 and the bottom two colormaps
are for two independent runs with RC-1/RC-2.

In the colormaps of Figs.  \ref{angclbc2}, \ref{angsegbc1} and \ref{angsegbc2} we see
there are patches of bright and dark pixels, the size of patches are larger for BC-2 compared to RC-2.
 As mentioned before, if $\langle c^{opp}(i,j) \rangle=0$ which is represented by
orange/deep yellow color in the colormap we cannot predict the angular positions
of the CLs/segment's CMs because of the reason explained above.
We can clearly  see that the colormaps from independent runs starting from different initial conditions look similar.

In figure \ref{angsegbc1}, comparing segment-CM colormaps in (a),(b) (for BC-1) with (c),(d) (for RC-1)
we do not find any difference in the nature of distribution of patches.  But as the number of CLs
increase as we go from BC-1 to BC-2 and RC-1 to RC-2 in Fig.\ref{angsegbc2}, we find differences in the pattern of
colormaps on comparing (a),(b) with (c),(d) corresponding to BC-2 and RC-2 CL sets, respectively.
In contrast, for color maps (a),(b) of Fig.\ref{angsegbc2}
we observe large patches of bright pixels as compared to the patches in (c),(d). Large patches
of bright/dark pixels for BC-2 suggest adjacent segments along the chain contour are on the same/opposite
hemispheres with respect to the CM of the coil.  The small patches of bright and dark pixels in (c),(d)
for RC-2 suggest more of random distribution of different segments.  The polymer is organized in both
BC-2 and RC-2 CL sets as colormaps from independent runs look similar, but the nature of the organization is different.
The reasons for large bright patches in the colormaps for CL-angular positions as shown in
Fig. \ref{angclbc2} is not same as for the colormaps in
Fig.\ref{angsegbc2}. The reasons for the difference has been explained previously for positional correlation colormaps.

\clearpage

\begin{figure}[!ht]
\includegraphics[width=0.95\columnwidth]{snapshot.eps}\\
\includegraphics[width=0.95\columnwidth]{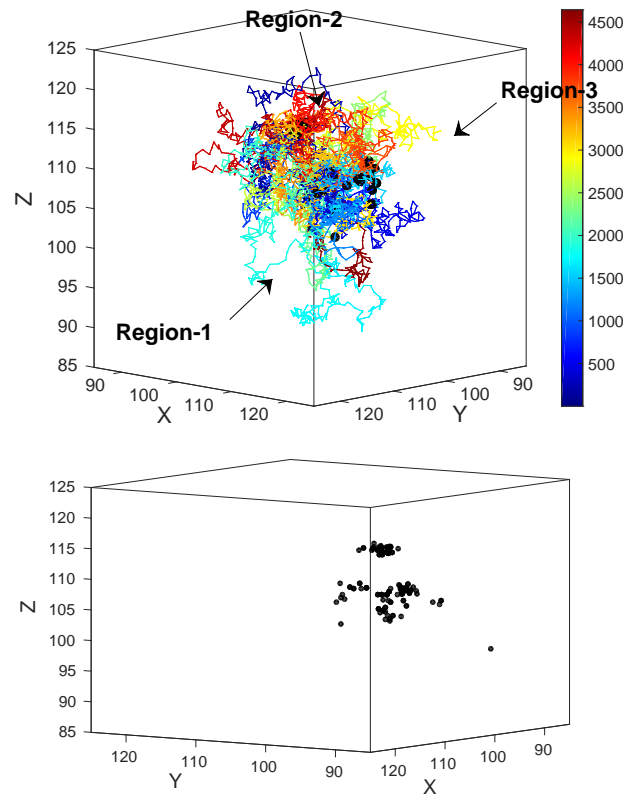}
\caption{\label{snapshot}
Representative snapshot from our simulation of DNA-polymer with BC-2 is shown in the top figure. The color bar on the right shows the color
which is used to represent the monomers numbered from $1$ to $4642$.
The black circles show the positions of CLs. The bottom figure shows
the position of CLs in space where we have removed the other monomers
for better visualization. The coordinates are the same as used for the snapshot above. In bottom snapshot we see there are approximately 4
clusters of CLs in space for BC-2 while the CLs are uniformly distributed in space for RC-2 (refer Supplementary section Fig. 24).
}
\end{figure}

Finally, we show a representative snapshot of the DNA-polymer in Fig.\ref{snapshot} (top).
The polymer is colored from blue to red as we go from monomer index 1 to 4642 along the contour.
This snapshot confirms what we have deduced from the previous figures
of positional and angular correlations. Large sections of the chain
are localized together in space. 
The snapshot confirms the kind of conformations expected from the colormaps
of angular correlation shown in Fig.\ref{angsegbc2}(a) and (b). For example, the section marked Region-1
representing monomers around 1750 (segment index 30) is diametrically opposite Region-3 with monomer
index 2990 (segment index 50).
 In Fig.\ref{angsegbc2} (a) we see the pixel corresponding to segment indices
(30,50) are black. The Region-2 represents monomer numbered around 4100, segment index 71. We can see the
pixels corresponding to segment indices (30,71) is yellow whereas pixels for (50,71) is white. The bottom
figure shows the CL distribution in space: only one of monomers out of the pair which constitutes a CL
has been plotted.

It is interesting to observe in Fig.\ref{snapshot} (bottom) that the CLs are clumped together in space in about four aggregates.
We believe that this helps in the mesoscale organization of the chain as multiple segments of the chain are pulled
towards the coil's center with multiple loops  on the periphery of the coil. The peripheral loops can
lead to relatively large fluctuations in the values of $I_1/I_3$ as seen in Fig.\ref{figI1}. This is further validated
by the Fig.\ref{probsegment2}(c), where we see a large number of segments are to be found in the outer region
with significant probabilities. Thus BC-2 set of cross-links leads to the reorganization of  the CLs in space such
that they form clusters in space with the possibility of polymer loops emanating from the CL-clusters in a rosette-like structure.
We interpret that loops from a particular CL cluster would be neighbours of
specific other polymer segments  due to the nature of arrangement, as opposed to spatial proximity to many segments as seen for
RC-2 in Fig.\ref{diff_colormap} while comparing colormaps for BC-2 and RC-2.

\begin{figure}[!ht]
\includegraphics[width=0.95\columnwidth]{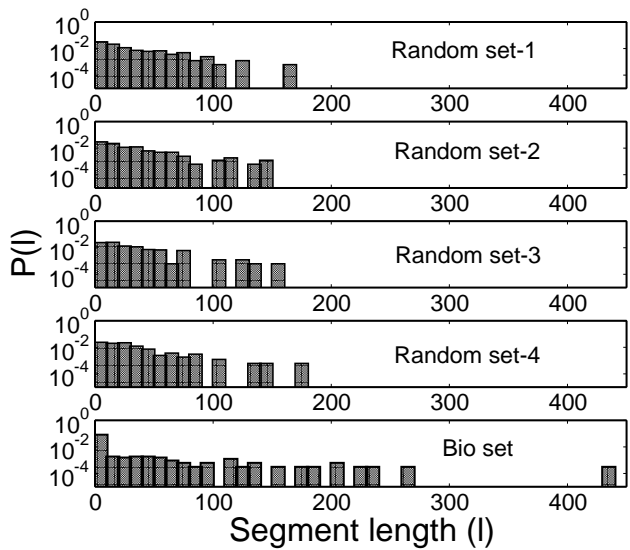}\\
\caption{\label{loop_dist}
Distribution of segment length between two adjacent CLs for BC-2 set and 4 different random CL sets corresponding to RC-2.
The x-axis shows the length of the segments between two CLs and the y-axis shows the frequency density of segment's length.
}
\end{figure}

We also calculate the distribution of length of segments between two adjacent CLs in the bio and 4
representative random sets. It is given in the Fig.\ref{loop_dist}. The distribution of lengths is a
fixed quantity once one chooses a particular CL set as an input to the simulation.
The number of segments of length $l$ between two adjacent CLs along the chain contour has been shown on x-axis (using a bin size $d\ell$ of 10 monomers) and the frequency density function (FDF) $P(l)$ is
plotted on the y-axis, where $P(l) =  N_S(l)/(N_S*d\ell)$. We denote the number of segments of  length $\ell$
by $N_S(l)$ and $N_S$ is the total number of segments between CLs. Thus $N_S=82*2 -1$ for RC-2 as each CL is constituted
of  a pair of monomers, and $l$ essentially is a count of the number of monomers between two CL-monomers along the contour.
For the bio-CL, one particular monomer is attached
with many other monomers (see supplementary material Table-1) hence there is a peak at segment length value $0$ to $10$.
We also observe in the randomly chosen CL sets after segment length $\approx150$ FDF is almost zero while
in the biologically obtained CL set there are a few segments till segment length $\approx400$. This shows in
the biological CL set there are several longer segments which can form bigger loops as compared to CLs chosen
in a random manner.

\section{Discussion}
The primary and new conclusions of our study is that if particular sets of monomers in a DNA-ring polymer are
held together by suitable proteins (cross-links at specific points in our model polymer),
it leads to an organization of the polymer coil.  The number of {\em effective} CLs is  82
for a ring polymer of 4642 monomers, or approximately $2$\% of the polymer chain. Moreover, the
monomers which are cross-linked in the bacterial DNA are not randomly chosen from the length of the contour
and lead to an organization of the ring polymer into a particular organization which is very different and distinct
compared to what is obtained using an equal number of random cross-links.
Of course, the DNA polymer undergoes local conformational fluctuations due to the
thermal energy but overall the structure is maintained in a statistical sense.
We can deduce the presence of distinctive mesoscale organization of DNA from the calculation
of three quantities: (a) radial distribution of segments, (b) positional correlations between
segments and (c) angular correlations between segments. Thus we have much more detailed
information of organization of different segments than can be obtained
from pair correlation function. We have used $159$ CLs for our simulation of DNA-polymer, but
these should be considered as only $82$ {\em effective} CLs. A minimal number of CLs are required
to be able to claim that there is a distinct organization of the DNA-polymer since we do not
obtain a well-defined structure with  $47$ bio-CLs (equivalently  $27$ {\em effective} CLs).
We can predict the 2-d arrangement of different
segments relative to each other with the statistical quantities obtained.
We find the clusters of CLs towards the center of the coil, these CLs are pulling different segments of the chain towards the center, and many loops on the periphery, which we interpret as the rosette-like structure. We have given a possible argument of how and why the  structure with relatively well localized DNA-polymer segments is achieved in a polymer, but a full understanding and systematic methodology of the choice of CL-positions from the view of polymer physics can be developed only in future, when we will have access to the larger number of contact maps of many DNAs
\section{Supplementary Material}
See supplementary material for the table of cross-linked monomers, 
colormaps of radial and angular correlations from additional runs and information of radial 
location of CLs and segments for BC-1 and RC-1.

\section{Acknowledgments}
We acknowledge the use of computational facilities provided by DBT Alliance, project numbers IA/I/12/1/500529,IA/I/11/2500290,
to C. Assisi, S. Nadkarni, M.S. Madhusudhan. We also acknowledge the use of a cluster bought using DST-SERB grant no. 
EMR/2015/000018 to A. Chatterji.  A. C. acknowledges funding support by  DST Nanomission, India under the Thematic Unit Program (grant no.: SR/NM/TP-13/2016).

\section{Appendix: Generation of contact frequency map}

In the field of bioinformatics, a sequence database is a  biological database which is 
a  collection of computerized  nucleic acid sequences. Paired-end sequencing allows researchers to 
sequence both ends of a DNA-fragment to generate high-quality, alignable sequence data. 
Paired-end sequencing facilitates detection of genomic reorganization and repetitive sequence elements. 

In a paired end-sequencing run, the distance between the alignments of the
two fragments is the length of the DNA fragment being sequenced. Aligners (software tools) use this
information to better align reads when faced with a read that align to multiple
regions such as those that may lie in a repeat region. To avoid this behavior, the reads
are aligned in single end mode while keeping track of the pairs. We have employed
the BWA \cite{farhat1} aligner to align reads as it has best sensitivity among short read
aligners.

The aligned reads are then binned at the desired resolution (or the minimum
distance between restriction sites). A 2D matrix with the required number of bins is
initialized. A large fraction of the reads in a 3C library were from fragments that were
not cross-linked and fall into the same bin or bins adjacent to each other. These read
pairs were filtered out. The counter in the bin with the coordinates indicated by the
alignment of each read in the pair is incremented for all the remaining reads. The
filled matrix gives the total number of contacts between different parts of the genome
and the resulting matrix is called the contact map.

To be able to compare between different runs, the contact map is normalized
so that effect of varying number of sequenced reads is accounted for. Each sum of the
number of contacts in each row and column in the matrix were normalized to 1. This
provides a normalized contact map, which can now be used to elucidate the 3D
structure of the genome and compare changes across different conditions.

Escherichia coli (E. coli) strain K12-MG1622 were obtained from German collection of
microorganisms and cell cultures at Leibniz institute (DSMZ).

The aligned reads were then binned at the desired resolution (or the minimum
distance between restriction sites). A 2D matrix with the required number of bins was
initialized. A large fraction of the reads in a 3C library were from fragments that were
not cross-linked and fall into the same bin or bins adjacent to each other. These read
pairs were filtered out. The counter in the bin with the coordinates indicated by the
alignment of each read in the pair is incremented for all the remaining reads. The
filled matrix gives the total number of contacts between different parts of the genome
and the resulting matrix is called the contact map.
To be able to compare between different runs, the contact map were
normalized so that effect of varying number of sequenced reads is accounted for. Each
sum of the number of contacts in each row and column in the matrix were normalized
to 1. This provides a normalized contact map, which can now be used to elucidate the
3D structure of the genome.

\bibliographystyle{apsrev4-1}
\bibliography{re_paper_tejal}
\clearpage
\pagebreak

\begin{center}
\textbf{\large Supplementary Materials}
\end{center}
\section{List of cross-linked monomers in our simulations.}
In the following table, we list the monomers which are cross-linked to model the constraints for
the DNA of bacteria Caulobacter Crecentus. Note that for random cross links (CL) set-1 and set-2 (RC-1, RC-2)
we have fewer number of CLs, as there are fewer {\em effective} CLs in the list of CLs.

In particular while counting the number of independent CLs, one should
pay special attention to the points listed below. As a consequence, $47$ CLs of BC-1 should
be counted as only $27$ independent CLs. Hence, we use just $26$ CLs in RC-1, when we compare organization
of RC-1 and  BC-1. Correspondingly, we have just $82$ CLs in RC-2, instead of $158$ in BC-2.  \\
\begin{itemize}
\item The rows corresponding to independent cross-links of set BC-1 are marked by $^*$, one can observe that the next row
of CLs are adjacent to the monomers marked just previously by $^*$. These cannot be counted as independent CLs.
\item The rows marked by $^+$ is not a independent CL at all, monomers $-$ and $-$ are trivially close to
each other by virtue of their position along the contour.
\end{itemize}
This table has been generated by analysis of raw data obtained from C. Cagliero et. al., Nucleic Acids Res, {\bf 41}, 6058-6071 (2013).

\begin{longtable*}{|l|l|l|l|l|l|l|l|l|l}
\hline
- & \multicolumn{2}{c|}{BC-1} &\multicolumn{2}{c|}{RC-1} &\multicolumn{2}{c|}{BC-2} & \multicolumn{2}{c|}{RC-2} \\
\hline
\midrule
Serial  & Monomer & Monomer & Monomer & Monomer & Monomer & Monomer &  Monomer & MOnomer  \\
no.  &  index-1 & index-2 &  index-1 & index-2&  Index-1 & Index-2 &  Index-1 & Index-2  \\
\hline
\hline
1 & 1$^*$ & 4642  & 1  & 4642  & 1  & 4642&      1   &  4642 \\
2 & 16$^*$ & 2515  & 3739 & 4531& 16  & 2515&     3739&  4531 \\
3 & 17 & 2516  & 3011 & 1610& 17  & 2516&     3011&  1610 \\
4 & 20$^*$ & 1051  & 2582 & 4367& 20  & 1051&     2582&  4367 \\
5 & 224$^*$& 2731  & 3370 & 1680& 20  & 3584&     3370&  1680 \\
6 & 225& 2731  & 556  & 2622& 21  & 1050&     556 &  2622 \\
7 & 226& 2730  & 1676 & 1426& 21  & 3584&     1676&  1426 \\
8 & 226$^*$& 3428  & 998  & 2741& 224 & 2731&     998 &  2741 \\
9 & 227& 2728  & 474  & 2233& 224 & 3429&     474 &  2233 \\
10 & 227& 2729  & 2522& 533 & 224 & 4208&     2522&  533  \\
11 & 228& 2727  & 1967& 2490& 224 & 4209&     1967&  2490 \\
12 & 228& 2728  & 2536& 616&  225 & 2730&     2536&  616  \\
13 & 229$^*$& 2727  & 769 & 4614& 225 & 2731&     769 &  4614 \\
14 & 271$^*$& 4509  & 2   & 2023& 226 & 2729&     2   &  2023 \\
15 & 272& 4508  & 3494& 2484& 226 & 2730&     3494&  2484 \\
16 & 275$^*$& 1300  & 2534& 1365& 226 & 3427&     2534&  1365 \\
17 & 280$^*$& 1051  & 3053& 2256& 226 & 3428&     3053&  2256 \\
18 & 291$^*$& 1051  & 3779& 2647& 226 & 4038&     3779&  2647 \\
19 & 316$^*$& 393  & 4199 & 4452& 226 & 4169&     4199&  4452 \\
20 & 317$^*$& 2172  & 2839& 1309& 227 & 2728&     2839&  1309 \\
21 & 382$^*$& 1469  & 1385& 449&  227 & 2729&     1385&  449  \\
22 & 383& 1469  & 4398& 371&  228 & 2727&     4398&  371  \\
23 & 527$^*$& 1529  & 522 & 1434& 228 & 2728&     522&   1434 \\
24 & 575$^*$& 1301  & 3676& 320 & 228 & 3946&     3676&  320  \\
25 & 609$^*$& 2515  & 178 &4317 & 229 & 2727&     178&   4317 \\
26 & 730$^*$& 3763  & 3220& 515 & 229 & 3424&     3220&  515  \\
27 & 731& 3764  & 527&  2992& 229 & 3947&     527&   2992 \\
28 & 732& 3765  &- & -&       229 & 3948&     2391&  1402 \\
29 & 733$^+$& 735  &- & -&        229 & 4172&     284&   4086 \\
30 & 733$^*$& 3766  & - & -&      229 & 4213&     4311&  2243 \\
31 & 1301$^*$& 3132  & - & -&     229 & 4214&     283&   1687 \\
32 & 1433$^*$& 1635 & - & -&      271 & 4509&     1599&  2420 \\
33 & 1434& 1634  & - & -&     272 & 1471&     4445&  3365 \\
34 & 1533$^*$& 3626  & - & -&     272 & 4508&     1523&  739  \\
35 & 1571$^*$& 3667  & - & -&     274 & 1301&     2721&  113  \\
36 & 1572& 3668  & - & -&     275 & 1300&     1371&  4360 \\
37 & 2728$^*$& 3945  & - & -&     275 & 1301&     4137&  2593 \\
38 & 2729& 3945  & - & -&     275 & 3130&     1026&  2807 \\
39 & 2730& 3943  & - & -&     276 & 2291&     3007&  2767 \\
40 & 3429$^*$& 3942  & - & -&     276 & 3130&     1576&  1282 \\
41 & 3471$^*$& 4177  & - & -&     276 & 3367&     3041&  3010 \\
42 & 3620$^*$& 3763 & - & -&      280 & 292 &     2558&  2709 \\
43 & 3620& 3764  & - & -&     280 & 1050&     2156&  3872 \\
44 & 3621& 3764  & - & -&     280 & 1051&     945 &  4229 \\
45 & 3622& 3765  & - & -&     291 & 1051&     4465&  2873 \\
46 & 3622& 3766  & - & -&     291 & 2760&     1943&  4488 \\
47 & 3623$^*$& 3766  & - & -&     315 & 393 &     4286&  881  \\
48 & - & -      & - & -&      316 & 391 &     3282&  3882 \\
49 & - & -      & - & -&      316 & 392 &     3555&  2445 \\
50 & - & -      & - & -&      316 & 393 &     1196&  40 \\
51 & - & -      & - & -&      317 & 392 &     1997&  3918 \\
52 & - & -      & - & -&      317 & 568 &     4178&  1595 \\
53 & - & -      & - & -&      317 & 569 &     678&   3768 \\
54 & - & -      & - & -&      317 & 1094&     3519&  164 \\
55 & - & -      & - & -&      317 & 1095&     2979&  4115 \\
56 & - & -      & - & -&      317 & 2172&     2871&  3747 \\
57 & - & -      & - & -&     382 & 1469&      3930&  4263 \\
58 & - & -      & - & -&     383 & 1468&      2787&  2654 \\
59 & - & -      & - & -&     383 & 1469&      1101&  831  \\
60 & - & -      & - & -&     393 & 567&       2785&  1485 \\
61 & - & -      & - & -&     393 & 1096&      3477&  1069 \\
62 & - & -      & - & -&     393 & 2171&      2345&  795 \\
63 & - & -      & - & -&     526 & 1529&      4037&  3848 \\
64 & - & -      & - & -&     527 & 1529&      395&   1040 \\
65 & - & -      & - & -&     527 & 1530&      328&   930  \\
66 & - & -      & - & -&     575 & 1301&      1926&  2551 \\
67 & - & -      & - & -&     576 & 3130&      4440&  1484 \\
68 & - & -      & - & -&     576 & 3367&      3799&  4456 \\
69 & - & -      & - & -&     581 & 1636&      4129&  837 \\
70 & - & -      & - & -&     581 & 1637&      1500&  1352\\
71 & - & -      & - & -&     582 & 1636&      3197&  947 \\
72 & - & -      & - & -&     608 & 2515&      263&   3435 \\
73 & - & -      & - & -&     609 & 2515&      2272&  277 \\
74 & - & -      & - & -&     688 & 1301&      4276&  702 \\
75 & - & -      & - & -&     730 & 3763&      3405&  978 \\
76 & - & -      & - & -&     731 & 3763&      388&   3658\\
77 & - & -      & - & -&     731 & 3764&      2796&  1022\\
78 & - & -      & - & -&     732 & 3621&      3411&  1122 \\
79 & - & -      & - & -&     732 & 3764&      861&   2185 \\
80 & - & -      & - & -&     732 & 3765&      3564&  1606 \\
81 & - & -      & - & -&     733 & 735 &      1860&  1447 \\
82 & - & -      & - & -&     733 & 3623&      904&   3577 \\
83 & - & -      & - & -&     733 & 3765&      - & -\\
84 & - & -      & - & -&     733 & 3766&      - & -\\
85 & - & -      & - & -&     734 & 3765&      - & -\\
86 & - & -      & - & -&     734 & 3766&      - & -\\
87 & - & -      & - & -&     738 & 1533&      - & -\\
88 & - & -      & - & -&     738 & 3626&      - & -\\
89 & - & -      & - & -&     782 & 2522&      - & -\\
90 & - & -      & - & -&     1051 & 3585&     - & -\\
91 & - & -      & - & -&     1208 & 1210&     - & -\\
92 & - & -      & - & -&     1269 & 1271&     - & -\\
93 & - & -      & - & -&     1301 & 1398&     - & -\\
94 & - & -      & - & -&     1301 & 2102&     - & -\\
95 & - & -      & - & -&     1301 & 2289&     - & -\\
96 & - & -      & - & -&     1301 & 3132&     - & -\\
97 & - & -      & - & -&     1301 & 3366&     - & -\\
98 & - & -      & - & -&     1301 & 3652&     - & -\\
99 & - & -      & - & -&     1397 & 2573&      - & -\\
100 & - & -      & - & -&    1397 & 3118&      - & -\\
101 & - & -      & - & -&    1433 & 1635&     - & -\\
102 & - & -      & - & -&    1434 & 1634&     - & -\\
103 & - & -      & - & -&    1435 & 1633&     - & -\\
104 & - & -      & - & -&    1469 & 2071&     - & -\\
105 & - & -      & - & -&    1470 & 2071&     - & -\\
105 & - & -      & - & -&    1470 & 2071&     - & -\\
106 & - & -      & - & -&    1470 & 2998&      - & -\\
107 & - & -      & - & -&    1470 & 3186&      - & -\\
108 & - & -      & - & -&    1470 & 4498&      - & -\\
109 & - & -      & - & -&    1470 & 4508&     - & -\\
110 & - & -      & - & -&    1470 & 4509&     - & -\\
111 & - & -      & - & -&    1471 & 4508&     - & -\\
112 & - & -      & - & -&    1533 & 3625&     - & -\\
113 & - & -      & - & -&    1533 & 3626&     - & -\\
114 & - & -      & - & -&    1571 & 3667&     - & -\\
115 & - & -      & - & -&    1572 & 3667&    - & -\\
116 & - & -      & - & -&    1572 & 3668&     - & -\\
117 & - & -      & - & -&    2726 & 4172&     - & -\\
118 & - & -      & - & -&    2727 & 4172&     - & -\\
119 & - & -      & - & -&    2728 & 3945&     - & -\\
120 & - & -      & - & -&    2728 & 4039&     - & -\\
121 & - & -      & - & -&    2728 & 4171&     - & -\\
122 & - & -      & - & -&    2729 & 3945&     - & -\\
123 & - & -      & - & -&    2729 & 4038&     - & -\\
124 & - & -      & - & -&    2730 & 3943&    - & -\\
125 & - & -      & - & -&    2730 & 4038&     - & -\\
126 & - & -      & - & -&    2731 & 3942&     - & -\\
127 & - & -      & - & -&    2731 & 4036&     - & -\\
128 & - & -      & - & -&    2732 & 3942&      - & -\\
129 & - & -      & - & -&    3424 & 4172&     - & -\\
130 & - & -      & - & -&    3426 & 3945&     - & -\\
131 & - & -      & - & -&    3426 & 4156&     - & -\\
132 & - & -      & - & -&    3427 & 3944&     - & -\\
133 & - & -      & - & -&    3427 & 4038&     - & -\\
134 & - & -      & - & -&    3428 & 3943&     - & -\\
135 & - & -      & - & -&    3428 & 3944&     - & -\\
136 & - & -      & - & -&    3429 & 3942&      - & -\\
137 & - & -      & - & -&    3471 & 4177&     - & -\\
138 & - & -      & - & -&    3472 & 4176&     - & -\\
139 & - & -      & - & -&    3472 & 4177&     - & -\\
140 & - & -      & - & -&    3619 & 3763&     - & -\\
141 & - & -      & - & -&    3620 & 3763&     - & -\\
142 & - & -      & - & -&    3620 & 3764&     - & -\\
143 & - & -      & - & -&    3621 & 3764&     - & -\\
144 & - & -      & - & -&    3621 & 3765&    - & -\\
145 & - & -      & - & -&    3622 & 3765&    - & -\\
146 & - & -      & - & -&    3622 & 3766&    - & -\\
147 & - & -      & - & -&    3623 & 3766&    - & -\\
148 & - & -      & - & -&    3623 & 3768&     - & -\\
149 & - & -      & - & -&    3942 & 4167&     - & -\\
150 & - & -      & - & -&    3942 & 4209&     - & -\\
151 & - & -      & - & -&    3943 & 4038&     - & -\\
152 & - & -      & - & -&    3944 & 4038&     - & -\\
153 & - & -      & - & -&    3944 & 4169&     - & -\\
154 & - & -      & - & -&    3944 & 4210&     - & -\\
155 & - & -      & - & -&    4036 & 4167&     - & -\\
156 & - & -      & - & -&    4036 & 4209&     - & -\\
157 & - & -      & - & -&    4037 & 4443&     - & -\\
158 & - & -      & - & -&    4041 & 4214&     - & -\\
159 & - & -      & - & -&    4172 & 4214&     - & -\\
\hline
\bottomrule
\caption{\label{tab:table1}
Table shows the list of pair of monomers which constitute the  CLs for {\em E.Coli}, these CLs are
used as an input to the simulation by constraining these monomers to be at a distance $a$ from each other.
The first monomer with label $1$ and the last monomer labelled $4642$ are linked together because the DNA is a ring polymer.
}
\end{longtable*}

\null\pagebreak \null\pagebreak

\newpage

\section{Radial location of CLs and segment's CM of {\em E.Coli}.}
In the main manuscript, we show the radial organization of different CLs and segment-CMs in Figs.6
and Fig.7, respectively for BC-2 and compare it with the DNA-polymer with CLs corresponding
to RC-2, which has the same number of effective CLs as in BC-2. In the following,
we give analogous plots with BC-1 and RC-1.
\begin{figure}[hbt]
\includegraphics[width=0.45\columnwidth]{e_contact1_inner.eps}
\hfill
\includegraphics[width=0.45\columnwidth]{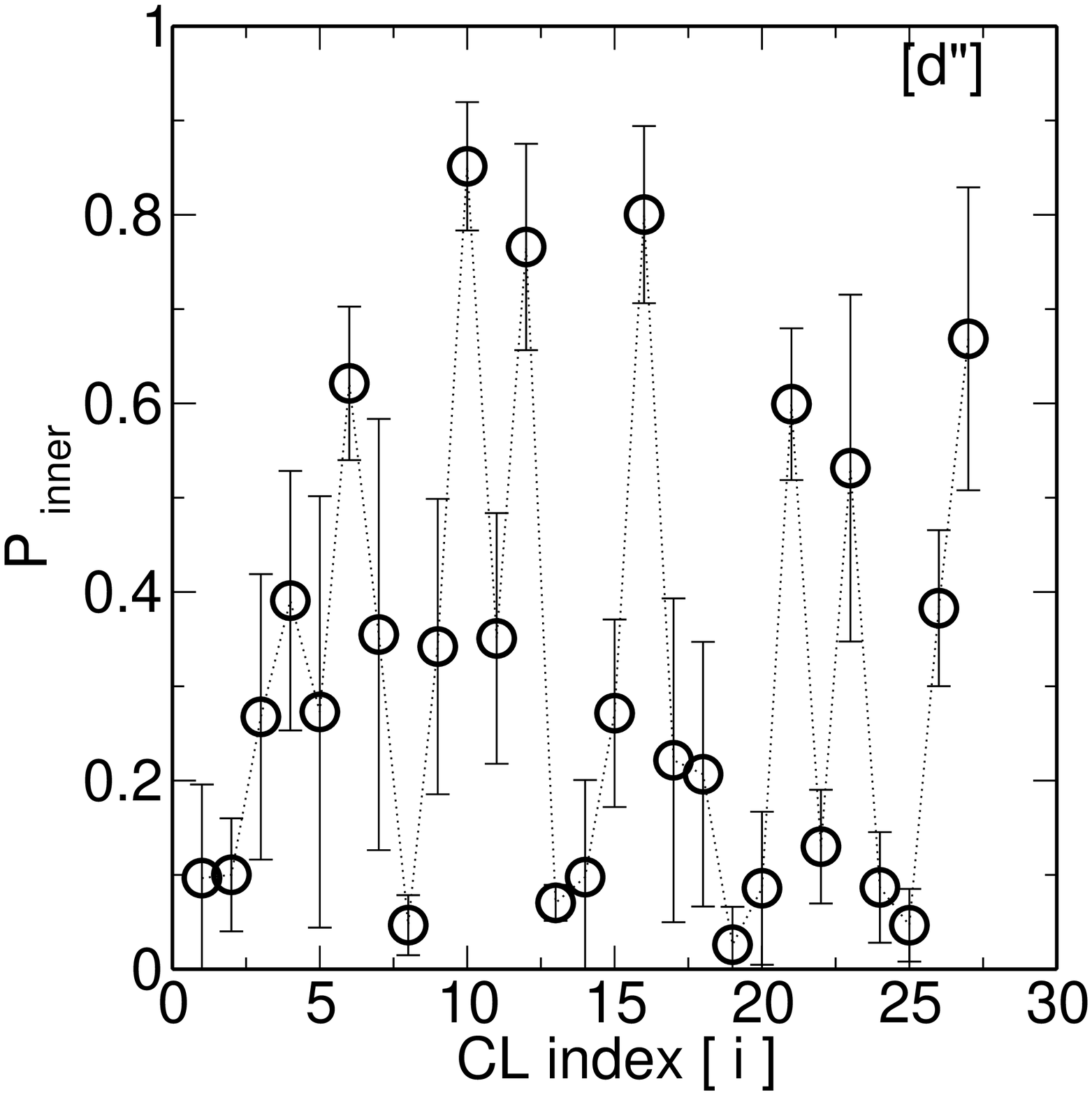} \\
\hfill
\vskip0.3cm
\includegraphics[width=0.45\columnwidth]{e_contact1_middle.eps}
\hfill
\includegraphics[width=0.45\columnwidth]{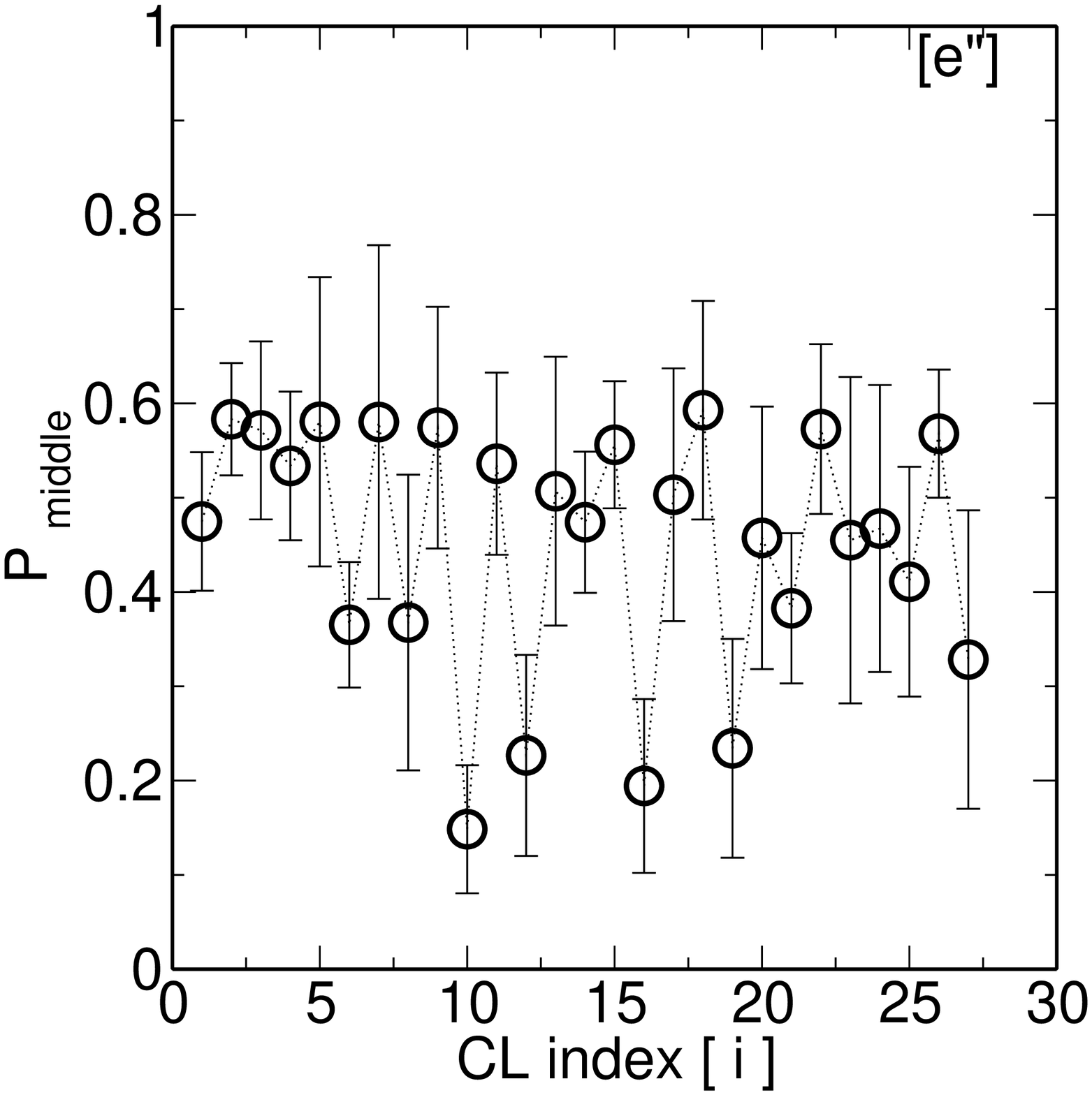} \\
\hfill
\vskip0.3cm
\includegraphics[width=0.45\columnwidth]{e_contact1_outer.eps}
\hfill
\includegraphics[width=0.45\columnwidth]{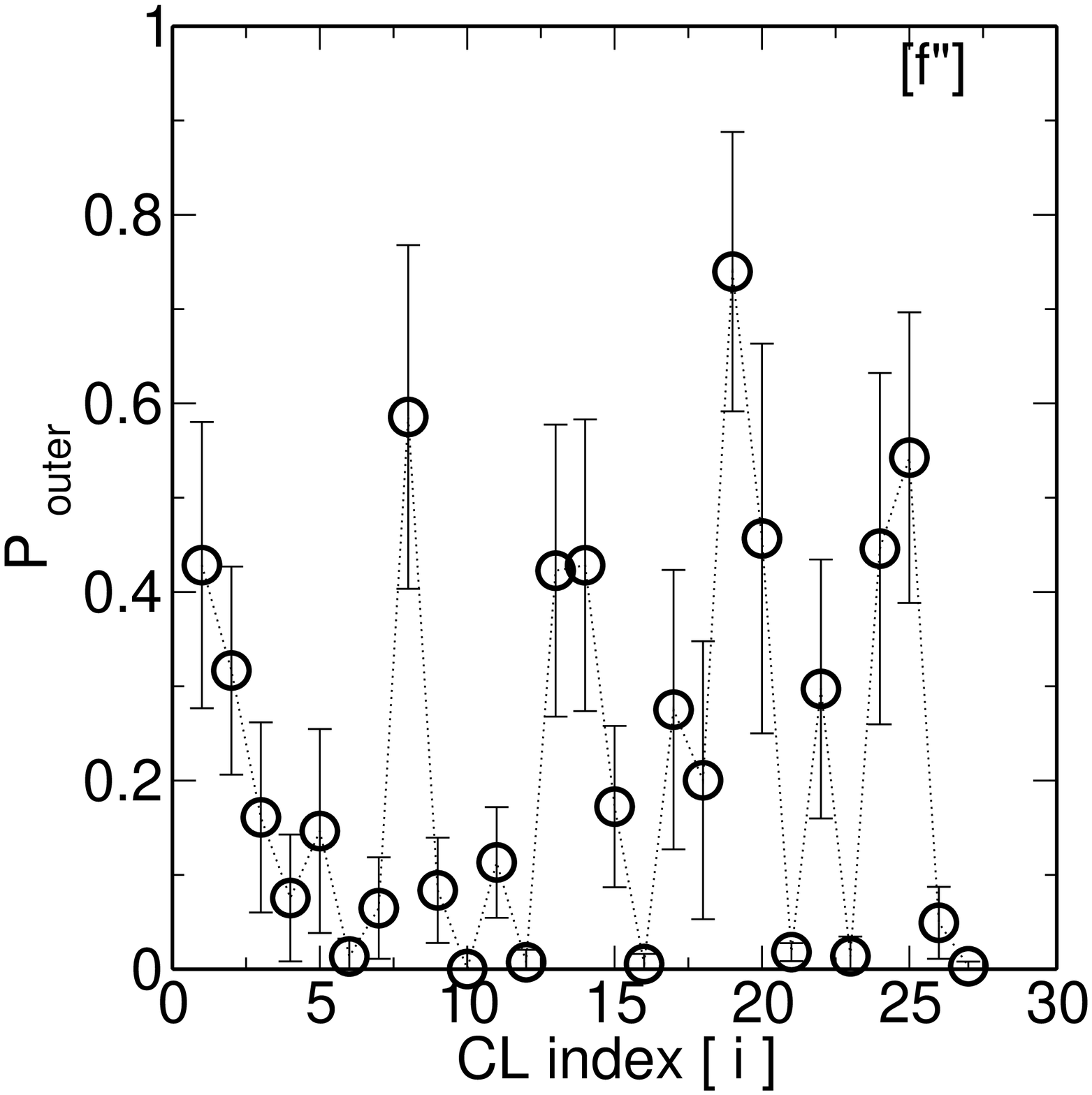}
\hfill
\vskip0.3cm
\caption{\label{Suppfigprobcl1}
Data for bacteria {\em E.Coli} with no. of CLs = BC-1 :
Subplots (a"), (b") and (c") show the probabilities of CLs to be
found in the inner, middle and outer region of DNA globule.
The x-axis is segment index, Each valuue of probability is the average over 9 independent initial conditions.
Eroor bar shows the standard deviation.
Subplots (d"), (e"), (f") are for RC-1. Each segment has 58 monomers, the dna-polymer has around 80 segments.
}
\end{figure}
\begin{figure}[hbt]
\includegraphics[width=0.49\columnwidth]{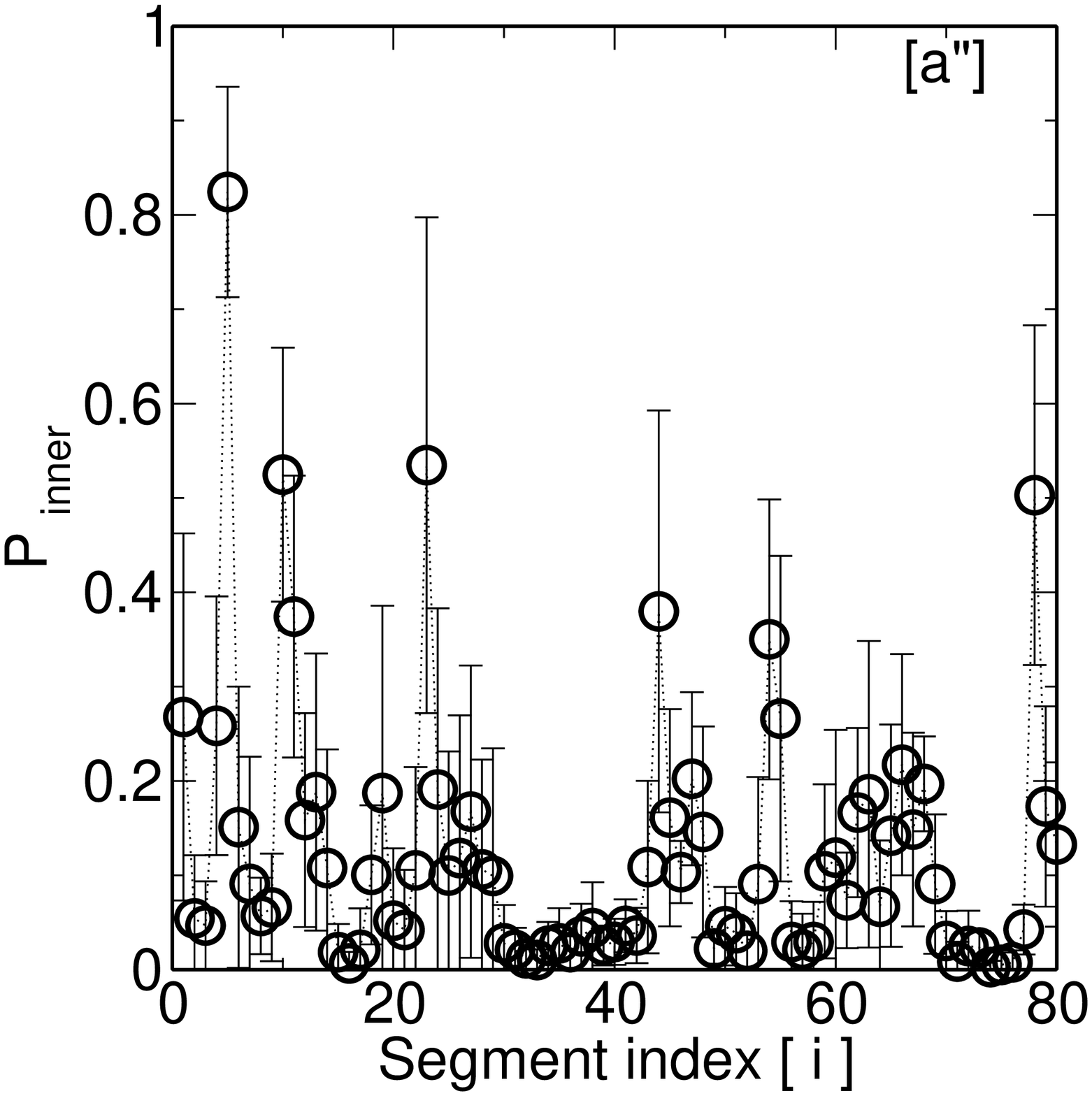}
\hfill
\includegraphics[width=0.49\columnwidth]{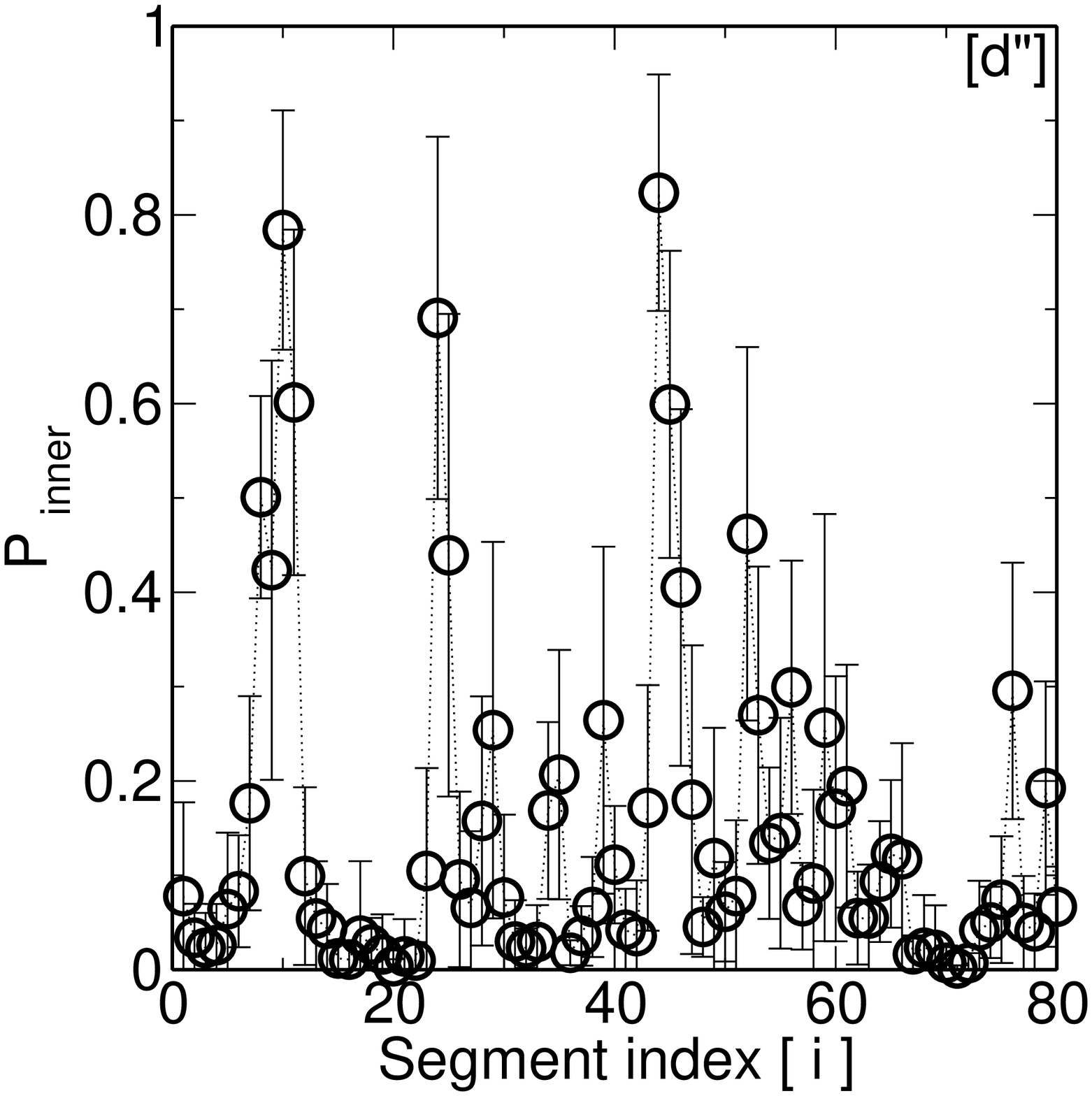}\\
\hfill
\vskip0.3cm
\includegraphics[width=0.49\columnwidth]{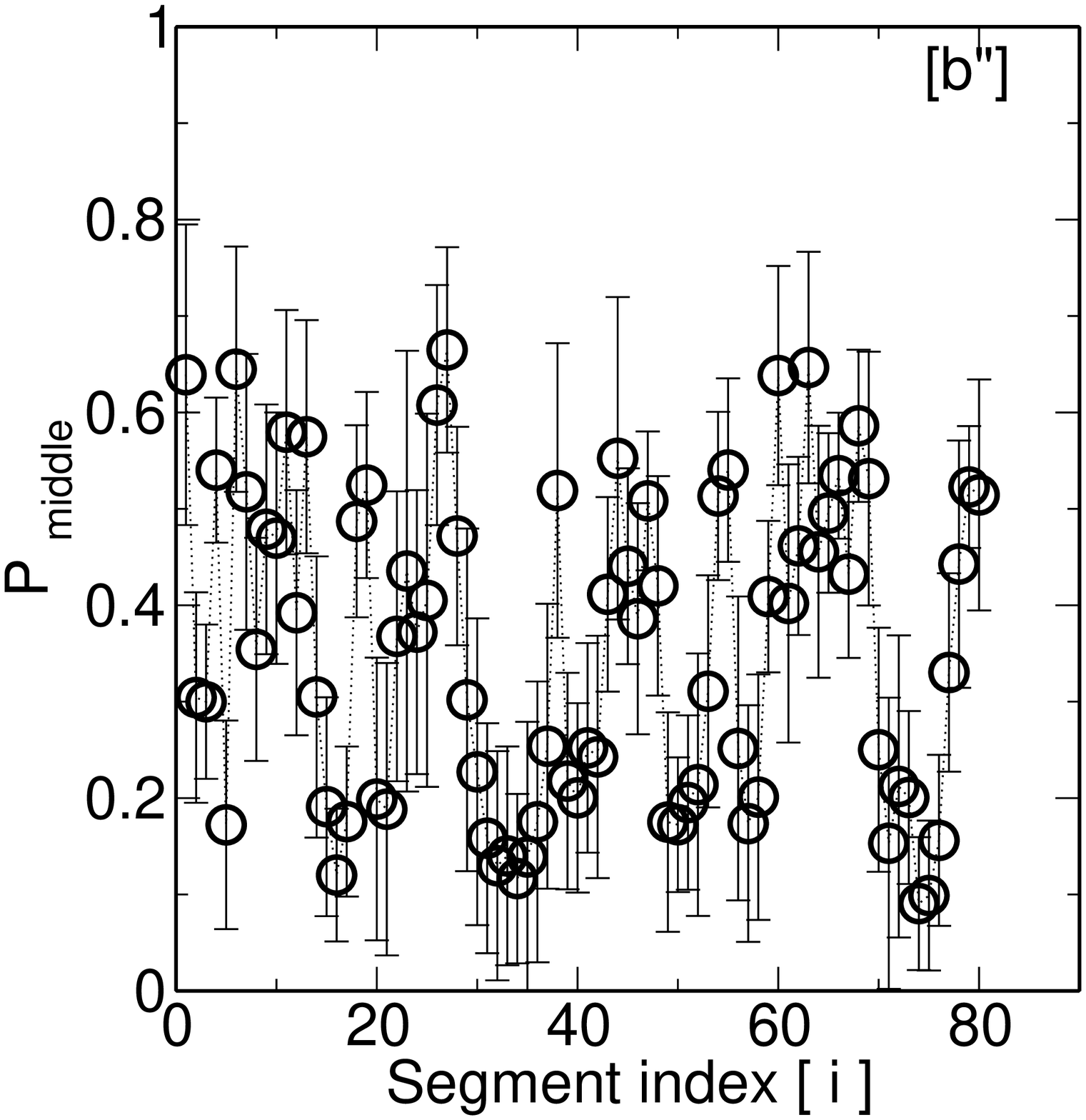}
\hfill
\includegraphics[width=0.49\columnwidth]{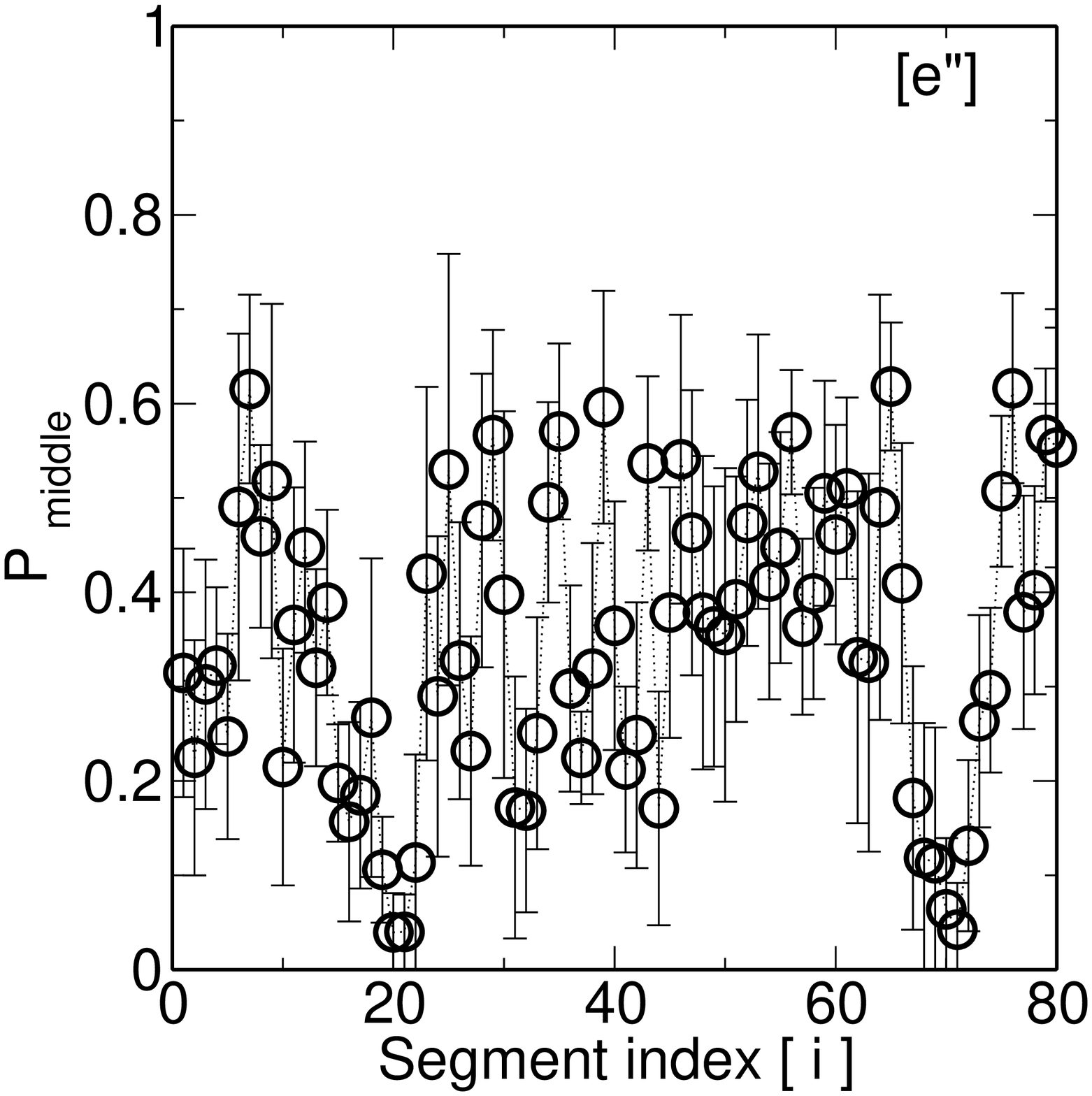} \\
\hfill
\vskip0.3cm
\includegraphics[width=0.49\columnwidth]{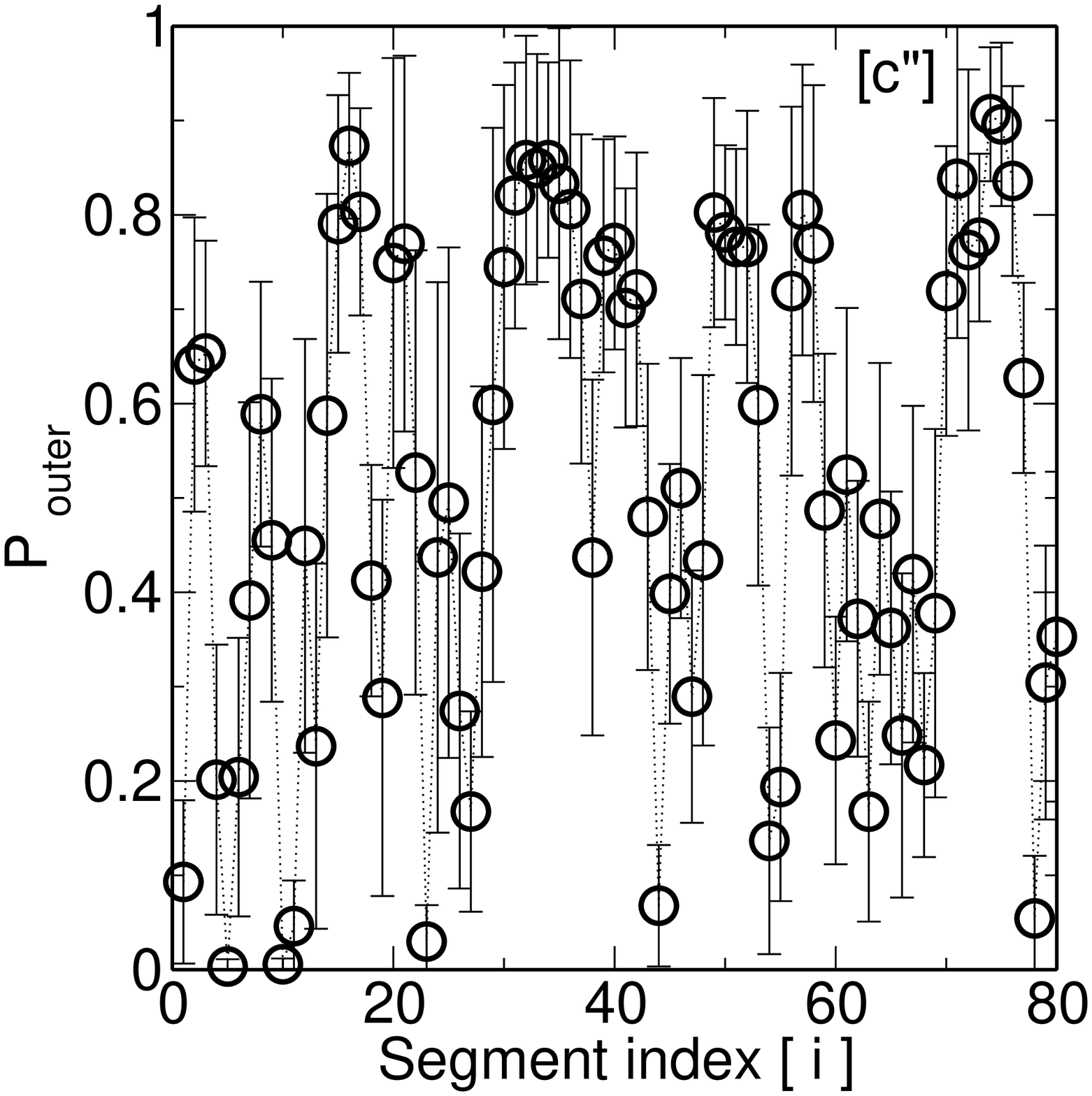}
\hfill
\includegraphics[width=0.49\columnwidth]{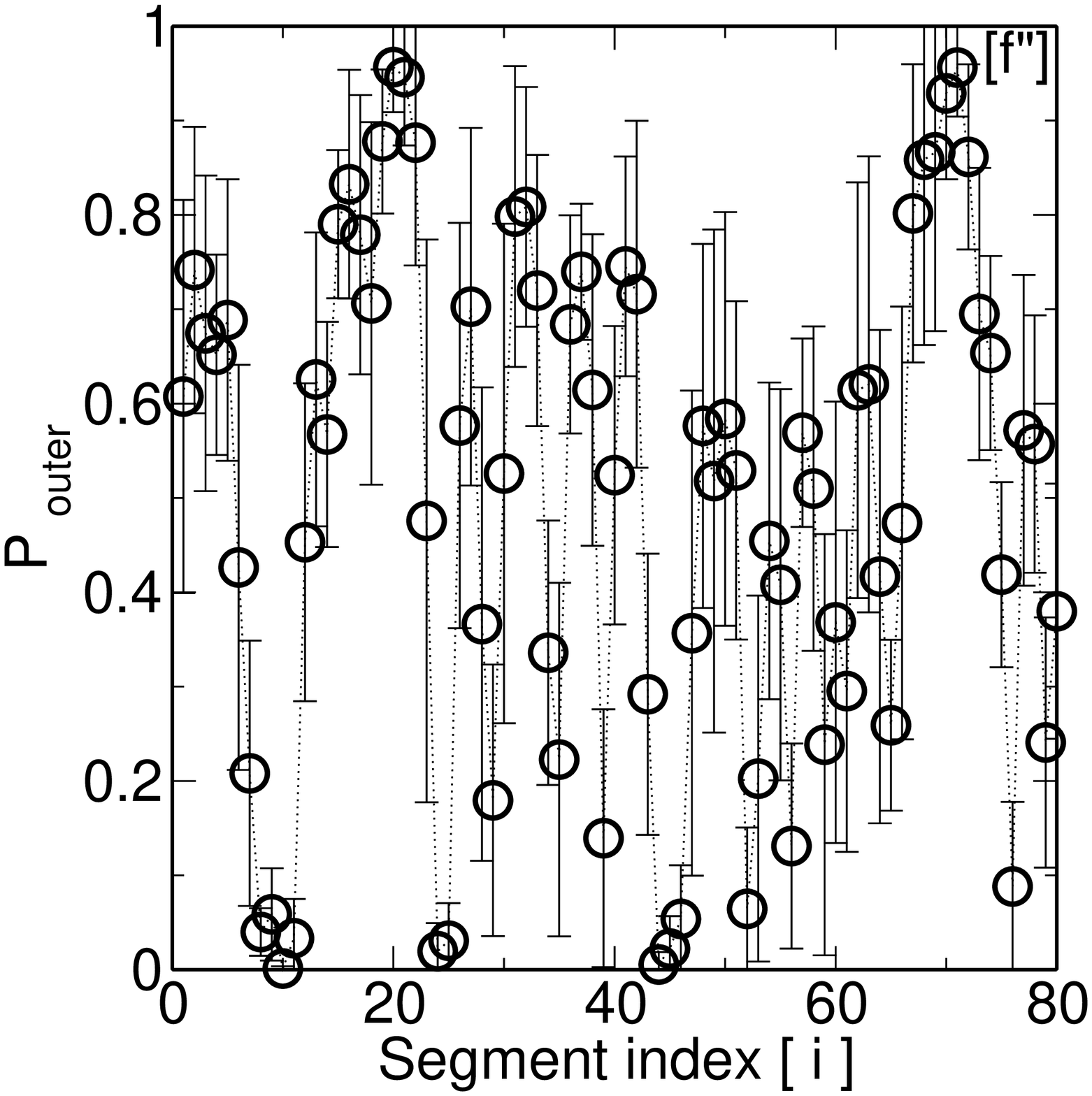} \\
\hfill
\vskip0.3cm
\caption{\label{Suppfigprobcl2}
Data for bacteria {\em E.Coli} with no. of CLs = BC-1 :
Subplots (a"), (b") and (c") show the probabilities of center of mass of polymer segments  to be found in the inner, middle and outer region of polymer globule.  The x-axis is segment index.
Each valuue of probability is the average over 9 independent initial conditions.
Eroor bar shows the standard deviation.
Subplots (d"), (e"), (f") are for RC-1. Each segment has 58 monomers, the dna-polymer has around 80 segments.
}
\end{figure}

\newpage
\clearpage

\section{ COLOR-MAPS for positional correlations}
\begin{figure}[!hbt]
\includegraphics[width=0.49\columnwidth]{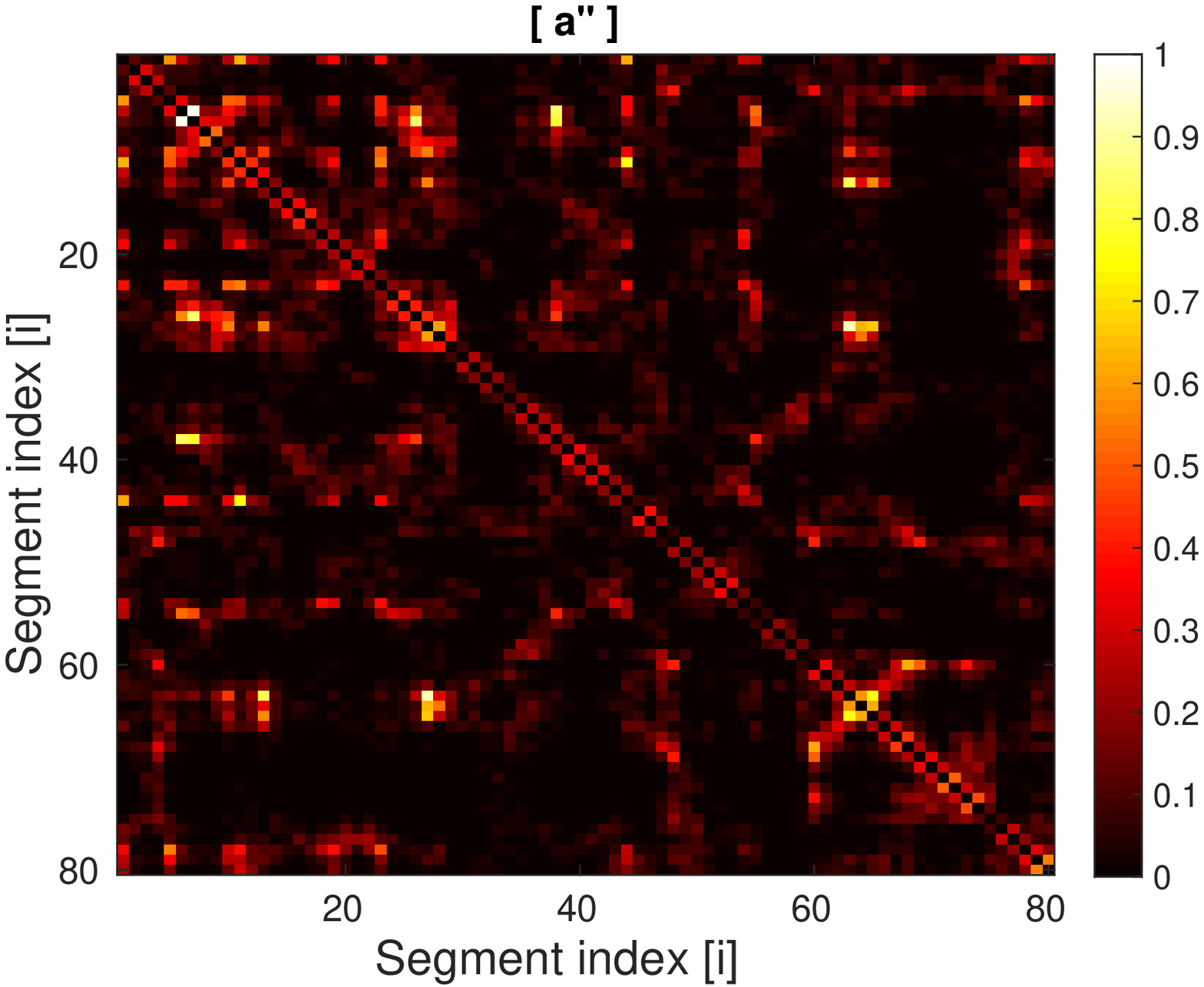}
\includegraphics[width=0.49\columnwidth]{bio_ec_i7_cp1_cm_dom_cor1_l1.eps} \\
\vskip0.05cm
\includegraphics[width=0.49\columnwidth]{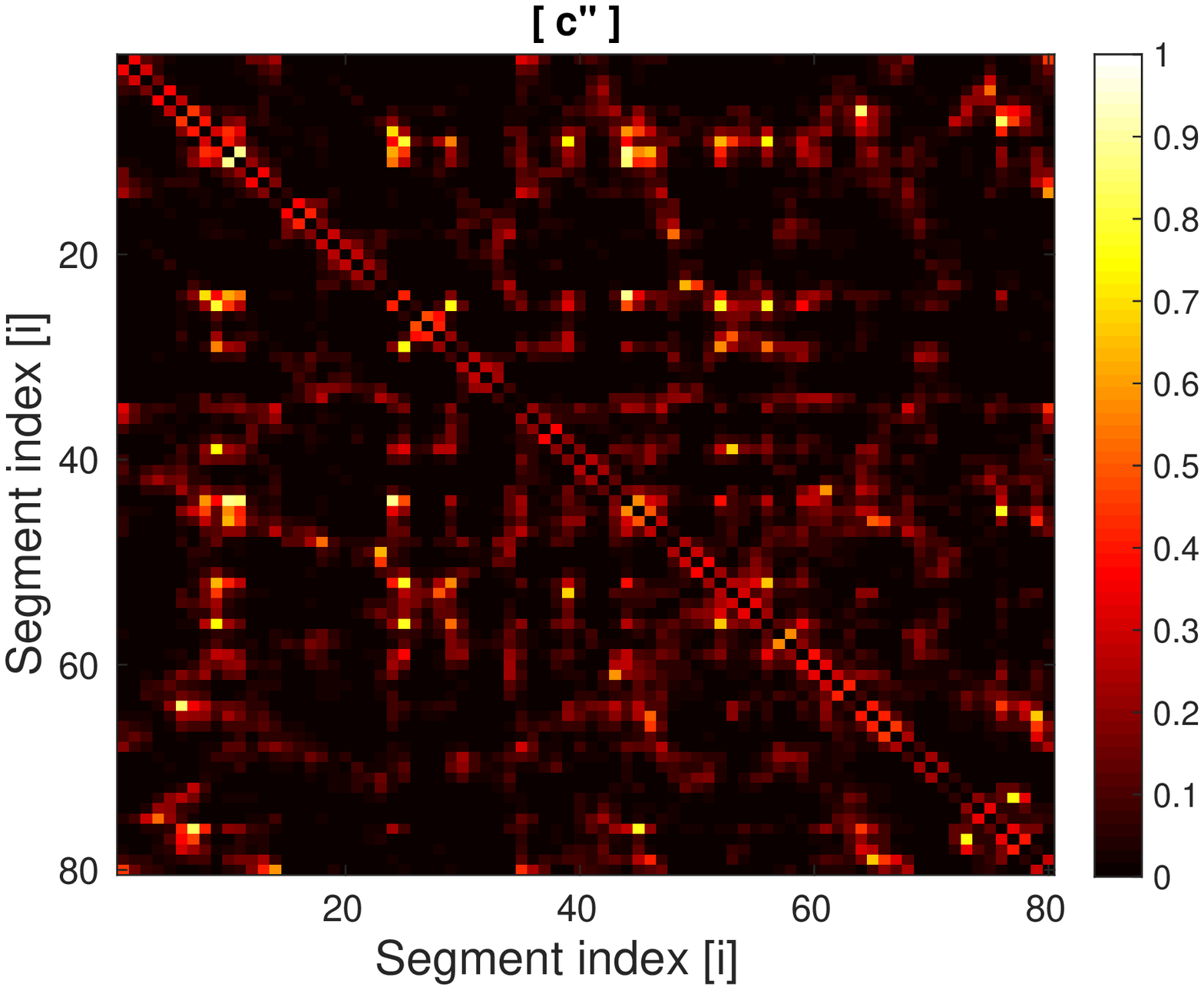}
\includegraphics[width=0.49\columnwidth]{r_bio_ec_i7_cp1_cm_dom_cor1_l1.eps} \\
\caption{\label{Suppfigprobcl4}
Colormaps show the probability of center of mass of the segments (i)
(58 monomer each) to be found within a distance $5\sigma$ of other segment's center of mass. The colormaps (a"), (b") and (c") and (d") are from two additional independent runs for BC-1 and RC-1, respectively.
}
\end{figure}

\begin{figure}[!hbt]
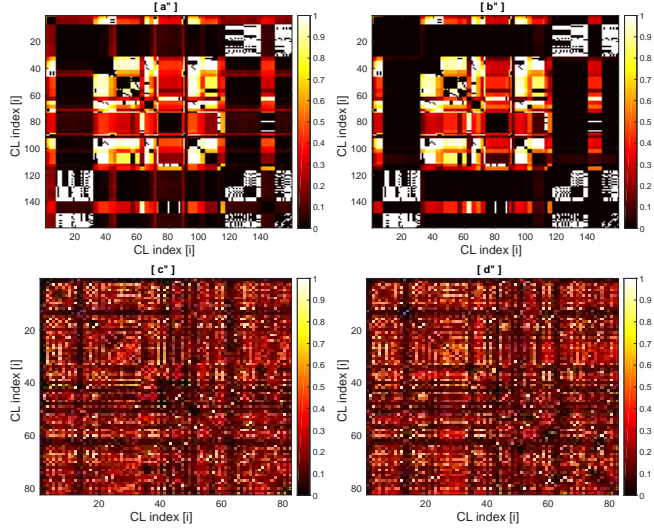

\includegraphics[width=0.49\columnwidth]{bio_ec_i5_cp4_dom_corr_l1.eps}
\includegraphics[width=0.49\columnwidth]{bio_ec_i7_cp4_dom_corr_l1.eps} \\
\vskip0.05cm
\includegraphics[width=0.49\columnwidth]{r_bio_ec_i5_cp4_dom_corr_l1.eps}
\includegraphics[width=0.49\columnwidth]{r_bio_ec_i7_cp4_dom_corr_l1.eps} \\
\caption{\label{Suppfigprobcl5}
Colormaps show the probability to find CLs $i$ and CLs $j$ within a distance of $5\sigma$.
The colormaps (a"), (b") and (c") and (d") are from two additional independent runs for BC-2 and RC-2, respectively.
}
\end{figure}

\begin{figure}[!hbt]
\includegraphics[width=0.49\columnwidth]{bio_ec_i5_cp4_cm_dom_cor1_l1.eps}
\includegraphics[width=0.49\columnwidth]{bio_ec_i7_cp4_cm_dom_cor1_l1.eps} \\
\vskip0.05cm
\includegraphics[width=0.49\columnwidth]{r_bio_ec_i5_cp4_cm_dom_cor1_l1.eps}
\includegraphics[width=0.49\columnwidth]{r_bio_ec_i7_cp4_cm_dom_cor1_l1.eps} \\
\hfill
\caption{\label{Suppfigprobcl6}
Colormaps show the probability of center of mass of the segments (i)
(58 monomer each) to be found within a distance $5\sigma$ of other segment's center of mass. The colormaps (a"), (b") and (c") and (d") are from two additional independent runs for BC-2 and RC-2, respectively.
}
\end{figure}

\clearpage

\section{ COLOR-MAPS for angular correlations}
\begin{figure}[!hbt]
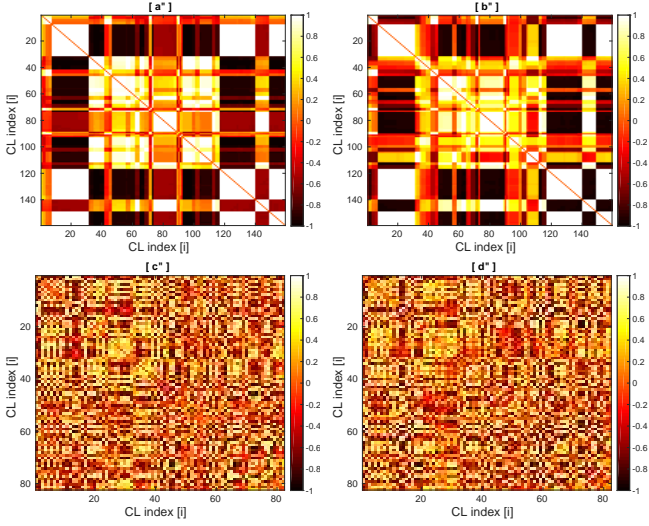

\includegraphics[width=0.49\columnwidth]{bio_ec_i5_cp4_near_contacts_l1.eps}
\includegraphics[width=0.49\columnwidth]{bio_ec_i7_cp4_near_contacts_l1.eps} \\
\vskip0.05cm
\includegraphics[width=0.49\columnwidth]{r_bio_ec_i5_cp4_near_contacts_l1.eps}
\includegraphics[width=0.49\columnwidth]{r_bio_ec_i7_cp4_near_contacts_l1.eps} \\
\caption{\label{Suppfigprobcl8}
Colormaps show the angular positions of different CL with respect to each other. The colormaps (a"), (b") and (c") and (d") are from two additional independent runs for BC-2 and RC-2, respectively.
}
\end{figure}

\begin{figure}[!hbt]
\includegraphics[width=0.49\columnwidth]{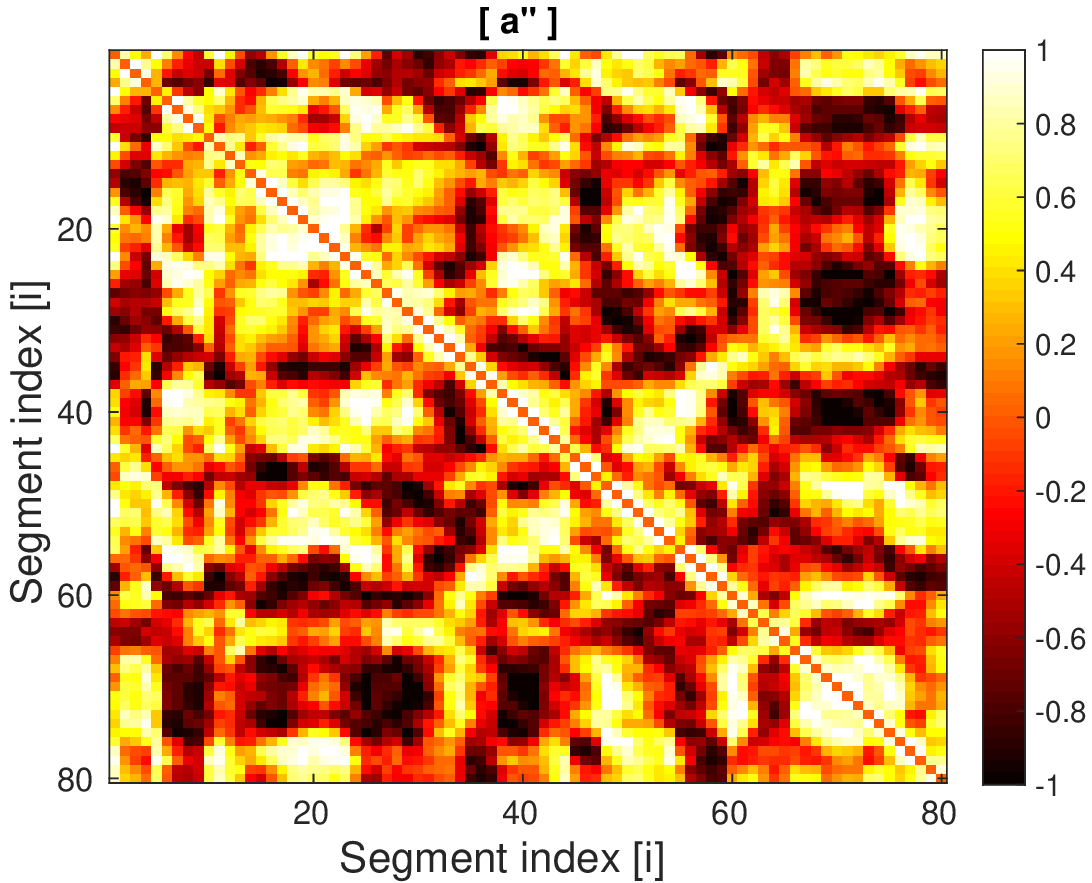}
\includegraphics[width=0.49\columnwidth]{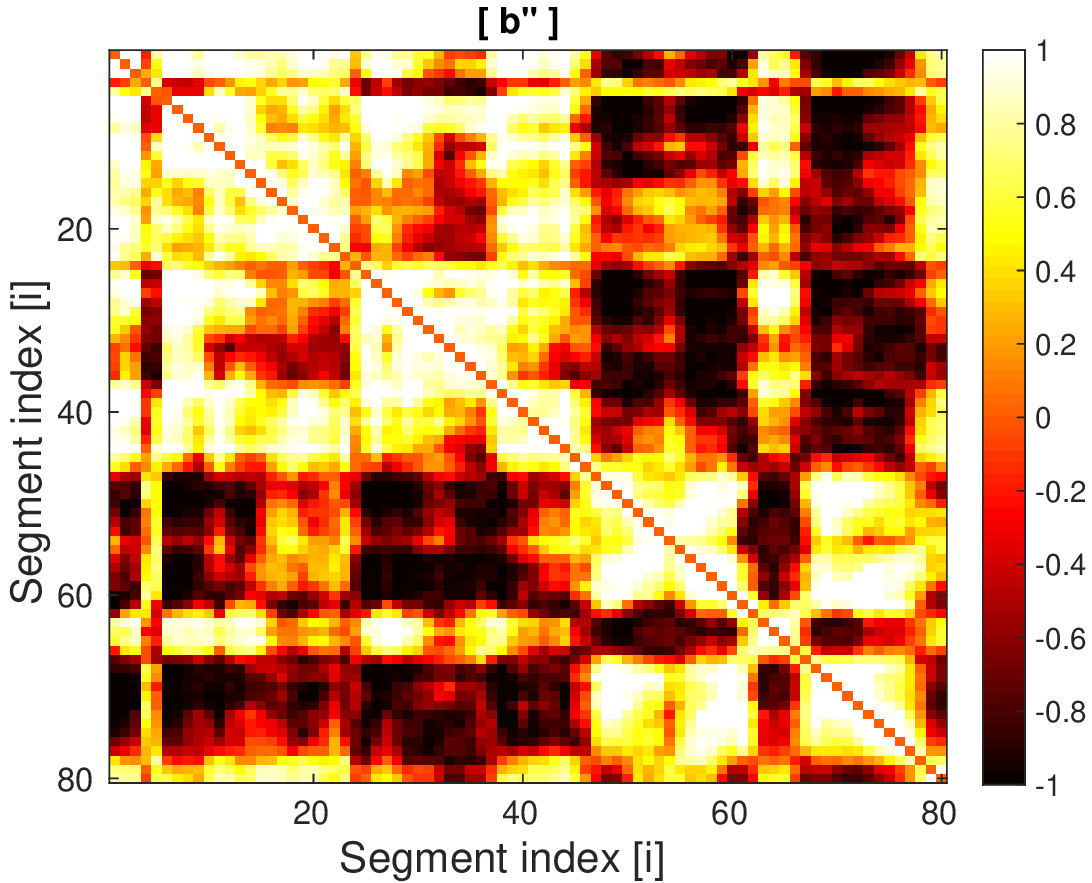} \\
\vskip0.05cm
\includegraphics[width=0.49\columnwidth]{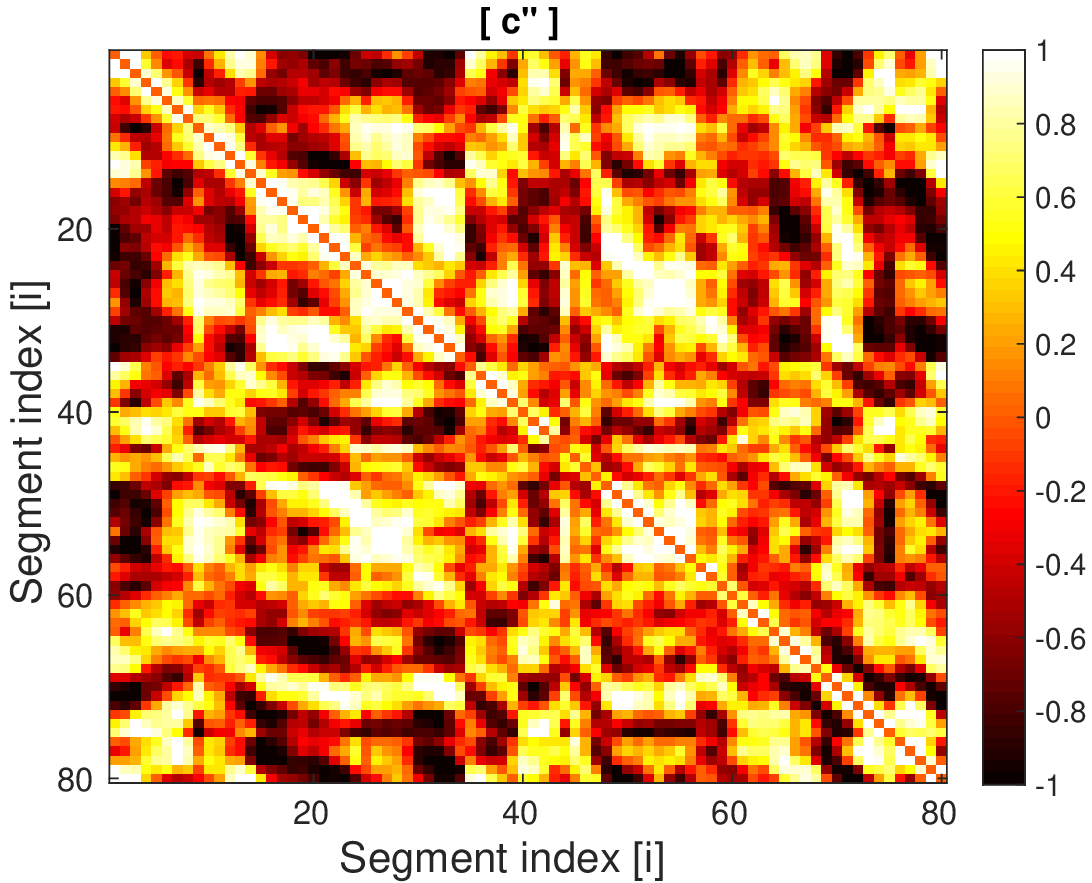}
\includegraphics[width=0.49\columnwidth]{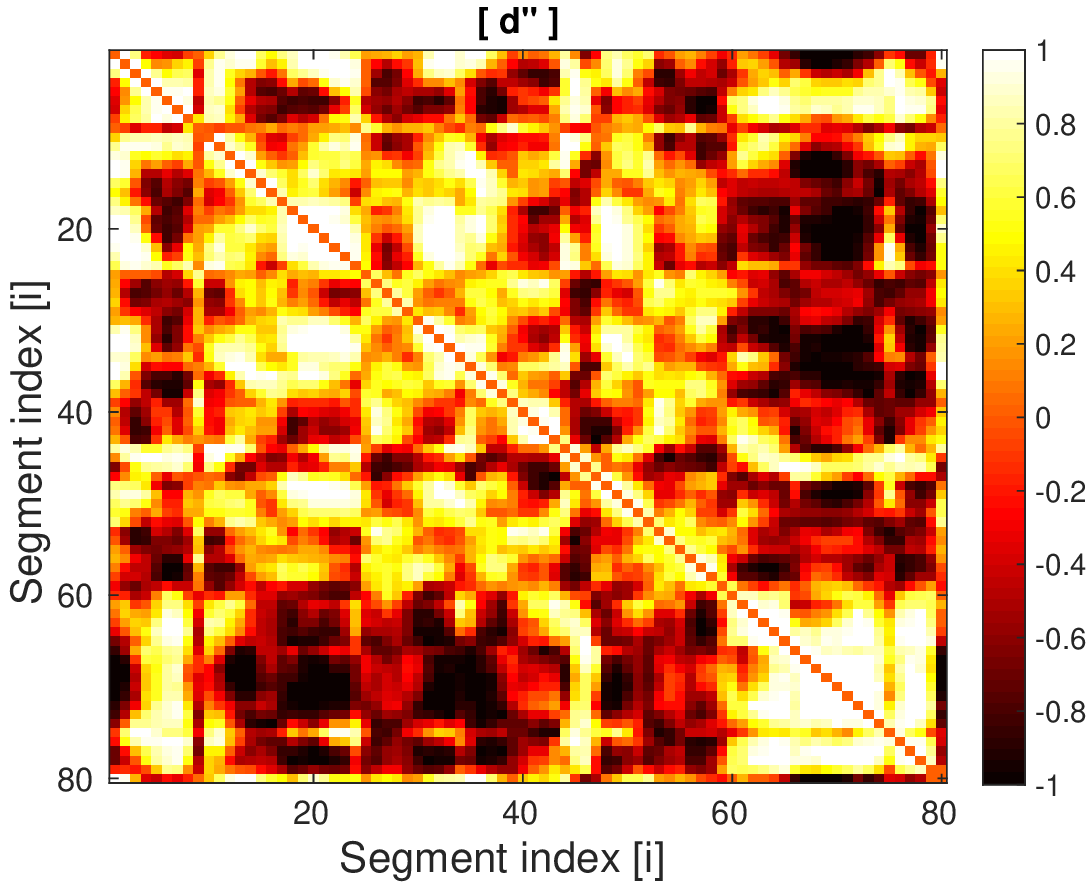} \\
\caption{\label{Suppfigprobcl9}
Colormaps show the angular positions of center of mass of the different segments with respect to each other. The colormaps (a"), (b") and (c") and (d") are from two additional independent runs for BC-1 and RC-1, respectively.
}
\end{figure}
\begin{figure}[!hbt]
\includegraphics[width=0.49\columnwidth]{bio_ec_i5_cp4_near_contacts_seqs_l1.eps}
\includegraphics[width=0.49\columnwidth]{bio_ec_i7_cp4_near_contacts_seqs_l1.eps} \\
\vskip0.05cm
\includegraphics[width=0.49\columnwidth]{r_bio_ec_i5_cp4_near_contacts_seqs_l1.eps}
\includegraphics[width=0.49\columnwidth]{r_bio_ec_i7_cp4_near_contacts_seqs_l1.eps} \\
\caption{\label{Suppfigprobcl10}
Colormaps show the angular positions of center of mass of the different segments with respect to each other. The colormaps (a"), (b") and (c") and (d") are from two additional independent runs for BC-2 and RC-2, respectively.
}
\end{figure}
\pagebreak
\section{Snapshot from simulation}
\begin{figure}[!hbt]
\includegraphics[width=0.70\columnwidth]{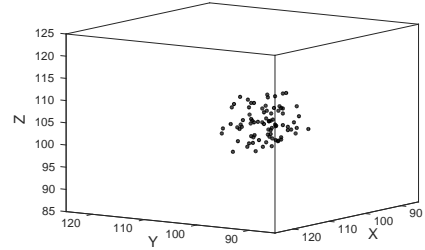} \\
\caption{\label{Suppsnapshot}
Representative snaphots from our simulations of the positions of CLs for RC-2.
}
\end{figure}
\clearpage
\end{document}